\documentclass[12pt,preprint]{aastex}

\shorttitle{LISM Dynamics}
\shortauthors{Redfield \& Linsky}

\begin{document}

\newcommand{\php}[0]{\phantom{--}}
\newcommand{\kms}[0]{km~s$^{-1}$}

\title{THE STRUCTURE OF THE LOCAL INTERSTELLAR MEDIUM IV: DYNAMICS,
MORPHOLOGY, PHYSICAL PROPERTIES, AND IMPLICATIONS OF CLOUD-CLOUD
INTERACTIONS\footnote{Based on observations made with the NASA/ESA
Hubble Space Telescope, obtained from the Data Archive at the Space
Telescope Science Institute, which is operated by the Association of
Universities for Research in Astronomy, Inc., under NASA contract NAS
5-26555.  These observations are associated with programs \#9525 and
\#10236.}}

\author{Seth Redfield\altaffilmark{2,3} and Jeffrey
L. Linsky\altaffilmark{4}}

\altaffiltext{2}{Department of Astronomy and McDonald Observatory,
University of Texas, Austin, TX 78712-0259; {\tt
sredfield@astro.as.utexas.edu}}
\altaffiltext{3}{Hubble Fellow.}
\altaffiltext{4}{JILA, University of Colorado and NIST, 
Boulder, CO 80309-0440; {\tt jlinsky@jila.colorado.edu}}

\begin{abstract}

We present an empirical dynamical model of the local interstellar
medium based on 270 radial-velocity measurements for 157 sight lines
toward nearby stars.  Physical-parameter measurements (i.e.,
temperature, turbulent velocity, depletions) are available for 90
components, or one-third of the sample, enabling initial
characterizations of the physical properties of LISM clouds.  The
model includes 15 warm clouds located within 15~pc of the Sun, each
with a different velocity vector.  We derive projected morphologies of
all clouds and estimate the volume filling factor of warm partially
ionized material in the LISM to be between $\sim$5.5\% and 19\%.
Relative velocities of potentially interacting clouds are often
supersonic, consistent with heating, turbulent, and metal-depletion
properties. Cloud-cloud collisions may be responsible for the
filamentary morphologies found in $\sim$1/3 of LISM clouds, the
distribution of clouds along the boundaries of the two nearest clouds
(LIC and G), the detailed shape and heating of the Mic Cloud, the
location of nearby radio scintillation screens, and the location of a
LISM cold cloud. Contrary to previous claims, the Sun appears to be
located in the transition zone between the LIC and G Clouds.

\end{abstract}


\section{INTRODUCTION}

In their now classical theoretical models for the interstellar medium
(ISM), \citet*{field69}, \citet{mckee77}, and
\citet{wolfire95a,wolfire95b} assumed the ISM to be in thermal and
steady-state equilibrium. In these models, three stable regimes
co-exist in pressure equilibrium: the cold neutral medium (CNM) with
temperature $T\geq 50$~K, the warm neutral (WNM) or ionized medium
(WIM) with $T \sim 8000$~K, and the hot ionized medium (HIM) with
$T\sim 1,000,000$~K.  These models include heating by ultraviolet (UV)
photons on grains and polycyclic aromatic hydrocarbon (PAH) molecules
and cooling by various forbidden lines and the hydrogen Lyman-$\alpha$
line. These models do not include gas flows or predict the expected
sizes of the various components. Given the low density of interstellar
gas and the presence of supernovae and strong stellar winds, one
expects that the gas will be far out of thermal and pressure
equilibrium and be highly dynamic.  Various reviews
\citep[e.g.,][]{cox05,mccray79} discuss these issues and highlight the
complexity of the ISM.

The Local Bubble (LB) is a region of low-density presumably hot gas
extending in all directions to hydrogen column densities $\log
N($\ion{H}{1}$) = 19.3$ \citep{lallement03}. Its shape is determined
by the onset of significant column density of \ion{Na}{1}, indicative
of a cold gas shell surrounding the LB.  Although the LB is irregular
in shape, it extends to roughly 100~pc from the Sun. For our purposes,
we consider the local interstellar medium (LISM) to consist of the
hot, warm, and cold gas located inside the LB.  The LISM gas has been
shaped by the supernovae explosions and winds of massive stars in the
Scorpio-Centaurus Association and ionized and heated by radiation from
hot stars and the Galactic UV background \citep[e.g.,][]{berghofer02},
and so should provide a useful test of interstellar gas properties in
our Galaxy and the assumptions that underlie theoretical models of the
ISM.  We can now study the LISM in detail because the ground-level
transitions of many neutral and ionized atoms present in the UV could
be observed with the high-resolution spectrographs on the {\it Hubble
Space Telescope} ({\it HST}).  With resolutions as high as $R \equiv
\lambda/\Delta\lambda = 100,000$ ($\Delta v = 3.0$ km s$^{-1}$), both
the Goddard High Resolution Spectrograph (GHRS) and the Space
Telescope Imaging Spectrograph (STIS) have obtained stellar spectra
containing numerous interstellar absorption lines. As described below,
these ultraviolet spectra, together with ground-based spectra in the
\ion{Ca}{2} H and K resonance lines, provide critical data for
sampling the kinematic and physical properties of warm interstellar
gas along 157 lines of sight.

The dynamical structure of the LISM has a direct influence on the
structure of the heliosphere around our solar system and astrospheres
surrounding other nearby stars.  The extent of the heliosphere
(astrosphere) is determined by the balance of momentum ($\rho v^2$)
between the outward moving solar (stellar) wind and the surrounding
interstellar medium.  Long-term variations in the solar wind strength
are not well known, but observations of astrospheres around young
solar analogs provide clues as to what kind of wind the Sun had in its
distant past.  The solar wind 3.5 billions years ago may have been
$\sim$35-fold stronger than it is today \citep{wood05let}.  In
contrast, density variations spanning 6 orders of magnitude are
commonly observed throughout the general ISM.  However, variations in
the dynamical properties of the surrounding ISM can also cause
significant variations in the structure of the heliosphere even
between clouds with little-to-no density variation.  Reviews of
heliospheric modeling include \citet{zank99} and \citet{baranov90},
and the detection of astrospheres around nearby stars is reviewed by
\citet{wood04}.  \citet{muller06} explore the response of heliospheric
models to various interstellar environments that exist in the LISM.
Significant heliospheric (or astrospheric) compression can impact
planetary albedos, atmospheric chemistry, and biological mutation
rates.  Reviews of the implications of heliospheric variability are
discussed by \citet{redfield06} and \citet{frisch06book}.

\citet{crutcher82} first noted that interstellar gas in the LISM flows
in roughly the same direction away from the center of the
Scorpio-Centaurus Association.  \citet{lallement92} then showed that
the flow of interstellar gas in the direction away from the Galactic
Center is consistent with a vector that differs somewhat from that of
the gas in the Galactic Center direction. They coined the term AG
Cloud for the former and G Cloud for the latter. The AG Cloud is now
called the Local Interstellar Cloud (LIC), since the Sun has been
presumed to be located just inside the LIC on the basis that the
velocity of neutral helium flowing into the heliosphere is consistent
with the LIC flow vector \citep{witte93}.  \citet{lallement95} argued
that the LISM has a complex velocity structure with at least seven
clouds located within 12 pc. Using a larger data set,
\citet*{frisch02} were able to identify 7 clouds in the LISM on the
basis of their kinematics. Using \ion{Na}{1} spectra of stars in the
Galactic anti-center hemisphere, \citet{genova03} identified 8 clouds
of presumably cold gas lying beyond 50~pc of the Sun with velocity
vectors very different from those identified in this paper. These
clouds may lie at the edge of the Local Bubble or beyond.

The present work expands on the earlier studies in two ways. First, we
analyze a much larger data set consisting of 270 individual velocity
components along 157 lines of sight through the LISM. Each velocity
component provides kinematical information (i.e., radial velocity) of
a parcel of gas that we can analyze together with other velocity
components to identify velocity vectors and morphologies of different
gas clouds. Second, high-resolution GHRS and STIS spectra for 55 of
these velocity components allow us to measure the widths of absorption
lines from atoms and ions of different atomic weight to determine the
temperature and turbulent velocity \citep{redfield04tt}, and for 65 of
the velocity components \ion{D}{1} \citep[an excellent proxy for
\ion{H}{1} in the LISM;][]{linsky06} is observed together with other
ions which can be used to calculate metal depletions
\citep[cf.,][]{redfield04sw}.  These measurements allow us to
determine some of the physical properties of the clouds in addition to
their kinematical and morphological properties.

In this paper, we use the term ``cloud'' to refer to a contiguous
parcel of interstellar gas inside the LISM with homogeneous
kinematical and physical properties. We determine the morphology of 15
such clouds (\S~\ref{sec:coll} and \S~\ref{sec:morph}) by assuming that
the interstellar gas flow inside each cloud is coherent and that the
clouds have sharp edges. An upper limit to the distance of each cloud
from the Sun is the distance to the nearest star whose spectrum shows
a velocity component consistent with the cloud's velocity vector. We
identify locations on the sky of possible cloud-cloud interactions and
check whether these locations are consistent with other phenomena
(\S~\ref{sec:collide}).  In subsequent papers we will describe how
these cloud-cloud interactions could explain radio scintillation
screens and the locations of cold clouds.  The results may be used to
test assumptions of sharp cloud boundaries, departures from coherent
flow, and search for evidence of shear, cloud rotation and expansion,
and alignment with magnetic fields \citep{cox03}.

\section{DATA ANALYSIS}

\subsection{\it Construction of LISM Observational Database}

The ability to assign large-scale dynamical flows to observed
projected radial velocities requires an extensive and densely sampled
observational database.  We have compiled the most comprehensive
high-spectral-resolution observational database from absorption line
transitions in the UV observed by {\it HST} and from the \ion{Ca}{2}
optical transition observed from the ground.  Transitions in the UV
are the most sensitive to the warm partially ionized clouds that
populate the LISM, while \ion{Ca}{2} is the transition most sensitive
to warm gas in the visible \citep{redfield06}.  Only six sight lines
have both UV and \ion{Ca}{2} interstellar absorption detections:
$\alpha$ Aql, $\alpha$ PsA, $\delta$ Cas, $\eta$ UMa, $\alpha$ Gru, and
$\epsilon$ Gru \citep[see references in][]{redfield02}.  Not all UV
velocity components are detected in \ion{Ca}{2}, but those that are
agree in observed velocity very well.  The absorption observed toward
$\alpha$ Gru provides the only example where the ultra-high-resolution
\ion{Ca}{2} observations resolve multiple components from absorption
identified as a single UV component \citep{crawford95,redfield02}.
About 26\% of the sight lines are observed in several ions.  Multiple
ion observations of the LISM along the same line of sight provide
independent-projected velocity measurements and provide additional
diagnostics of the physical properties of the material (e.g.,
depletion, temperature, ionization, etc).  Even though only 1--3
velocity components are identified per sight line, moderately high
spectral resolution is required to adequately separate and resolve
individual absorbers with similar-projected velocities.

Our database, which includes 270 individual velocity components along
157 sight lines, is derived from: (1) the complete high-resolution UV
database of {\it HST} observations of LISM absorption toward stars
within 100\,pc \citep[see][and references
therein]{redfield02,redfield04sw}, which represents 55\% of the
velocity components in our sample, (2) the high-resolution \ion{Ca}{2}
absorption measurements toward stars within 100\,pc
(\citealt{frisch02}; and references within \citealt{redfield02}),
which represent 32\% of the components in our sample, and (3) the
moderate-resolution UV database of {\it HST} observations of LISM
absorption toward stars within 100\,pc (\citealt*{wood96};
\citealt{wood00,wood05sup}), representing the remaining 13\% of the
components in our sample.  Physical-parameter measurements (i.e.,
temperature, turbulent velocity, depletions) are available for 90
components, or one-third of the sample.

All absorption is assumed to be caused by the LISM.  Contamination of
the absorption database by absorption caused by edge-on circumstellar
disks, although possible, is highly unlikely.  Not only are nearby
stars with circumstellar material rare, the requirement of an edge-on
orientation further limits the likelihood of observing such systems.
Only a handful of such systems, which show circumstellar absorption,
are known within 100\,pc.  The most prominent example is $\beta$ Pic,
whose spectrum shows stable absorption at the stellar rest frame and
variable absorption components, both due to circumstellar gas
\citep{hobbs85,brandeker04}, in addition to a LISM component, resolved
from the circumstellar material only in the heaviest (i.e., narrowest)
ions, such as \ion{Fe}{2} \citep{lallement95,redfield02}.  Only two
other stars in the LISM database have known edge-on circumstellar
disks: (1) $\beta$ Car \citep{lagrangehenri90} in which only the
\ion{Na}{1} absorption feature was observed to vary, whereas the
\ion{Ca}{2} absorption is relatively steady, and match the UV
observations \citep*{redfield07lismdd,redfield02}, and (2) AU Mic
\citep*{kalas04} which shows no circumstellar absorption in H$_2$
\citep{roberge05,france07} or other UV lines, including Lyman-$\alpha$
\citep{wood05sup} and the single UV observation does not allow for any
constraint on the constancy of the observed absorption
\citep{redfield02}.  Therefore, we have retained these absorption
features in the LISM database, but their removal does not
significantly change the velocity vectors determined for the clouds
for which they are members.

We focus on using \ion{Ca}{2} because it primarily traces warm LISM
gas, whereas \ion{Na}{1} primarily traces cold gas not common in the
LISM.  However, \ion{Na}{1} is occasionally detected in absorption
toward nearby stars (e.g., \citealt*{blades80}; \citealt{vallerga93};
\citealt*{welty94}; \citealt{welsh94}).  Approximately a third (49/157
= 31\%) of the sight lines in our sample also have \ion{Na}{1}
observations, of which LISM absorption is detected along only 16 lines
of sight.  About one-half of the \ion{Na}{1} detections are toward
stars within 50 pc, so although relatively uncommon within 100 pc,
\ion{Na}{1} absorption is not significantly dominated by the cold gas
located near the edge of the Local Bubble \citep{lallement03}.
Therefore, a \ion{Ca}{2} absorption component that is associated with
a \ion{Na}{1} component at the same velocity is not necessarily
indicative of distant gas.  Practically all LISM \ion{Na}{1}
absorption components have companion absorption components in
\ion{Ca}{2} at the same velocity, which indicates that the cold gas
detected by the \ion{Na}{1} absorption is physically associated with
the warm gas detected by the \ion{Ca}{2} absorption, and not separate
clouds at coincident velocities.  It appears that much of the cold gas
in the LISM is associated directly with warmer gas and these
structures share a common velocity vector.

Our combined database samples the sky unevenly because the sight lines
were often selected to observe UV bright stars or for purposes other
than measuring LISM absorption.  Although this is the densest-sampled
UV/optical database of LISM absorption to date, there remain
significant regions of poor sampling in both space and distance.  The
distribution of sight lines is shown for all of our dynamical cloud
structures in Figures~\ref{fig:lic}--\ref{fig:cet}.  The various
symbols used to signify the sight lines, as well as a discussion of
the morphology of the derived clouds, is provided in
Section~\ref{sec:morph}.  The median angular distance from one sight
line to its nearest neighbor is 6.6 degrees, ranging from observations
of binary stars with angular separations of $\sim$10 arcsecs (e.g.,
$\alpha$ Cen A and B, $\alpha$ CMa A and B), to the poorly sampled
region near $l=137^{\circ}$ and $b=49^{\circ}$ where the maximum
nearest neighbor separation is 21.5 degrees.  Areas of poor sampling
limit our ability to detect dynamical cloud structures, as many sight
lines through the same collection of gas are required to determine an
accurate velocity vector.  Poor sampling also limits our ability to
estimate distances to structures in these regions.

\subsection{\it Criteria for Identifying an Interstellar Cloud}

\begin{figure}
\includegraphics[angle=90,width=6.9in]{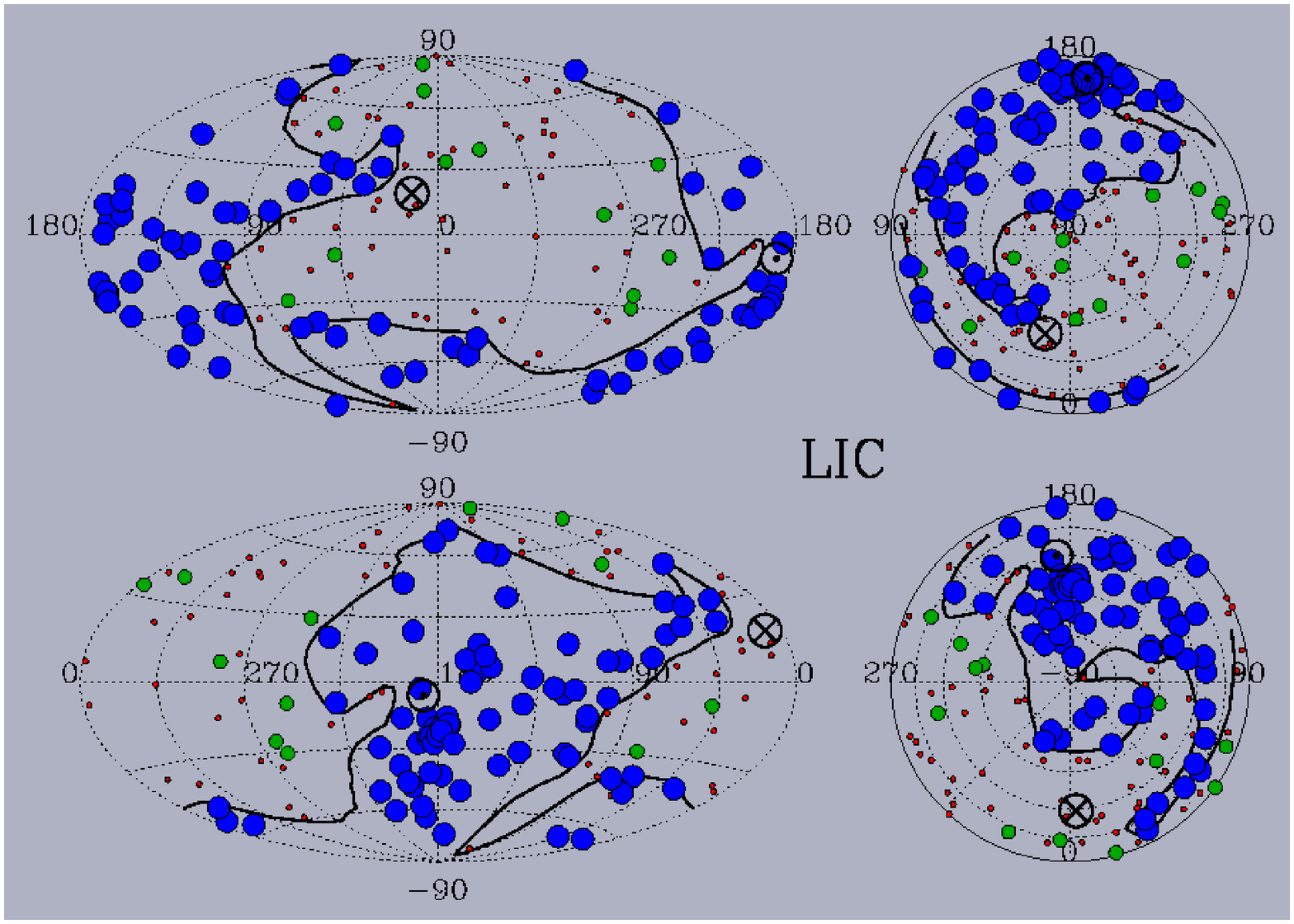}
\caption{Four projections of the LIC boundaries in Galactic
coordinates.  All sight lines used in our analysis are displayed.  The
large blue symbols indicate sight lines used to calculate the LIC
velocity vector.  The medium-sized green symbols indicate sight lines
with projected velocities that are consistent with the estimated
vector, but are considered part of another cloud (see \S~2.2), while
the small red symbols indicate lines of sight that are inconsistent
with the calculated velocity vector.  The boundaries of the LIC are
drawn to encompass all consistent sight lines (i.e., blue symbols),
while avoiding all other lines of sight (i.e., red and green symbols)
.  The upwind heliocentric direction of the velocity vector is
indicated by the $\otimes$ symbol, while the downwind heliocentric
direction is indicated by the $\odot$ symbol.  The four projections
from upper left and moving counter-clockwise are: a Hammer projection
of Galactic coordinates with the Galactic center in the middle, a
Hammer projection of Galactic coordinates with the Galactic
anti-center in the middle, a Lambert projection from the south
Galactic pole, and a Lambert projection from the north Galactic
pole. \label{fig:lic}}
\end{figure}

\begin{figure}
\includegraphics[angle=90,width=6.9in]{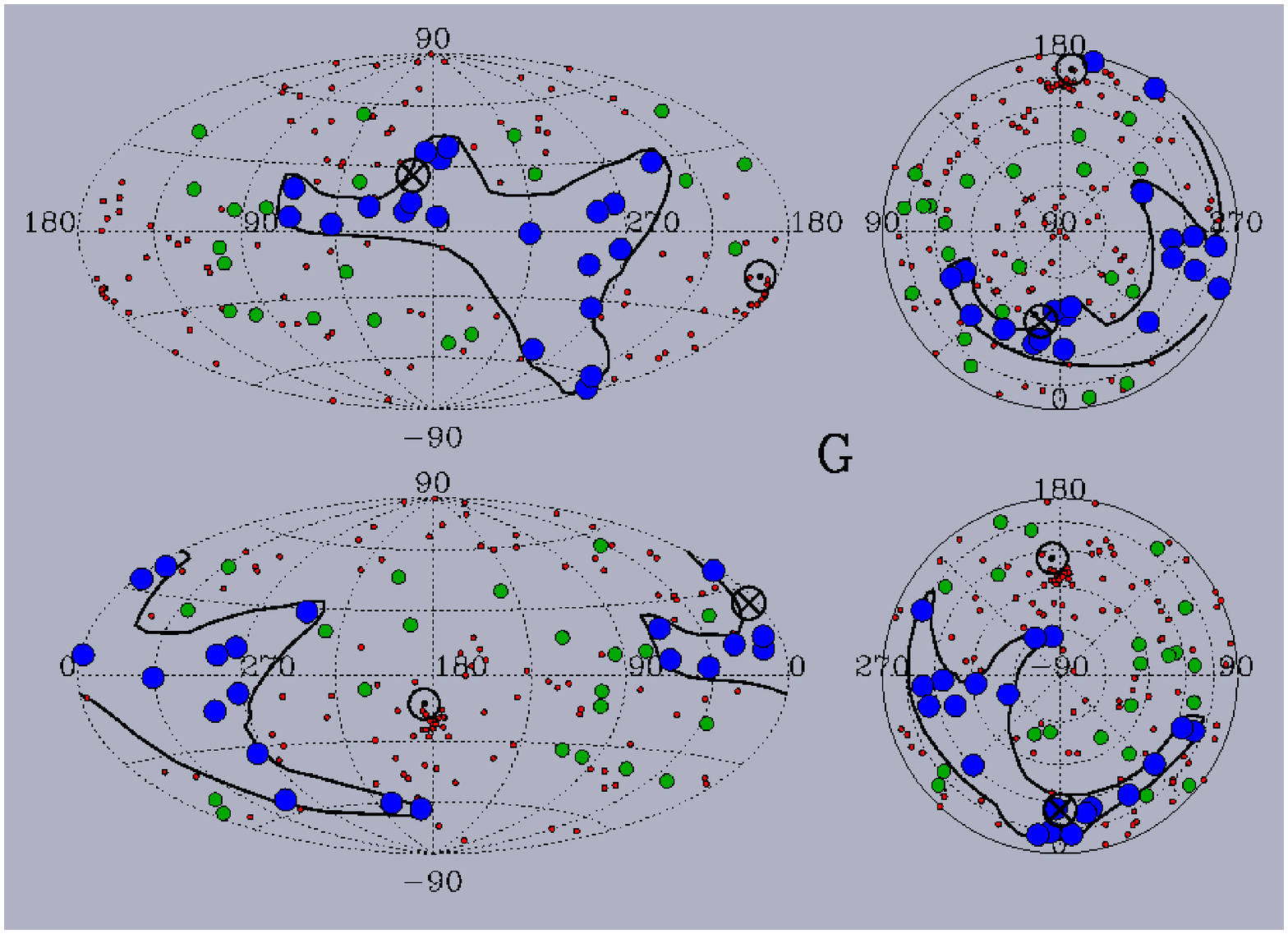}
\caption{Same as Figure~\ref{fig:lic} but for the G Cloud. \label{fig:g}}
\end{figure}

\begin{figure}
\includegraphics[angle=90,width=6.9in]{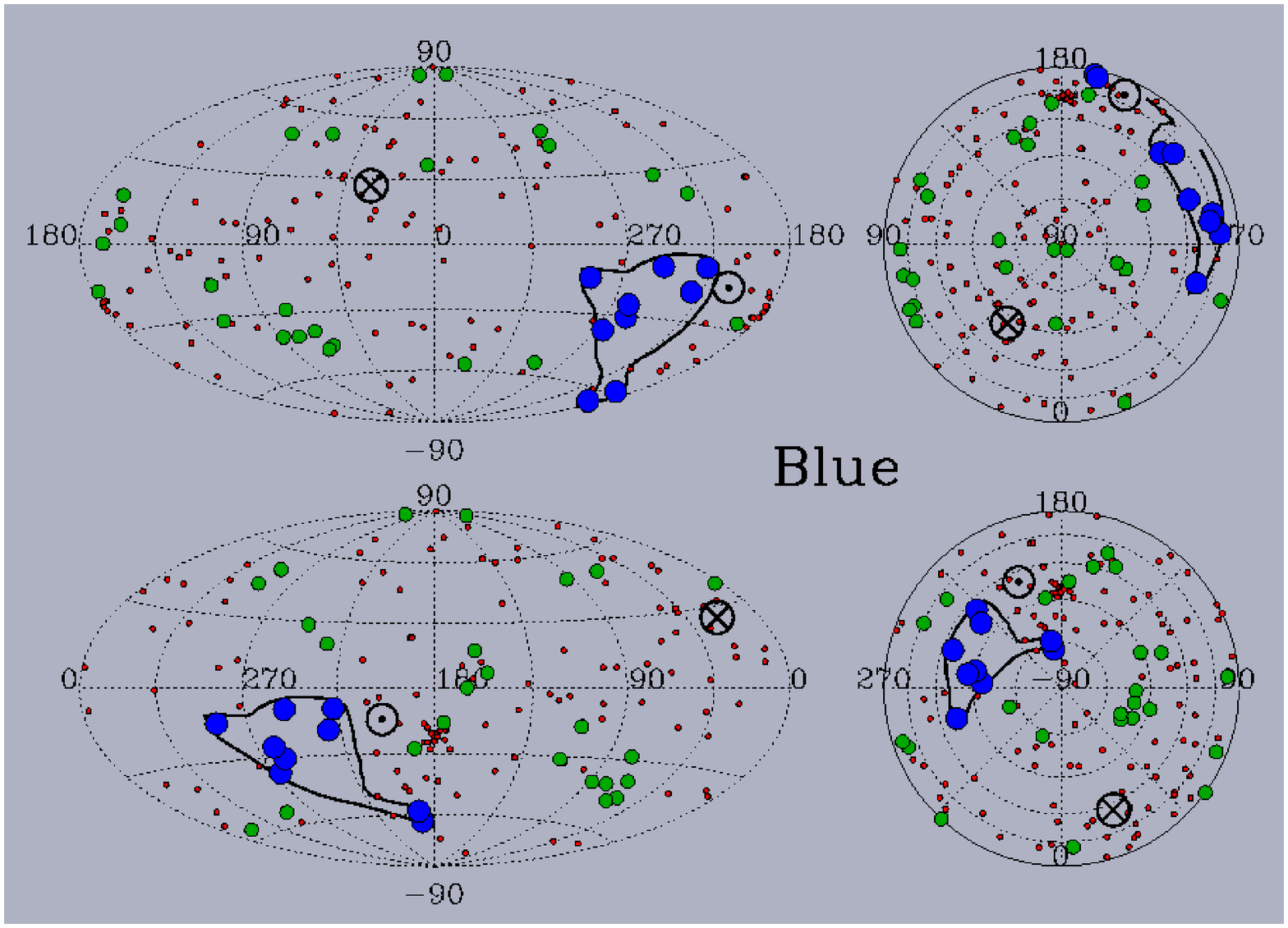}
\caption{Same as Figure~\ref{fig:lic} but for the Blue Cloud. \label{fig:blue}}
\end{figure}

\begin{figure}
\includegraphics[angle=90,width=6.9in]{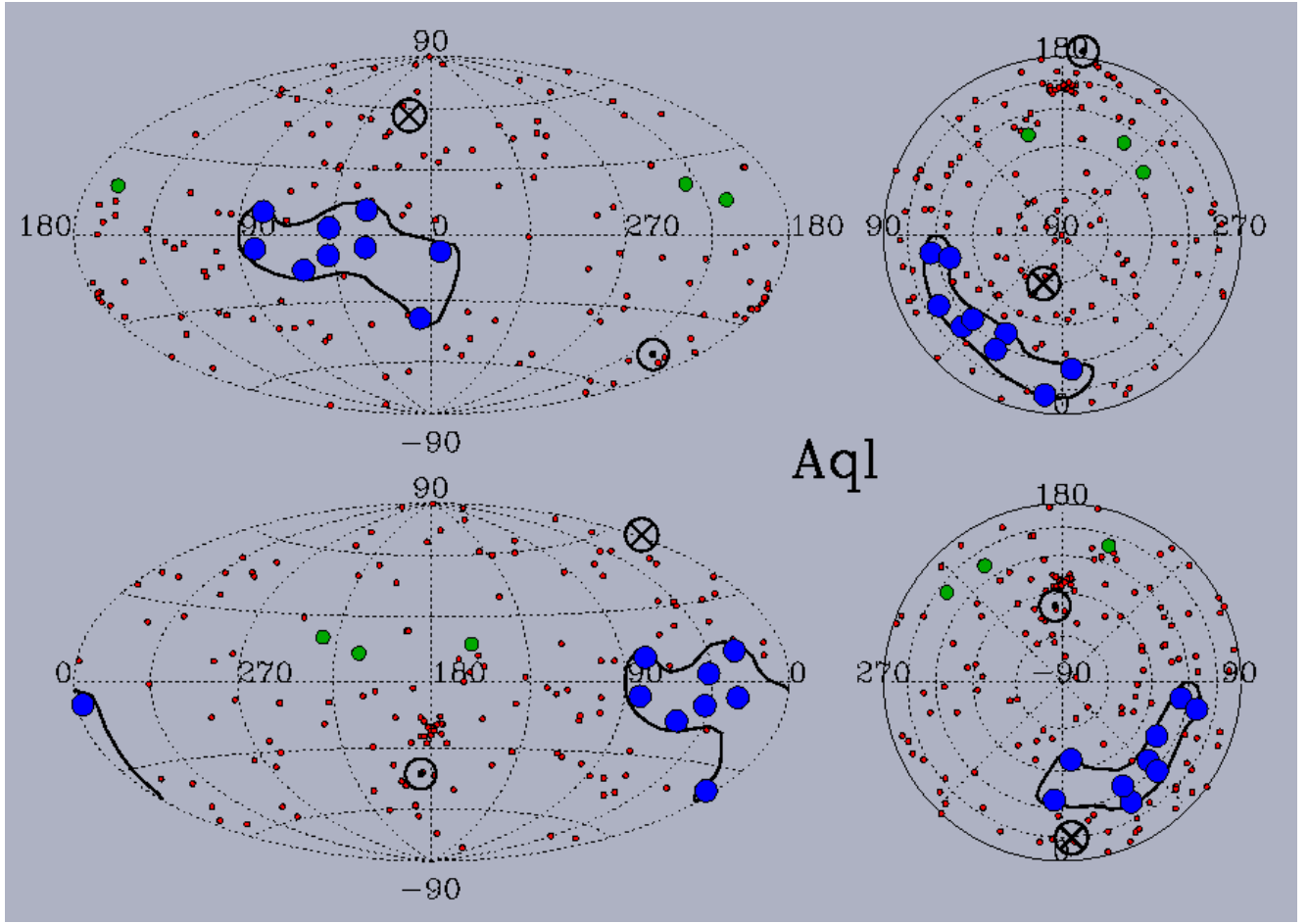}
\caption{Same as Figure~\ref{fig:lic} but for the Aql Cloud.  Note the
small number of coincident velocities (i.e., green medium-sized
symbols).  The Aql Cloud velocity vector is significantly different
than the average LISM flow, but successfully characterizes 9 closely
spaced sight lines.\label{fig:aql}}
\end{figure}

\begin{figure}
\includegraphics[angle=90,width=6.9in]{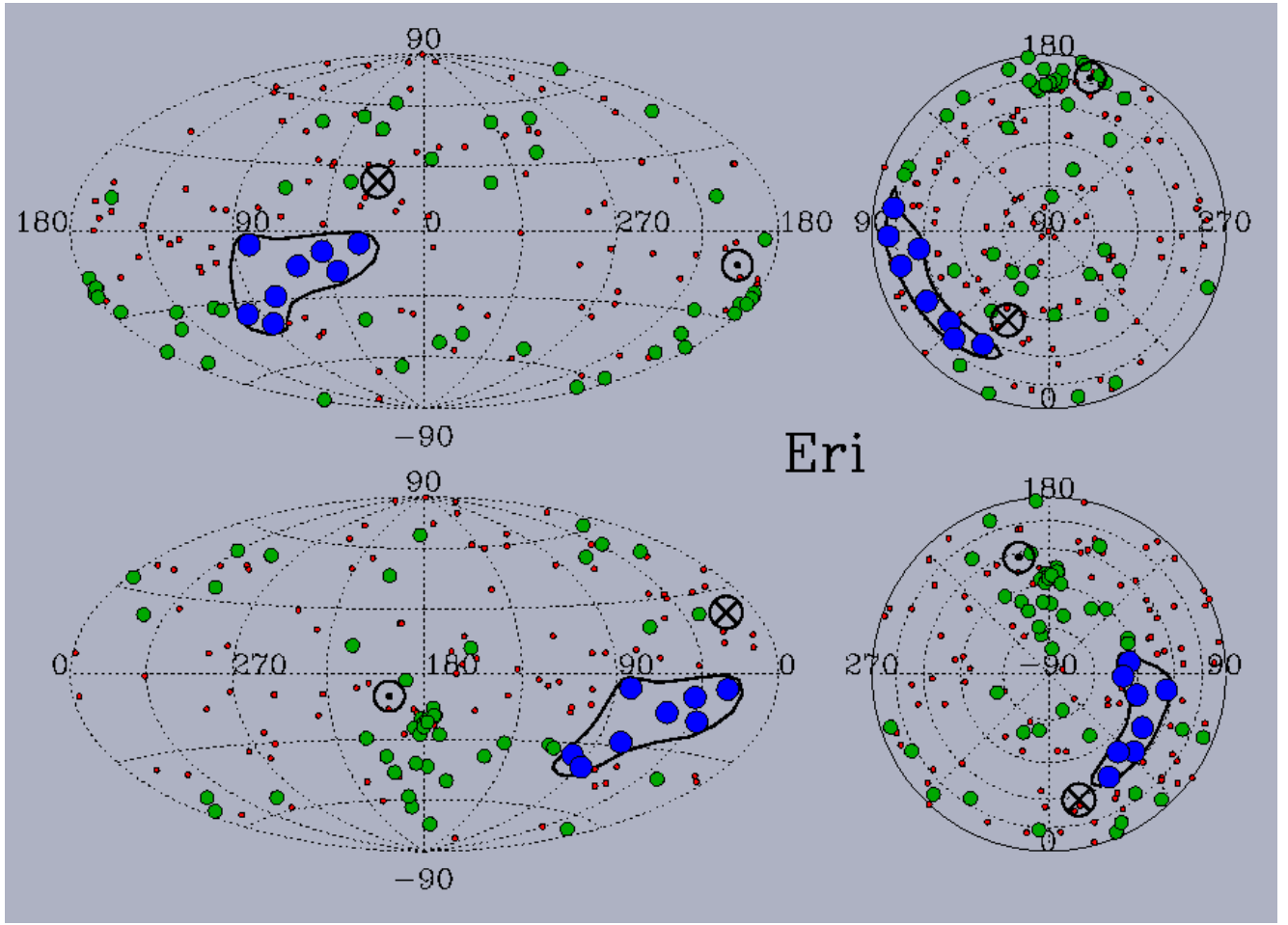}
\caption{Same as Figure~\ref{fig:lic} but for the Eri Cloud. \label{fig:eri}}
\end{figure}

\begin{figure}
\includegraphics[angle=90,width=6.9in]{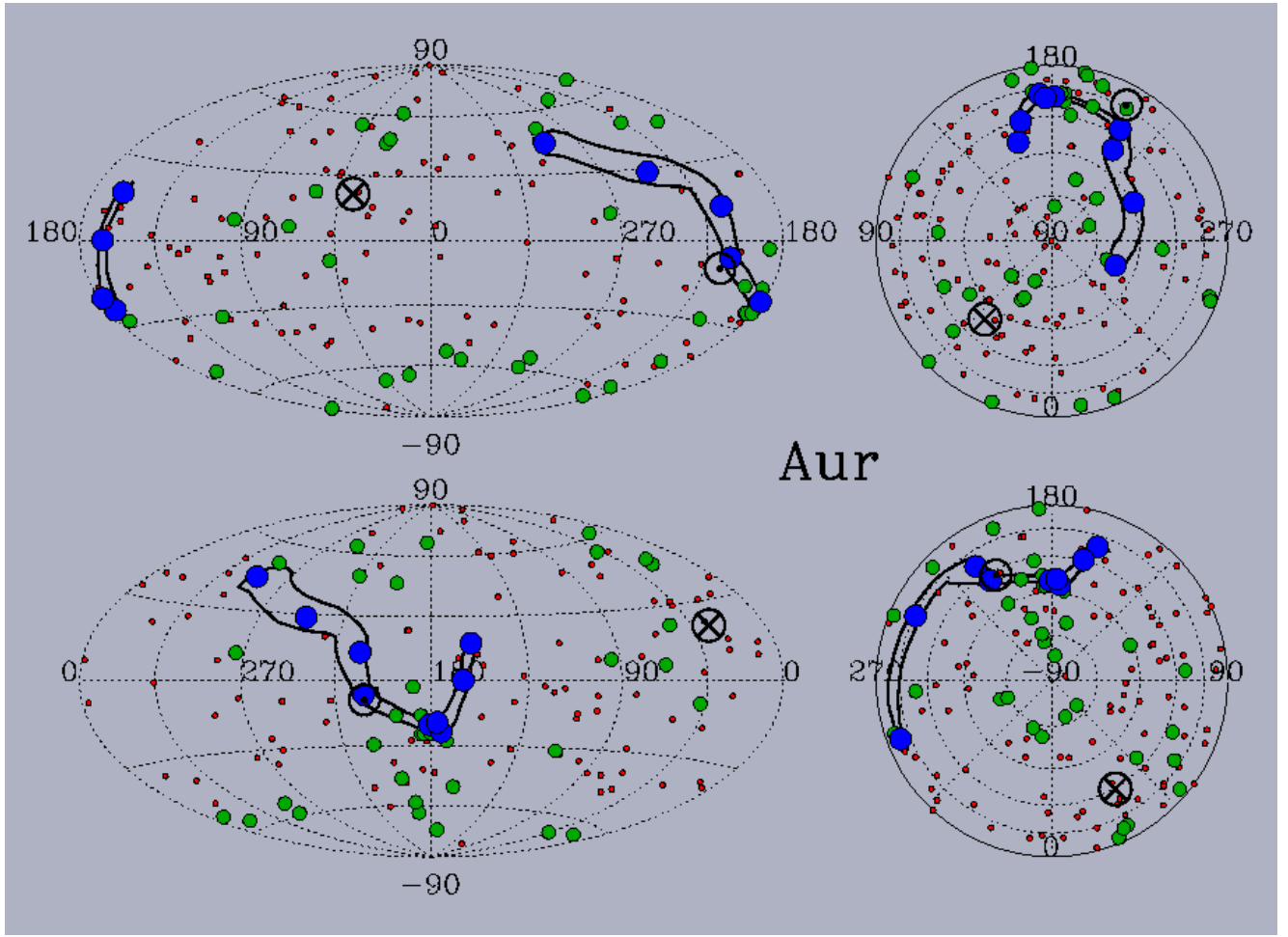}
\caption{Same as Figure~\ref{fig:lic} but for the Aur Cloud. \label{fig:aur}}
\end{figure}

\begin{figure}
\includegraphics[angle=90,width=6.9in]{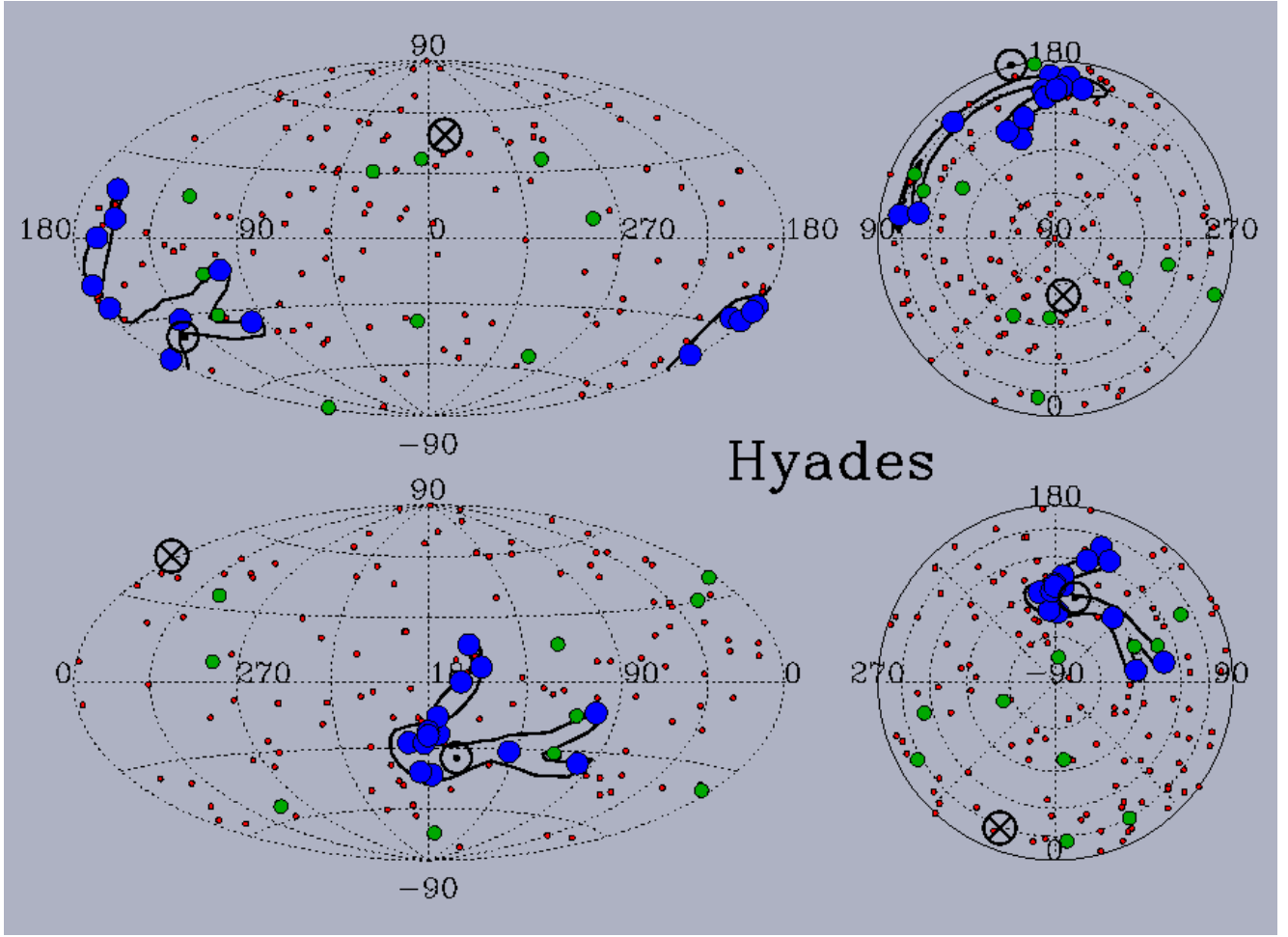}
\caption{Same as Figure~\ref{fig:lic} but for the Hyades Cloud. \label{fig:hyades}}
\end{figure}

\begin{figure}
\includegraphics[angle=90,width=6.9in]{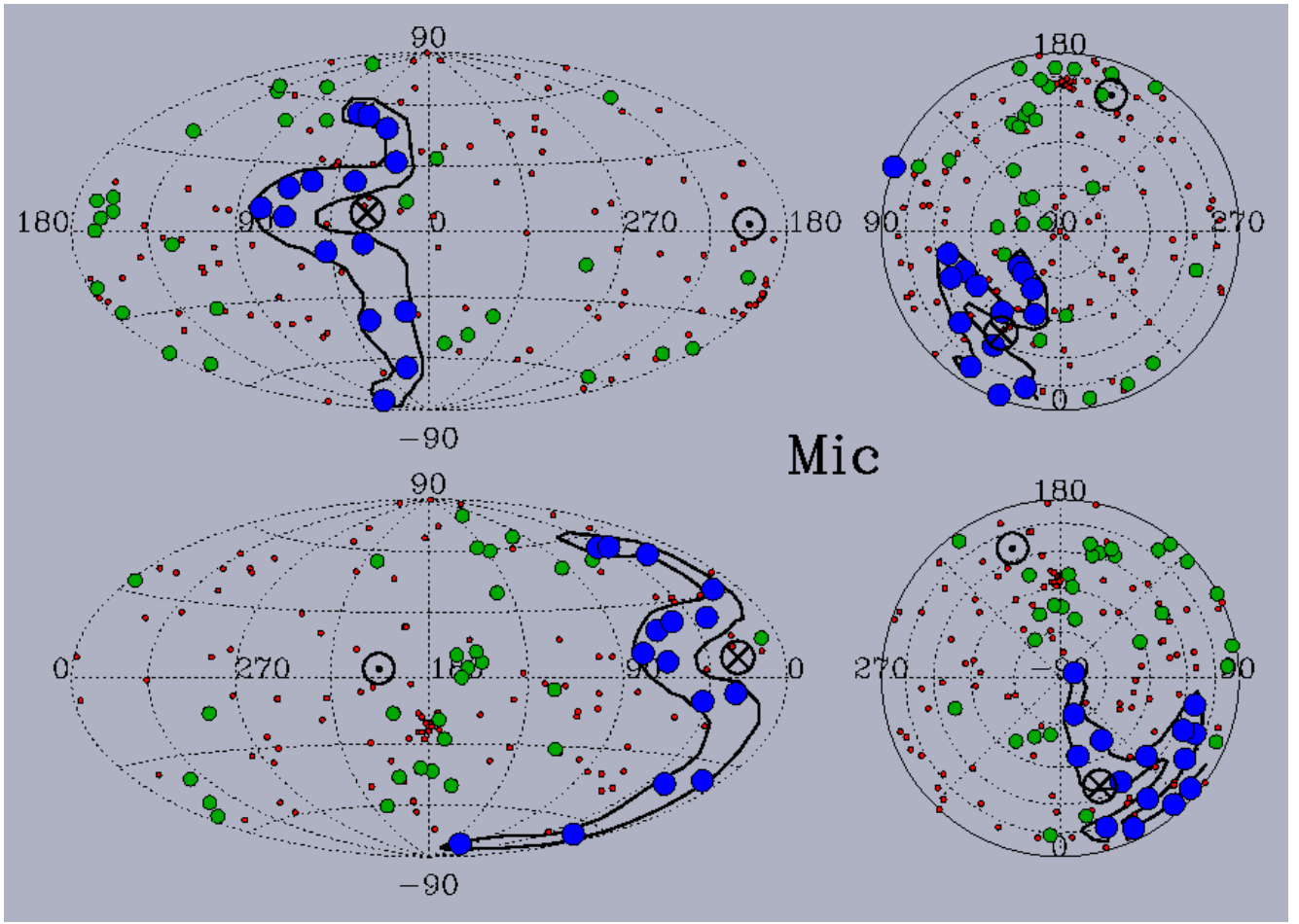}
\caption{Same as Figure~\ref{fig:lic} but for the Mic Cloud. \label{fig:mic}}
\end{figure}

\begin{figure}
\includegraphics[angle=90,width=6.9in]{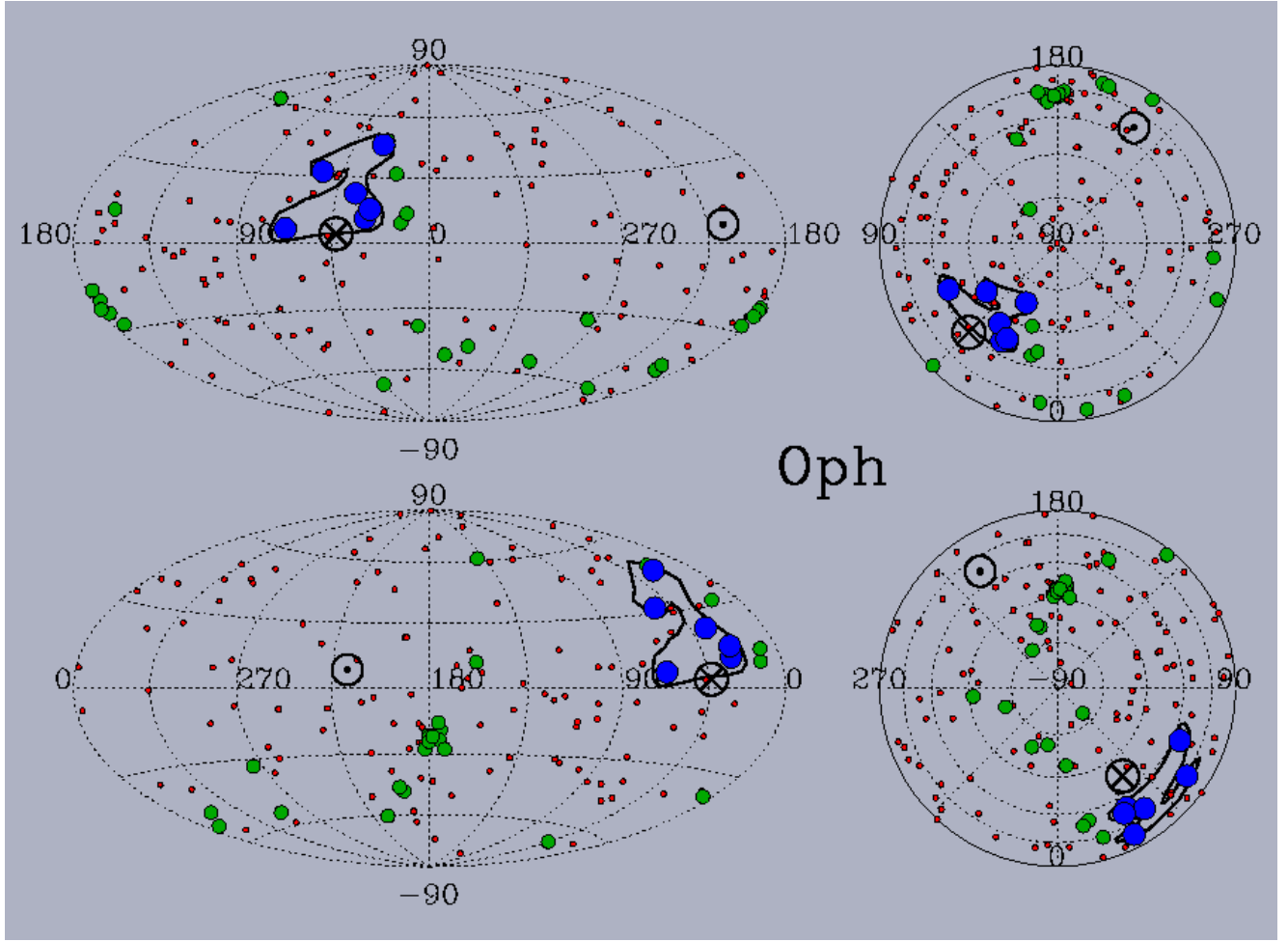}
\caption{Same as Figure~\ref{fig:lic} but for the Oph Cloud. \label{fig:oph}}
\end{figure}

\begin{figure}
\includegraphics[angle=90,width=6.9in]{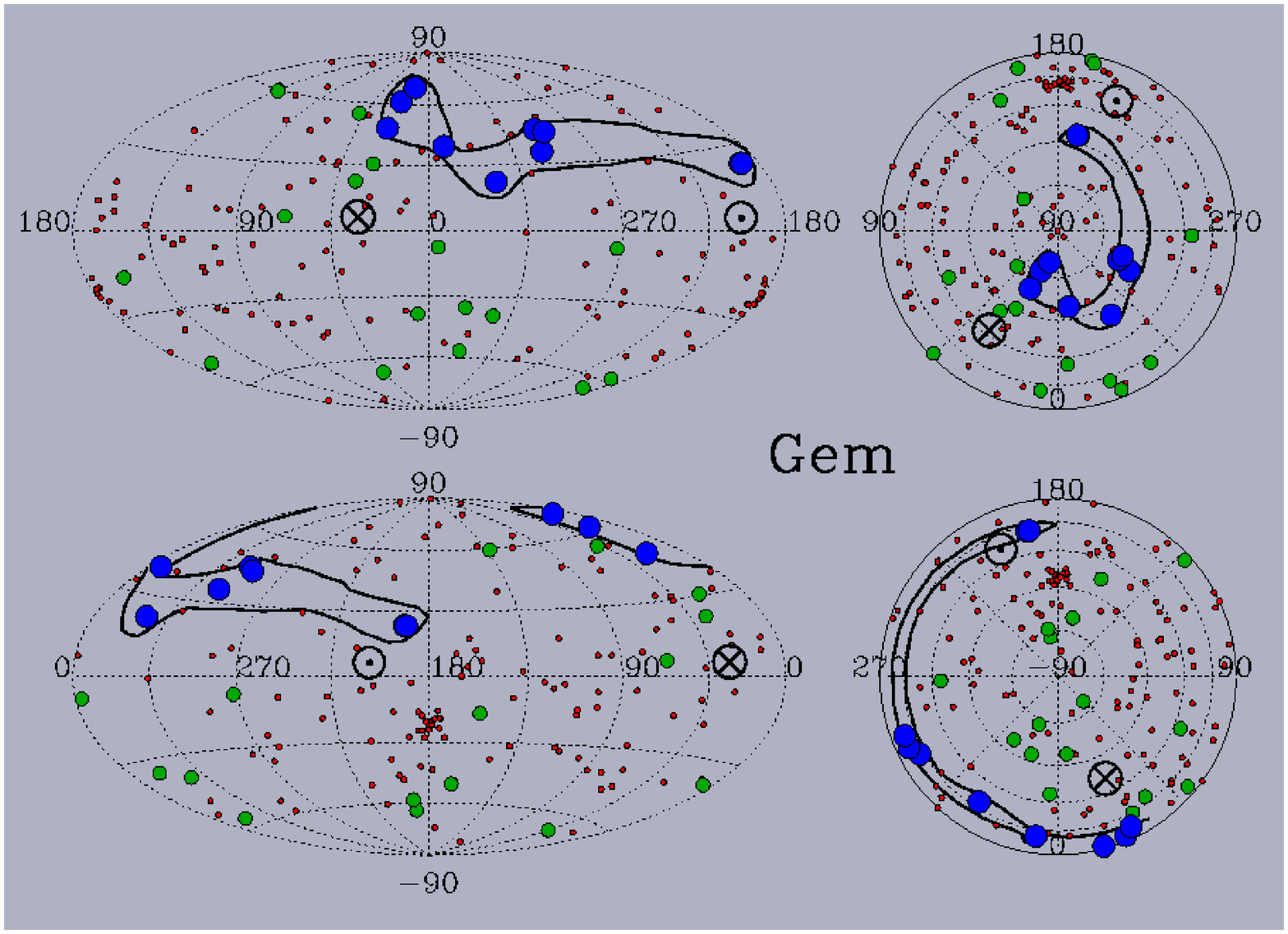}
\caption{Same as Figure~\ref{fig:lic} but for the Gem Cloud. \label{fig:gem}}
\end{figure}

\begin{figure}
\includegraphics[angle=90,width=6.9in]{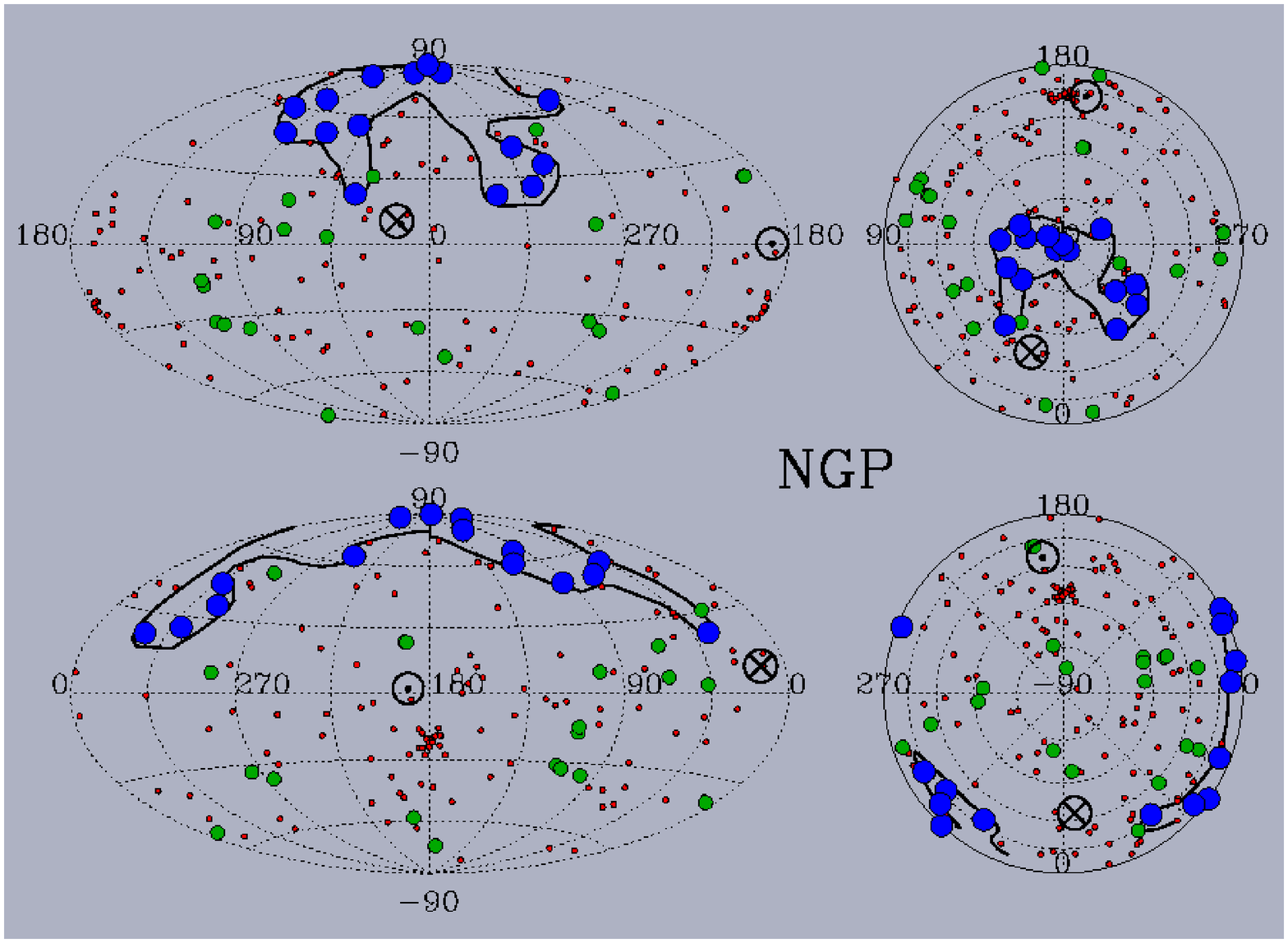}
\caption{Same as Figure~\ref{fig:lic} but for the NGP Cloud. \label{fig:ngp}}
\end{figure}

\begin{figure}
\includegraphics[angle=90,width=6.9in]{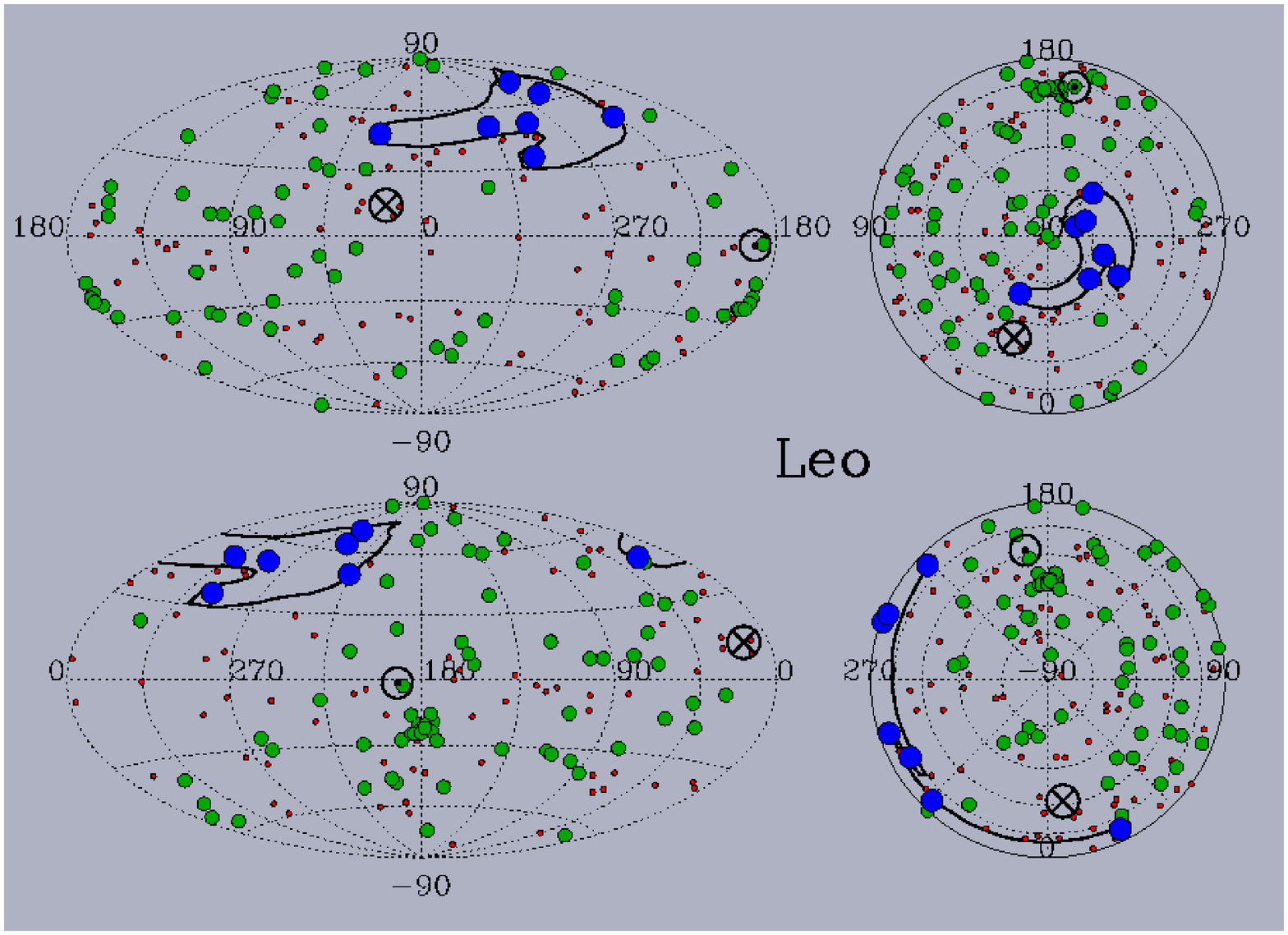}
\caption{Same as Figure~\ref{fig:lic} but for the Leo Cloud. \label{fig:leo}}
\end{figure}

\begin{figure}
\includegraphics[angle=90,width=6.9in]{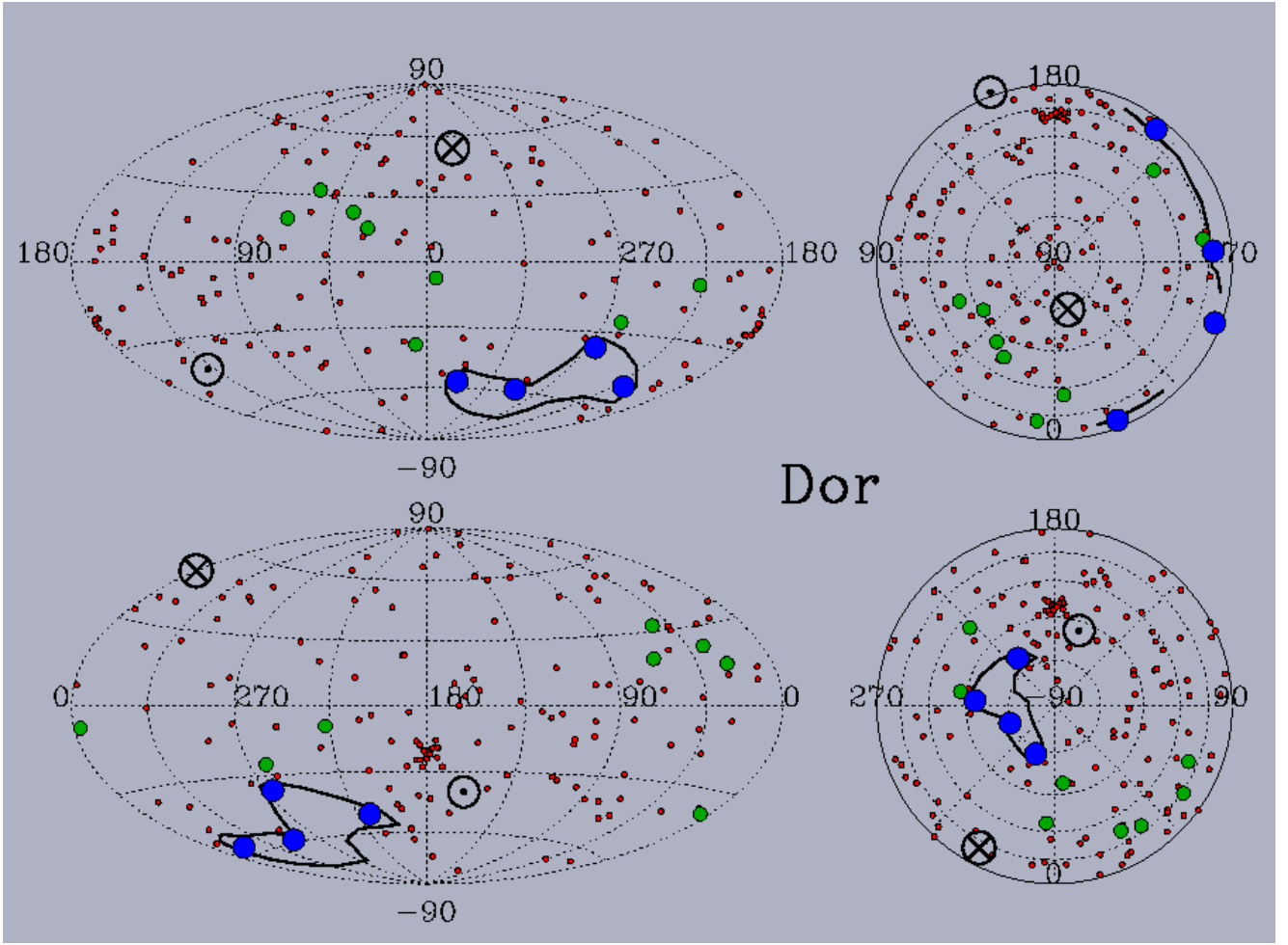}
\caption{Same as Figure~\ref{fig:lic} but for the Dor Cloud. \label{fig:dor}}
\end{figure}

\begin{figure}
\includegraphics[angle=90,width=6.9in]{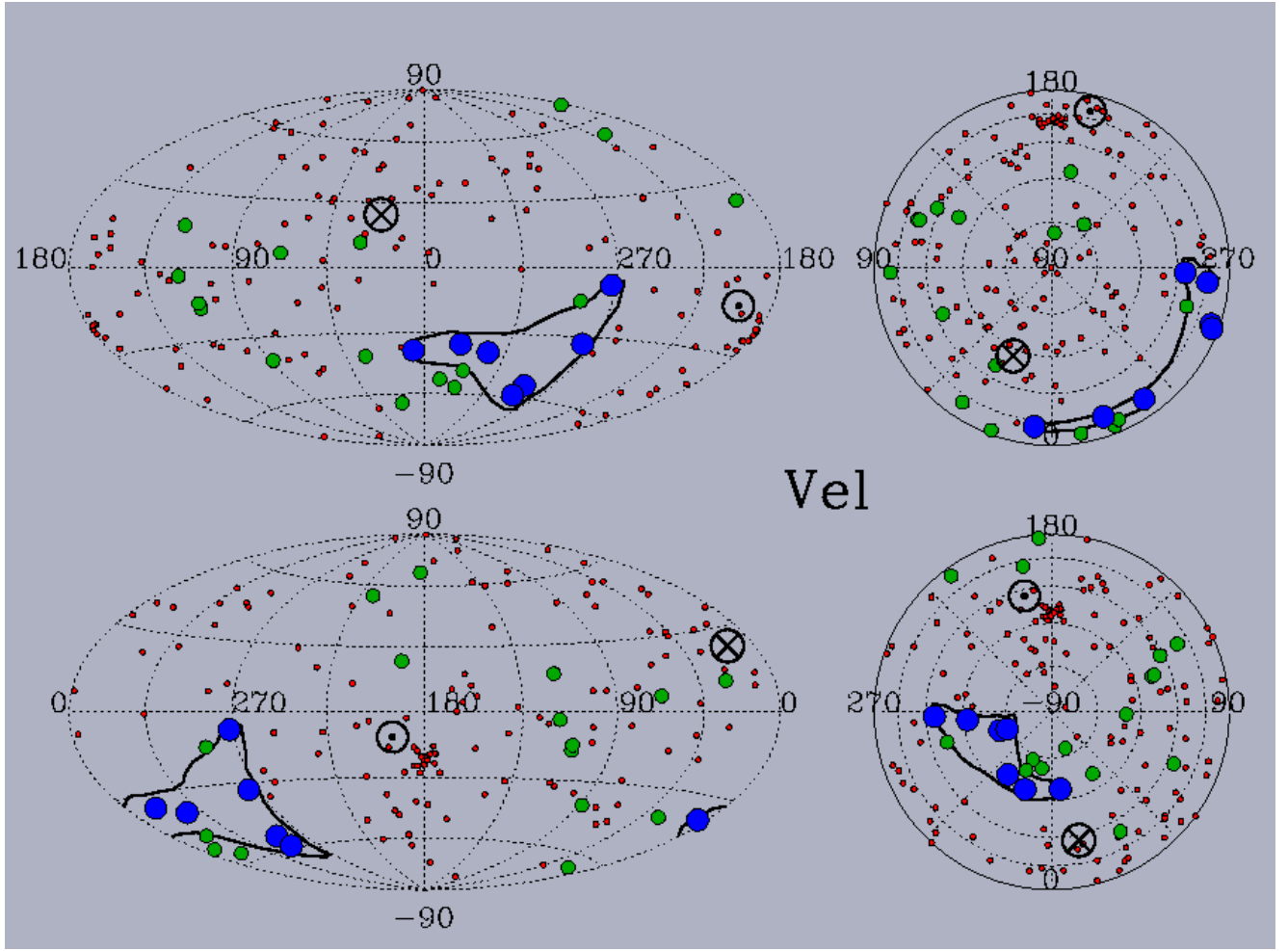}
\caption{Same as Figure~\ref{fig:lic} but for the Vel Cloud. \label{fig:vel}}
\end{figure}

\begin{figure}
\includegraphics[angle=90,width=6.9in]{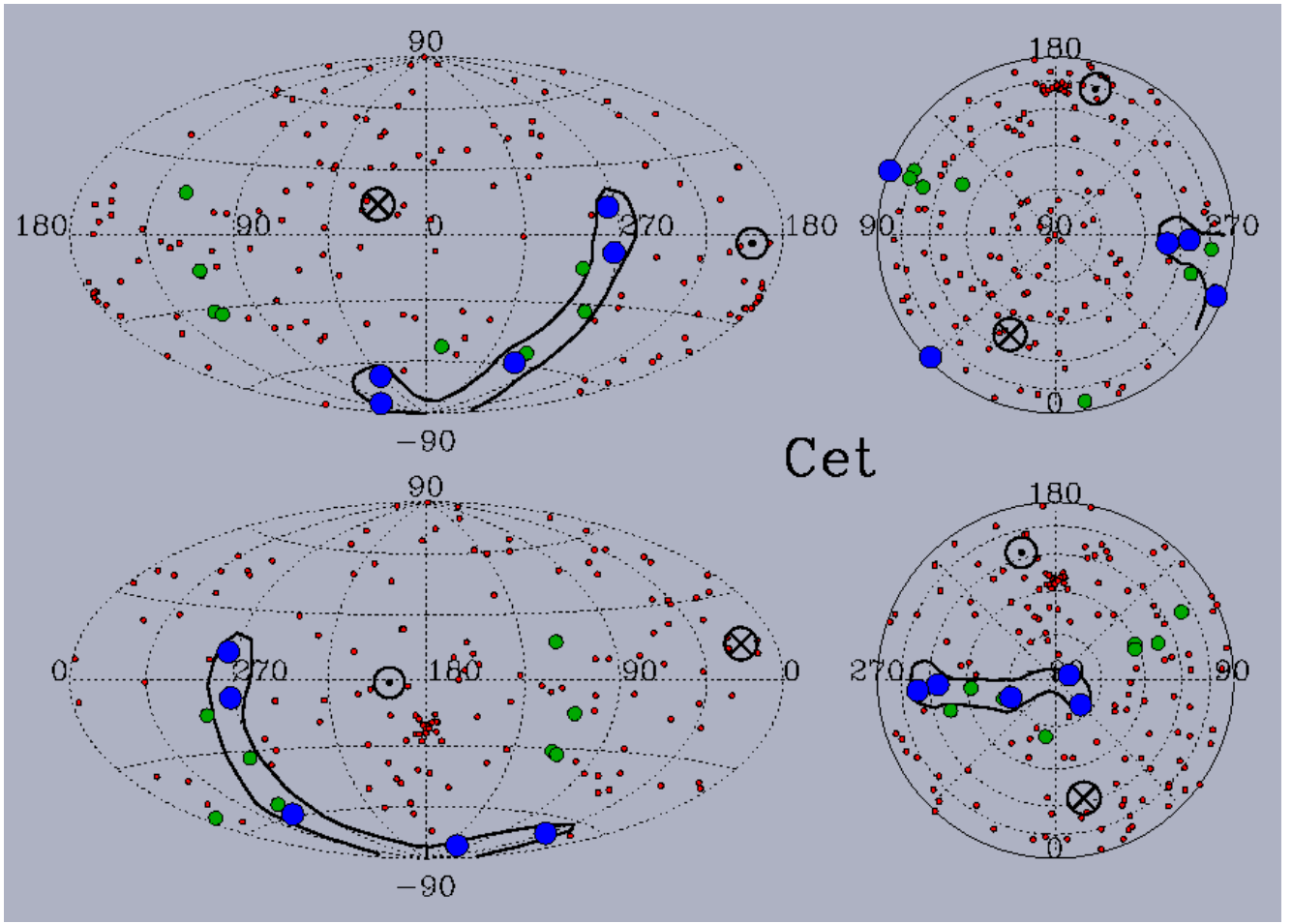}
\caption{Same as Figure~\ref{fig:lic} but for the Cet Cloud.  Note that the boundaries of the Cet Cloud include a couple sight lines that were not used in the velocity vector calculation which nonetheless have consistent projected velocities (i.e., green symbol sight lines), and therefore may traverse Cet Cloud material.
\label{fig:cet}}
\end{figure}

We began our search for identifiable structures in the LISM with the
properties of the LIC as our prototype. As shown by
\citet{lallement92} and by \citet{redfield00}, the LIC moves as if it
were a rigid structure, that is the observed radial velocities toward
nearby stars over a wide range of Galactic coordinates are consistent
with a single velocity vector.  The scatter of the measured radial
velocities about the mean vector is generally less than 1~\kms, which
is similar to the absolute velocity precision of STIS echelle
data. \citet{redfield00} constructed a three-dimensional model for the
LIC based on absorption-line data for 32 lines of sight.  The edge of
the cloud was determined by the measured \ion{H}{1} column density
along each line of sight and the assumption that the \ion{H}{1} number
density is the same throughout the LIC.  This simple assumption cannot
be readily tested and could be far from the truth.  Thus the true
shape of the LIC is not well determined and the question of its edge,
whether it be sharp or gradual, is unknown. Although
\citet{redfield00} concluded that the LIC is roughly spherical in
shape, the shape of interstellar clouds often appears to be
filamentary based on an abundance of observations of nearby
\citep{frisch83} and distant \citep[e.g.,][]{graham95} filamentary
structure in the ISM, presumably organized by magnetic fields
\citep*{jackson03}. We have therefore not assumed any {\it a priori}
shape for the clouds in the LISM. In practice, we have followed a few
simple rules in identifying interstellar clouds.
Figures~\ref{fig:lic}--\ref{fig:cet} show the spatial distribution and
projected boundaries of the resulting dynamical clouds, and
Tables~\ref{tab:licmem}--\ref{tab:cetmem} list sight line membership
of the clouds.

\begin{enumerate}
\item We determine the three-dimensional heliocentric velocity vector
(three free parameters: the velocity magnitude [$V_0$] and the
direction in Galactic coordinates [$l_0$, $b_0$]) that best fits the
radial velocity database, where $V_r = V_0 (\cos b \cos b_0 \cos
(l_0-l) + \sin b_0 \sin b)$.  Here $V_0$ is $>$0 for downwind
directions and $<$0 for upwind directions.  The first application of
this procedure to the entire database yields a velocity vector
consistent with absorption due to the LIC.  Since LIC absorption is
seen over much of the sky, the LIC should have the greatest number of
observed sight lines and dominate the dynamical fit of the entire
database.  We next delete the velocity component that most
significantly disagrees with the predicted projected velocity for LIC
absorption and then recompute the velocity vector that best fits the
remaining points.  This procedure is continued until a satisfactory
fit to the data is derived. Our criterion for ending the iteration
process is that the removal of next most discrepant data point does
not significantly reduce the goodness-of-fit measure, $\chi_{\nu}^2$,
as determined using the F-test, where $\chi_{\nu}^2 \equiv
\chi^2/\nu$, and $\nu$ are the number of degrees of freedom, and
$\chi^2 \equiv \sum\left[(V_r({\rm obs}) - V_r({\rm
pred}))/{\sigma_{V_r({\rm obs})}}\right]^2$ \citep{bevington92}.

\item The next step is to apply the requirement of contiguity: we
assume that the LIC does not have any detached pieces that have
acceptable radial velocities but cannot be sensibly connected to the
rest of the LIC because there are lines of sight between the two
regions that do not show radial velocities consistent with the LIC
velocity vector.  Because of the similarity of the different velocity
vectors of LISM gas, coincident projected velocities of two or more
dynamical structures is common. (Those sight lines that have
components consistent with the vector but not spatially contiguous are
displayed as medium-sized green symbols in
Figures~\ref{fig:lic}--\ref{fig:cet}).  Although limiting the
definition of LISM clouds to spatially coherent structures may
prohibit the identification of complex morphologies, it has the
advantage of preventing the merging of distinct dynamical structures
with similar velocity vectors.  We draw a first approximation of the
LIC shape (see Figure~\ref{fig:lic}) consistent with all data points
lying within 1$\sigma$ of the predicted value if derived from
high-spectral-resolution data and within 3$\sigma$ if drawn from the
moderate-resolution data, with a contiguous morphology.  We require at
least 4 velocity components to constitute a distinct dynamical
structure.

\item At this stage we reintroduce previously deleted velocity
components that are consistent in velocity and continuity with the new
vector. Particularly for the smallest clouds, some velocity components
were prematurely removed from the fit at the earliest stages.  In
addition, it is at this point that other sight line properties, if
available, are compared in order to avoid assigning clearly different
collections of gas with a coincidental velocity vector.  When several
cloud vectors predict a similar projected velocity for a particular
sight line, we assign the velocity component to the cloud with the
nearest neighboring line of sight that is uniquely a cloud member.

\item At this point, we have a nominal assignment of sight lines to a
particular dynamical cloud.  An iterative reevaluation of sight line
membership of previously determined dynamical clouds is performed and
occasionally a reassignment of cloud membership is made, although
this was relatively rare.  The process is then repeated for the
remaining unassigned velocity components.  This iterative velocity
vector technique is most successful at identifying clouds comprised of
a large number of components (e.g., LIC and G clouds), that subtend
large angles on the sky (e.g., NGP and Mic clouds), or are
significantly different dynamically than the average LISM flow (e.g.,
Blue and Aql clouds).  However, it has difficulty identifying compact
dynamical clouds defined by only a handful of sight lines.  In order
to search for these kinds of clouds, we began the process with a
preselected subset of sight lines which are either spatially grouped
in a region without an identified dynamical cloud, or have a common
velocity difference from the general LISM flow, a technique used by
\citet{frisch02}.  Those that produced a satisfactory velocity vector
and survived the constraint of continuity resulted in the
identification of some of our smallest clouds (e.g., Dor and Oph
clouds).

\end{enumerate} 

\subsection{\it Collection of Warm Nearby Interstellar Clouds\label{sec:coll}}

We were able to fit rigid velocity vectors for 15 clouds in the
LISM\footnote{Projected and transverse velocities can be calculated
for any sight line at
http://cobalt.as.utexas.edu/$\sim$sredfield/LISMdynamics.html.}.
Absorption component membership and properties are given for each
cloud in Tables~\ref{tab:licmem}--\ref{tab:cetmem}, and the velocity
vectors and goodness-of-fit metrics, $\chi^2_{\nu}$, are given in
Table~\ref{tab:vecs}.  The names of clouds are either historical
(e.g., LIC: \citealt{mcclintock78}; G: \citealt{lallement92}; Blue:
\citealt{gry95}; Hyades: \citealt{redfield01}; and North Galactic Pole
(NGP): \citealt{linsky00}), or based on constellations that dominate
the area of the sky coincident with the cloud location.  The sight
line members are listed in order of distance to the target star, along
with the observed projected velocity of LISM absorption, the deviation
($\sigma$) from the predicted projected velocity of LISM absorption,
any other LISM properties along the line of sight (e.g.,
$N($\ion{H}{1}$)$, $T$, $\xi$, $D({\rm Fe})$, and $D({\rm Mg})$), and
a list of other LISM clouds that could possibly explain the observed
absorption component.  The deviation between the observed and
predicted projected velocity is given as, $\sigma = (|v_0 -
v_{\star}|) / \sigma_v$, where $v_0$ is the predicted projected
velocity of absorption, $v_{\star}$ is the observed velocity, and
$\sigma_v$ is the error in the observed velocity, where we have
imposed a minimum $\sigma_v$ of 1~km~s$^{-1}$ for all high resolution
data, and a minimum $\sigma_v$ of 3~km~s$^{-1}$ for all medium
resolution data.  The list of other clouds that could possibly explain
the observed absorption component were required to meet slightly
relaxed constraints from those imposed for cloud membership, such that
those clouds listed in the last column of
Tables~\ref{tab:licmem}--\ref{tab:cetmem} are within 10$^{\circ}$ of
the sight line and predict a projected velocity within 3$\sigma$ of
the observed velocity.

Three clouds, NGP, Oph, and Cet, have $\chi^2_{\nu} > 3$, indicating a
relatively poor match between our rigid velocity vector and the
observed projected velocities.  We believe that the high
$\chi^2_{\nu}$ values for these clouds indicate departures from the
assumption of rigidity rather than the existence of several cloudlets
with very similar velocity vectors.  For example, the NGP and Oph
clouds are relatively compact collections of many sight lines, 15 and
6, respectively, which supports a genuine connection between the
absorbing material despite the poor fit to a rigid velocity vector.
Likewise, although the Cet cloud is filamentary and comprised of only
5 sight lines, its high velocity makes it unlikely that a set of
random velocities, for a contiguous group of sight lines, would all be
consistent with such an extreme velocity vector.

\begin{figure}
\plotone{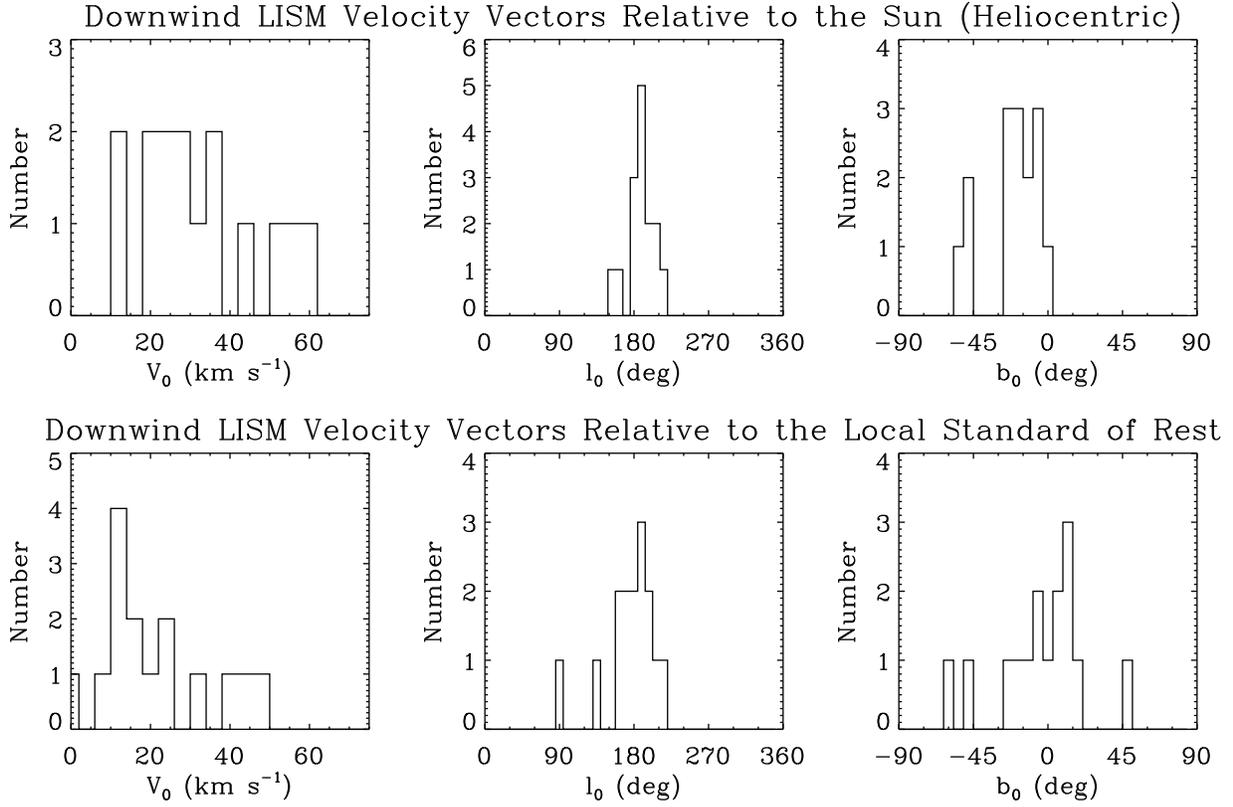}
\caption{Distributions of fit parameters, downwind heliocentric ({\it
top}) velocity ($V_0$), and direction in Galactic coordinates ($l_0$,
$b_0$) for 15 clouds identified within 15 pc.  The distribution of
downwind velocity and direction relative to the Local Standard of Rest
(LSR) is shown in the {\it bottom} panels, where the upstream solar
motion relative to the LSR ($V_{\odot} = 13.4$ km~s$^{-1}$; $l_{\odot}
= 207.7^{\circ}$; $b_{\odot} = -32.4^{\circ}$), was derived by
\citet{dehnen98}.  The bin sizes in $V_0$ are 4 km~s$^{-1}$,
9$^{\circ}$ in $l_0$, and 6$^{\circ}$ in $b_0$.  All velocity vectors
appear to be driven in the same direction, although at a range of
velocities.
\label{fig:solhist}}
\end{figure}

The distributions of velocity amplitude ($V_0$) and direction in
Galactic coordinates ($l_0$, $b_0$) for the 15 clouds are given in
Figure~\ref{fig:solhist} in both the solar rest frame (heliocentric)
and relative to the local standard of rest (LSR).  The vector
solutions for all clouds have similar directions, suggesting that
there is a common history or dynamical driver for all the warm LISM
clouds, but there is a wide range of velocity amplitudes suggesting
the presence of shocks in the LISM \citep[cf.][]{mccray79}.  In
particular, five clouds have velocity components that differ
significantly from the mean value: the Blue and Hyades clouds have
$V_0 < 15$ km~s$^{-1}$, and the Aql, Dor, and Cet clouds have $V_0 >
50$ km~s$^{-1}$.

\begin{figure}
\epsscale{1.}
\includegraphics[angle=-90,width=6.9in]{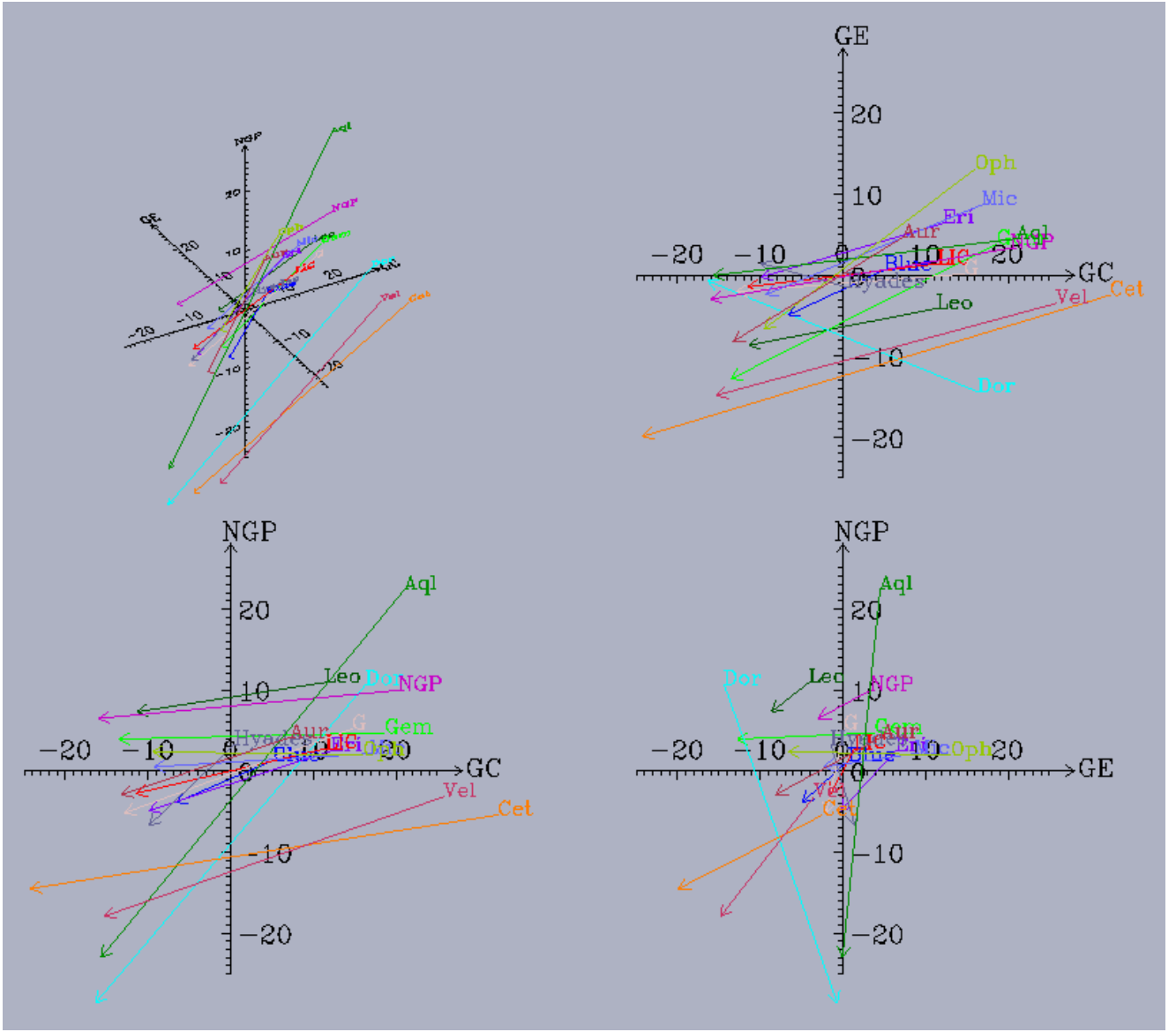}
\caption{Heliocentric velocity vectors of all 15 clouds.  The vectors
are centered in the direction of the center of the cloud and at the
distance of the closest star with the cloud's absorption velocity
and point downwind.  The Sun is moving
in roughly the opposite direction as the LISM clouds.  Starting from
the top left and moving clockwise, the plots are viewed from $l =
230^{\circ}$ and $b = 45^{\circ}$, the North Galactic Pole ($b =
90^{\circ}$), the Galactic Center ($l = 0^{\circ}$ and $b =
0^{\circ}$), and Galactic East ($l = 90^{\circ}$ and $b = 0^{\circ}$).
\label{fig:vecs}}
\end{figure}

Figure~\ref{fig:vecs} shows the projections of the three-dimensional
velocity vector solutions along different Galactic axes.  The location
of the center of each vector is placed in the direction of the center
of the cloud at the distance of the closest star with the cloud's
absorption velocity.  This figure likewise demonstrates that the 15
velocity vectors are all variations on the same theme, in that they
all are aligned in approximately the same direction; however,
significant differences do exist between individual velocity vectors.

About 18.8\% of the velocity components in our database cannot be
assigned to any of our 15 derived velocity vectors.  All unassigned
components are listed in Table~\ref{tab:unass}.  Many of these
velocity components may represent more distant LISM clouds that
subtend smaller fractions of the sky and are probed by too few sight
lines to derive a unique velocity vector.  Approximately 90\% of the
51 unassigned absorbers are toward stars beyond 15 pc.  Only five stars
within 15 pc contain unassigned absorption components, and all 
have an unidentified absorption component in addition to absorption
from identified nearby clouds.  

The first attempt to fit a rigid velocity vector to absorption lines
from nearby stars was made by \citet{crutcher82}.  Seven stars that
were presumed to be within 100 pc, and which were dominated by a
single velocity component in moderate resolution ($R \sim $80,000)
\ion{Ti}{2} observations by \citet{stokes78}, were used to solve for a
single LISM velocity vector, $V_0 = 28$ km~s$^{-1}$, $l_0 =
205^{\circ}$, and $b_0 = -10^{\circ}$.  Two of the seven stars turn
out to have {\it Hipparcos} distances $>$100 pc, while 4 of the
remaining 5 have high-resolution UV or \ion{Ca}{2} observations and
are included in our database.  However, all of these stars show clear
evidence for multiple components in high-resolution spectra.  Although
\citet{crutcher82} was able to derive the general LISM flow direction,
this work demonstrates that analysis of the dynamical structure of the
LISM requires: (1) high-spectral-resolution observations of ions
sensitive to LISM material, (2) accurate distances to the background
stars, and (3) a much larger number of sight lines to disentangle the
complicated spatial and kinematic structure of local material.

In a series of papers, including \citet{lallement92} and
\citet{lallement95}, Lallement et al.\ used high-resolution ($R \sim
$110,000) \ion{Ca}{2} observations and UV observations of \ion{Mg}{2}
and \ion{Fe}{2} of $\sim$16 stars to derive two rigid velocity vectors
that encompassed significant areas of the sky.  The velocity vectors
were associated with the LIC, where they derive a solution, $V_0 =
25.7$ km~s$^{-1}$, $l_0 = 186.1^{\circ}$, and $b_0 = -16.4^{\circ}$,
and the G Cloud, with a solution of $V_0 = 29.4$ km~s$^{-1}$, $l_0 =
184.5^{\circ}$, and $b_0 = -20.5^{\circ}$.  The LIC and G Cloud
velocity vectors that we derive from 5 times the number of lines of
sight are similar (see Table~\ref{tab:vecs}).  For the LIC, we
calculated a heliocentric velocity vector with a magnitude of 23.84
$\pm$ 0.90 \kms\ flowing toward Galactic coordinates $l=187\fdg0 \pm
3\fdg4$ and $b=-13\fdg5 \pm 3\fdg3$.  Our LIC vector is within
$\sim$1$\sigma$ of the direction proposed by \citet{lallement92} but
1.8 km~s$^{-1}$ smaller in amplitude. Our G vector is almost identical
to their previous determination.  This agreement, derived from a much
larger sample, demonstrates not only the reliability of the analysis,
but also the ability to derive accurate velocity vectors from a
relatively small number of sight lines.

\citet{frisch02} derived a bulk flow vector, essentially the average
velocity vector consistent with 96 velocity components from 60 stars
within 132 pc.  As we found with our larger database, the fit to all
of the velocity components leads to a solution that approximates the
LIC velocity vector, since LIC absorption dominates observations of
nearby stars, although G Cloud absorption also contributes
significantly.  \citet{frisch02} assumed the direction of that bulk
flow vector for all LISM clouds (except the LIC), then identified
compact collections of absorbers that show common velocity magnitude
departures from the bulk flow velocity.  Our Blue Cloud vector,
derived from 10 velocity components, matches well with that calculated
by \citet{frisch02} from only 2 velocity components.  However, the
other cloud vectors discussed by \citet{frisch02} are not obviously
comparable to the vectors that we derive.  In particular, the
remaining clouds only differ from the bulk flow by
$\leq$3$\sigma$. Therefore, several of the velocity components
identified in \citet{frisch02} are included as members of the LIC in
our calculation.  Although the directions of velocity components in
the LISM are similar, the assumption that they are identical can
hinder the identification of distinct dynamical structures.

{\it In situ} measurements derived from neutral helium
\citep[e.g.,][]{witte04}, pick-up ions \citep[e.g.,][]{gloeckler04},
and backscattered UV emission
\citep[e.g.,][]{vallerga04,lallement04back}, provide information on
the interstellar flow vector that our solar system is presently
encountering.  \citet{mobius04} summarize the results of these
experiments and provide the weighted mean values for the flow vector,
$V_0 = 26.24\pm 0.45$ km~s$^{-1}$, $l_0 = 183.4^{\circ}\pm
0.4^{\circ}$, and $b_0 = -15.9^{\circ}\pm 0.4^{\circ}$.  Although
technically this vector is consistent within 3$\sigma$ of both the LIC
and G Cloud vectors, the {\it in situ} velocity is intermediate
between the velocities of the LIC and G clouds by 2.4$\sigma$ and
2.9$\sigma$, respectively.  Previous studies
\citep[e.g.,][]{lallement92} have concluded that the flow in the
heliosphere is at the LIC velocity.  \citet{redfield00} and others
have argued that the solar system is located near the edge of the LIC
and is moving toward the G Cloud.  However, the new lower LIC velocity
amplitude that we now derive suggests that the {\it in situ}
measurements could be sampling an interaction region between the
faster-moving G Cloud material and the slower-moving LIC cloud
material, see Table~\ref{tab:vecs} and \S~\ref{sec:licg}.  This
conclusion is supported by the $6303 \pm 390$~K temperature of
interstellar gas in the heliosphere \citep{mobius04}, which, like the
velocity, is intermediate between the temperature of the LIC cloud gas
($7500\pm 1300$~K) and the G Cloud gas ($5500\pm 400$~K), although the
differences have a lower significance of 0.9$\sigma$ and 1.4$\sigma$,
respectively.  Additional temperature measurements, and therefore a
refinement of the mean temperature, of both the LIC and G clouds would
be possible with observations of multiple ions along additional sight
lines.  Currently, such measurements are available along only 29 sight
lines, 24 of which probe the LIC or G cloud material
\citep{redfield04tt}.  If the heliosphere is now located in a
transition zone between the clouds, then we predict that long term
{\it in situ} measurements will gradually approach the G Cloud
velocity and temperature.

\section{MORPHOLOGY OF WARM CLOUDS IN THE LISM \label{sec:morph}} 

Figures~\ref{fig:lic}--\ref{fig:cet} show the morphologies of each
cloud encompassing the sight lines consistent with the 15 rigid
velocity vectors derived from the LISM database\footnote{Probable
cloud membership based on the projected cloud morphologies, can be
calculated for any sight line at
http://cobalt.as.utexas.edu/$\sim$sredfield/LISMdynamics.html.}.  The sight
lines utilized in the velocity vector calculations are indicated by
the large blue symbols, while sight lines with consistent projected
velocities, but assigned to other clouds, are shown by the medium
green symbols, and those sight lines with observed projected
velocities that are inconsistent with the velocity vector fit are
indicated by the small red symbols.  The projected morphology of each
cloud is drawn to include all sight lines used in the velocity fit,
while avoiding all lines of sight that are inconsistent with the
velocity vector. Each figure shows the cloud morphology from four
different directions in Galactic coordinates.

Although the filamentary nature of the clouds could be exaggerated in
some cases because of the low spatial sampling and our requirement of
cohesion, approximately a third of the clouds have projected
morphologies that are clearly filamentary.  It is possible that a
couple of the ``compact'' morphologies may actually be filamentary,
but due to a chance orientation, are projected as a compact cloud on
the sky.  However, it would be highly unlikely that the orientations
of many ``compact'' clouds would be precisely aligned along the line
of sight to hide the true morphology of the clouds.  Determining the
true morphology of these clouds, regardless of orientation, requires a
database with high spatial {\it and} distance sampling.  The
orientations of the observed filamentary clouds are not similar, which
argues against an association with a global magnetic field that may
thread through the LISM.  Instead, the filamentary regions, which
generally trace the boundary between the LIC and G clouds, may
indicate regions of cloud-cloud interactions, where the rigid velocity
structure is disrupted and potentially shocked by the collision of two
adjacent clouds.

\section{PHYSICAL PROPERTIES}

Table~\ref{tab:sumprop} summarizes basic physical properties (e.g.,
coordinates of the cloud center, upper limits to the cloud's distance,
projected surface area, weighted mean temperature, turbulent velocity,
and depletion of iron and magnesium) of the 15 clouds.  The surface
area on the sky is simply the surface area of the projected boundaries
shown in Figures~\ref{fig:lic}--\ref{fig:cet}.  The temperature and
turbulent velocities are derived from comparisons of the measured
Doppler widths of absorption lines of elements with different atomic
masses (e.g., deuterium and iron) using the relation, $b^2 = 2kT/m +
\xi^2$ \citep{redfield04tt,wood96}.  The depletions, $D(X) = \log
\left(X/H\right) - \log \left(X/H\right)_{\odot}$, are calculated
using the \citet*{asplund05} solar abundances, where the hydrogen
abundance is typically calculated from \ion{D}{1} and converted to
\ion{H}{1} using the remarkably constant LISM D/H ratio of $1.56 \pm
0.04 \times 10^{-6}$ \citep{linsky06}.  The depletions do not take
into account partial ionization of hydrogen, which is likely
important, or neutral or doubly ionized magnesium and iron, which are
likely much less important since they are not expected to become a
dominant ionization stage of either element
\citep[cf.][]{slavin02,lehner03}.  The weighted mean and $1\sigma$
uncertainty of the mean are listed for all physical properties with
measurements on multiple sight lines.  No significant correlation
appears to exist between any of the physical properties listed and
cloud morphology.  In the following discussion of physical properties
of individual clouds, we consider only the nine clouds that have more
than 1 sight line with physical measurements.

\subsection{\it Distance Limits}

Although a detailed look at the distances of the 15 LISM clouds is
beyond the scope of this paper, we can immediately place distance
constraints based on the distance of our background sources and
provide some insight into the three dimensional structure of the LISM.
In Table~\ref{tab:sumprop}, we list the distance of the closest star
with absorption from the cloud, which provides an upper limit to the
distance of the cloud.  All of the clouds lie within 15 pc, and half
lie within $\sim$5 pc, which is much smaller than the volume of the
Local Bubble, but consistent with the large projected surface area
that these clouds subtend.  The distribution of many clouds with a
range of dynamical properties in such a small volume, makes collisions
between clouds a real possibility.  The implications of such
interactions are explored in \S~\ref{sec:collide}.

Although the G Cloud has a more stringent distance upper limit than
the LIC, since the temperature and velocity of the interstellar
material that is flowing into the solar system is consistent with
early estimates of LIC material \citep{mobius04}, implying that the
Sun is currently inside the LIC (although, see \S~\ref{sec:licg} for a
detailed discussion of this topic).  However, since LIC absorption is
not observed in all directions (e.g., toward the Galactic Center) and
since the Sun is moving in roughly the direction of the Galactic
Center and the G Cloud (Figure~\ref{fig:all}), the heliosphere has
been thought to be at the very edge of the LIC \citep{redfield00}.



\subsection{\it Volume Filling Factor of the Warm LISM \label{sec:fill}}

We have not yet created a full three-dimensional morphological model
of the LISM, but with a few assumptions and a simple toy model, we can
estimate the volume filling factor of the warm partially ionized gas
in the LISM.  First, we assume that all of the warm LISM material is
located within 15 pc of the Sun.  Although the LISM is often
considered to be the volume of material within the Local Bubble, which
extends out to roughly 100 pc in all directions, it seems that most of
the warm material is located only a short distance from the Sun. This
is shown, for example, in Figure~14 in \citet{redfield04sw}, where the
average number of absorbers per unit distance levels off at a distance
of $\sim$15 pc.  No significant correlation exists between the
observed line width or column density and the distance of the
background star.  Therefore, it is unlikely that unrecognized line
blends along more distant sight lines are the cause of the observed
leveling off of the average number of absorbers, but is indicative of
the true distribution of warm gas in the LISM.  Based on Figure~14 of
\citet{redfield04sw}, there are on average $\sim$1.7 absorbers per
sight line.  Therefore, the projected surface area of all LISM clouds
should total $\sim$1.7$ \times 4\pi$.

Initially, we assume that all of the warm LISM clouds are similar in
size and and density to the LIC.  We assume that all warm clouds have
a radius of 1.5 pc and a total hydrogen density of 0.2 cm$^{-3}$,
obtained from measurements of the \ion{He}{1} volume density streaming
into the solar system from \citet{gloeckler04}, the \ion{H}{1} to
\ion{He}{1} ratio of the LISM from \citet{dupuis95}, and the
three-dimensional model of the LIC from \citet{redfield00}.  In
addition, the assumed radius and density result in a full cloud
hydrogen column density of $\sim$2$ \times 10^{18}$ cm$^{-2}$, which
matches the typical observed hydrogen column density
\citep{redfield04sw}.  

We ran 1000 simulations of cloud distributions, for a range of total
warm clouds ($n$) in the LISM from $n = $ 1--100 clouds, where all
clouds were randomly centered at distances from 0--15 pc.  Since
clouds are not allowed to overlap, only one cloud can surround the
solar system, that is, a LIC analog.  In order to more closely match
the model with the observed LISM, we have assumed the LIC projected
surface area of 18270 square degrees for the LIC analog, see
Table~\ref{tab:sumprop}, instead of 41250 square degrees.  The
remaining clouds projected surface areas were calculated based on
their size, geometry, and distance.  In the spherical cloud scenario,
the solution to an average of 1.7 absorbers per sight line leads to
$\sim$55 clouds within 15 pc, and a volume filling factor of warm
partially ionized material of $\sim$5.5\%.

We also ran a suite of simulations varying the geometry (i.e.,
ellipsoids with a range of aspect ratios from 1.33:1 to 10:1, both
flattened (i.e., pancakes) and elongated (i.e., cigars)) and fraction
of ellipsoid to spherical (which ranged from 0.3 to 1.0).  The
orientation of all clouds were determined randomly.  Non-spherical
geometries naturally lead to larger volume filling factors and fewer
clouds, since a greater total volume can be produced with fewer clouds
without necessarily increasing the projected surface area.  The volume
filling factors that resulted ranged from 5.5\% to 19\%.

\begin{figure}
\epsscale{1.1} 
\plottwo{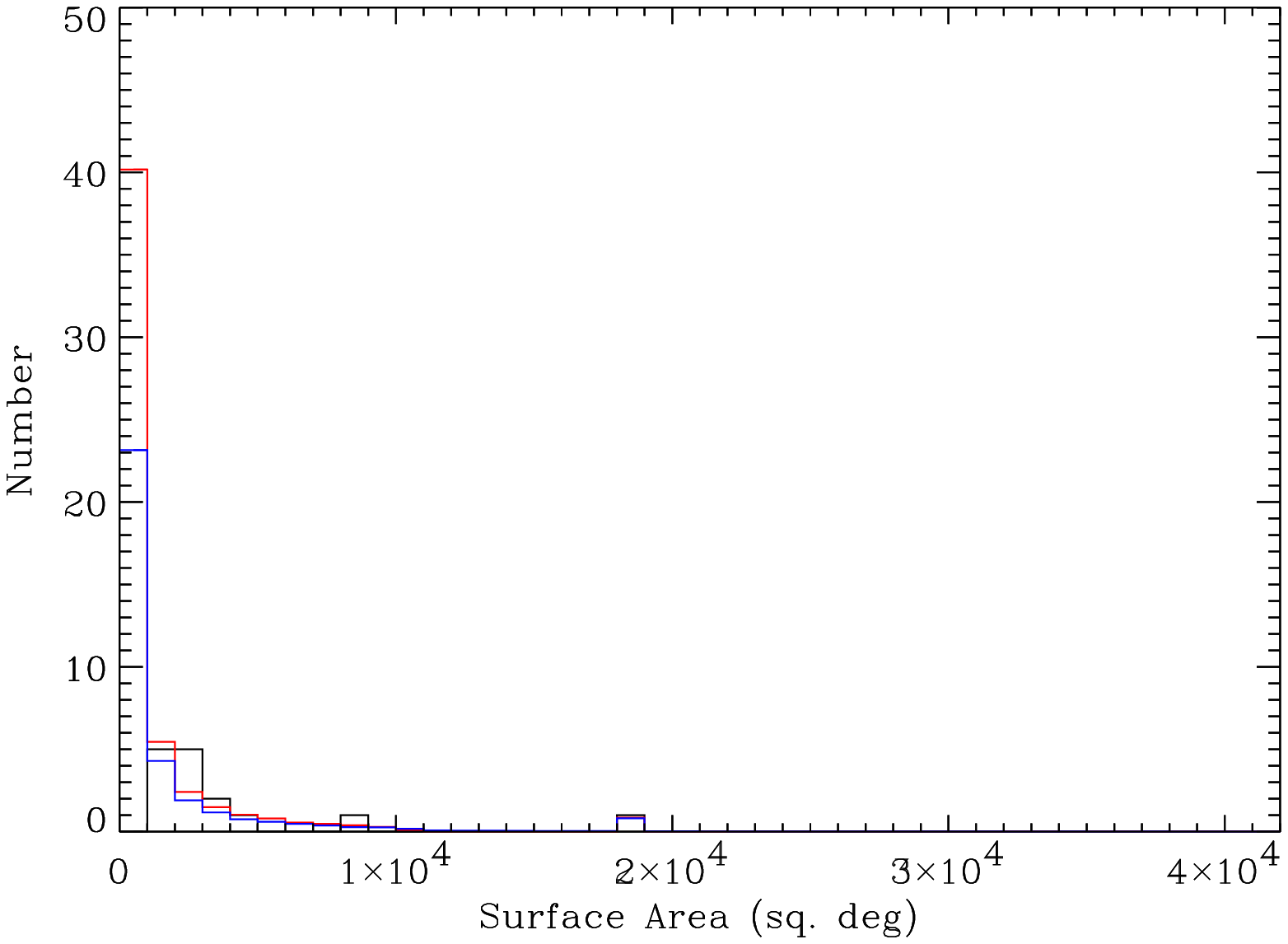}{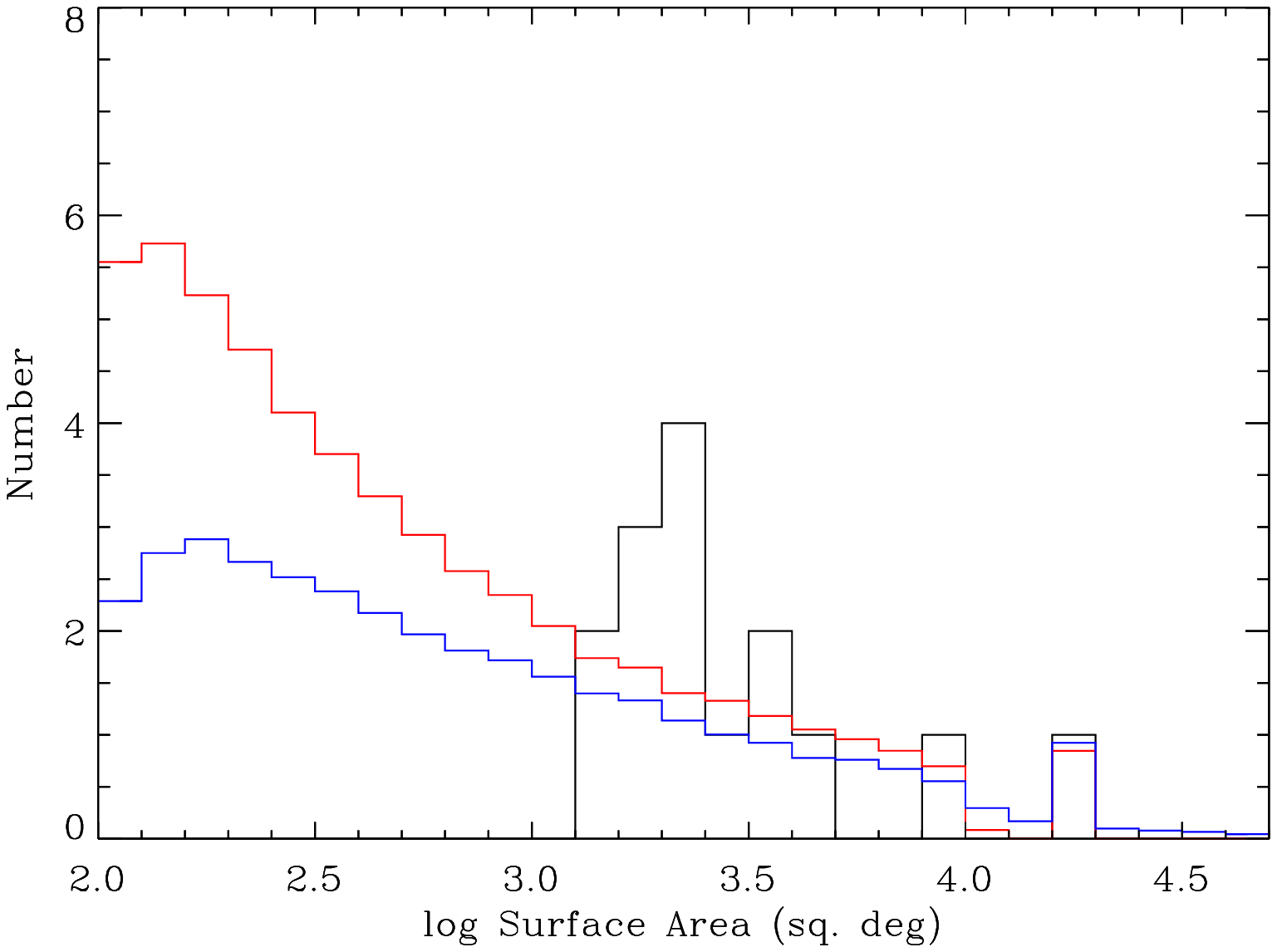}
\caption{Histogram (black) of the observed angular areas of the 15
nearby dynamical clouds, based on their projected morphologies shown
in Figures~\ref{fig:lic}--\ref{fig:cet}.  The histograms on the {\it
left} have bin sizes of 1000 square degrees, while the histograms on
the {\it right} have bin sizes of 0.1 dex in logarithmic square
degrees.  The red histogram indicates the average distribution of
angular areas of the simple model of 55 randomly distributed spherical
LIC-like clouds within 15 pc of the Sun, discussed in
Section~\ref{sec:fill}.  This model leads to a volume filling factor
of $\sim$5.5\%.  The blue histogram shows another simple simulation of
35 randomly oriented clouds in which half were ellipsoids with aspect
ratios of 10:1.  Although this simulation also reproduces the observed
projected surface areas of the 15 large clouds fairly well, the volume
filling factor in this case is $\sim$19\%.
\label{fig:area}}
\end{figure}

\begin{figure}
\includegraphics[angle=-90,width=6.9in]{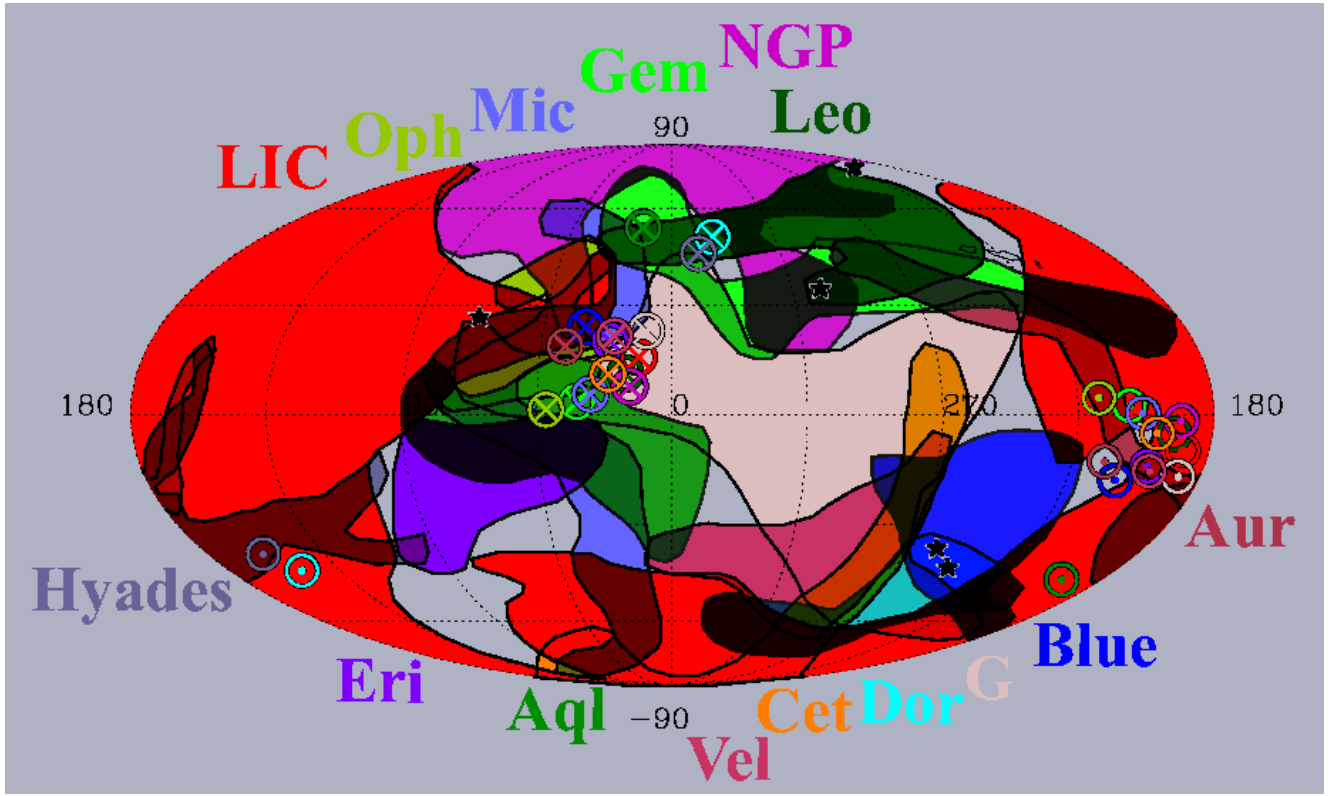}
\caption{All dynamical cloud morphologies are overlaid and colorcoded
as in Figure~\ref{fig:vecs}.  The upwind heliocentric direction of the
velocity vector for each cloud is indicated by the $\otimes$ symbol,
while the downwind heliocentric direction is indicated by the $\odot$
symbol.  The Sun is moving approximately antiparallel to the LISM
clouds.  The star symbols indicate sight lines of radio scintillation
sources, and the series of three small clouds centered at $l =
222^{\circ}$ and $b = 44^{\circ}$ are the \ion{H}{1} contours from
\citet{heiles03b}, of the cold cloud recently identified to be within
the Local Bubble by \citet{meyer06}.
\label{fig:all}}
\end{figure}

Figure~\ref{fig:area} compares the projected surface area distribution
of our sample with two of our idealized simulations.  The distribution
of observed projected surface areas matches fairly well with that
predicted for both the $\sim$55 spherical LIC-like clouds within 15
pc, and a simulation of 35 clouds, half of which are ellipsoids with an
elongated aspect ratio of 10:1.  There is an observational bias toward
detecting the nearest clouds with the largest projected surface areas.
We restricted our dynamical cloud modeling to collections of gas that
had at least 4 sight lines with which to determine a velocity vector.

About 18.8\% of absorbers are not accounted for in our 15 cloud
dynamical model of the LISM. The missing absorbers may represent
detections of more distant and smaller projected surface area clouds.
If we assume that we have detected all clouds with a $\log$ surface
area $>$ 3.1 ($\sim$1260 square degrees), which is the lower limit of
the LISM clouds with measured velocity vectors (see
Table~\ref{tab:sumprop}), we can estimate the number of sight lines
that probe ``undetected'' clouds, or clouds with a $\log$ surface area
$<$ 3.1, (see Figure~\ref{fig:area}).  Our toy models have 10.2--11.9
clouds with $\log$ surface area $>$ 3.1, slightly lower, but similar
to the 15 observed clouds.  The total projected surface area for all
clouds with $\log$ surface area $>$ 3.1 range from 55290 to 60430
square degrees in our simulations.  This matches well the total
projected surface area of the observed clouds (57830 square degrees).
With an estimate of the total projected surface area of ``undetected''
clouds, and the assumption that all sight lines are uniformly
distributed, we can predict the percentage of observed components that
will be left over, after those associated with the large nearby clouds
are removed.  The percentage of unassociated velocity components in
the simulations are between 14.5\% and 21.5\%, which matches closely
the percentage of components in our database (18.8\%) that are
unassigned, which allows for the possibility that the absorbing
material associated with these components is indeed located within
$\sim$15\,pc, even though the background star is much further away.

\section{WHERE CLOUDS COLLIDE \label{sec:collide}}

\subsection{\it The ``Ring of Fire'' Around the G Cloud}

Figure~\ref{fig:all} shows the projected morphologies of all 15
clouds, which collectively cover more than 90\% of the sky.  The LIC
and G Clouds clearly dominate the sky, and contain large areas where
one or the other is the only absorber along the line of sight.  At the
boundaries of the LIC and G clouds, several overlapping absorbers are
typically present, particularly near $l$ from 40$^{\circ}$ to
80$^{\circ}$ and $b$ from --15$^{\circ}$ to +30$^{\circ}$, as well as
near $l$ from 270$^{\circ}$ to 320$^{\circ}$ and $b$ from
+20$^{\circ}$ to +50$^{\circ}$ and from --70$^{\circ}$ to
--30$^{\circ}$.  These areas at the boundaries of the LIC and G clouds
may be dynamical interaction zones, which produce ``new'' clouds with
significantly different kinematic properties.  We refer to the active
boundary of the G Cloud where the G and LIC clouds may be colliding as
the ``Ring of Fire'', in analogy to the Pacific Ocean ``Ring of Fire''
where dominant tectonic plates (here interstellar clouds) interact,
resulting in a highly dynamic interaction zone that gives rise to
earthquakes and volcanos (here interstellar shocks, heating, or
turbulent flows).

\begin{figure}
\begin{center}
\includegraphics[angle=-90,width=4.9in]{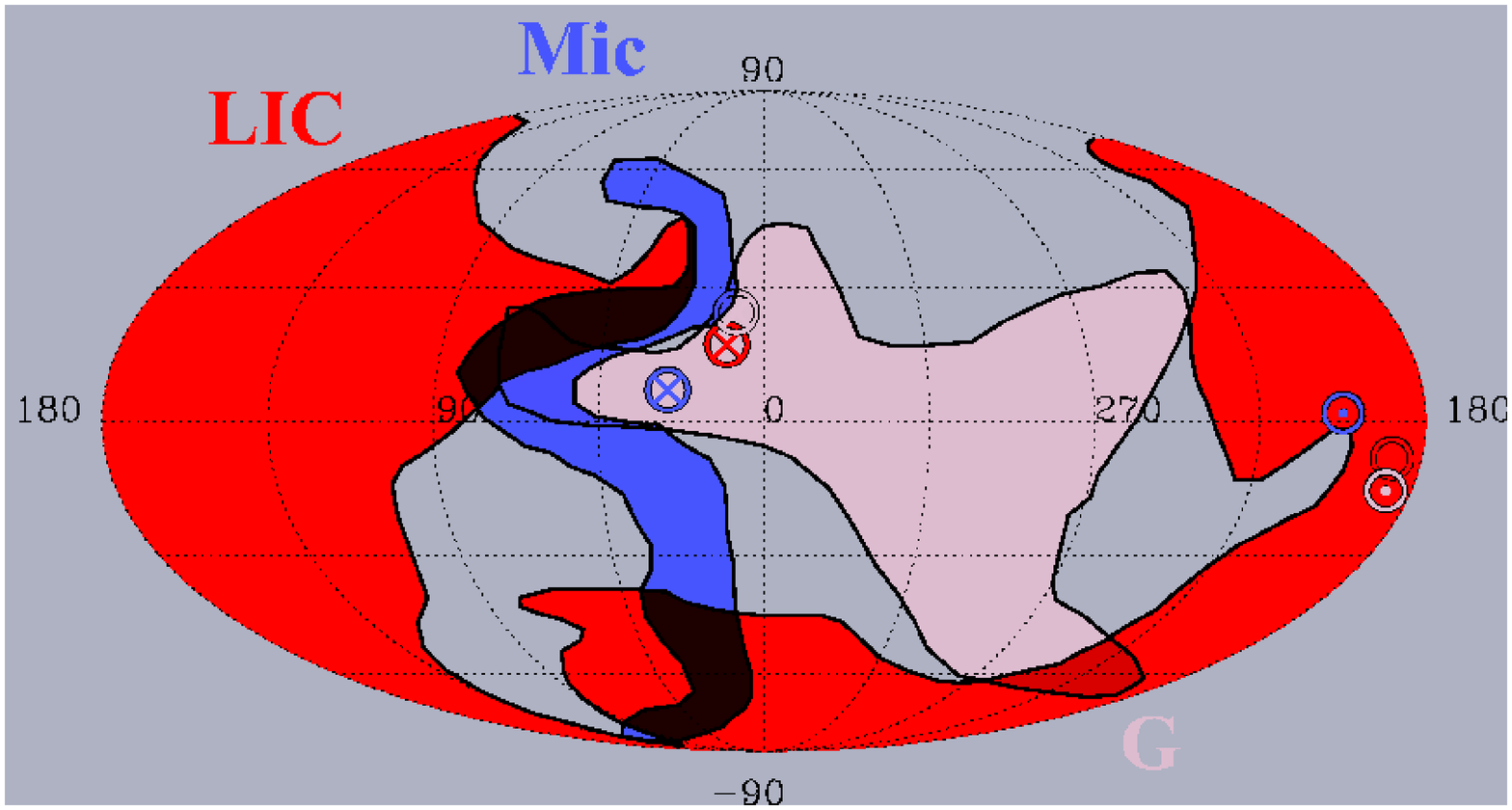}
\includegraphics[angle=-90,width=4.9in]{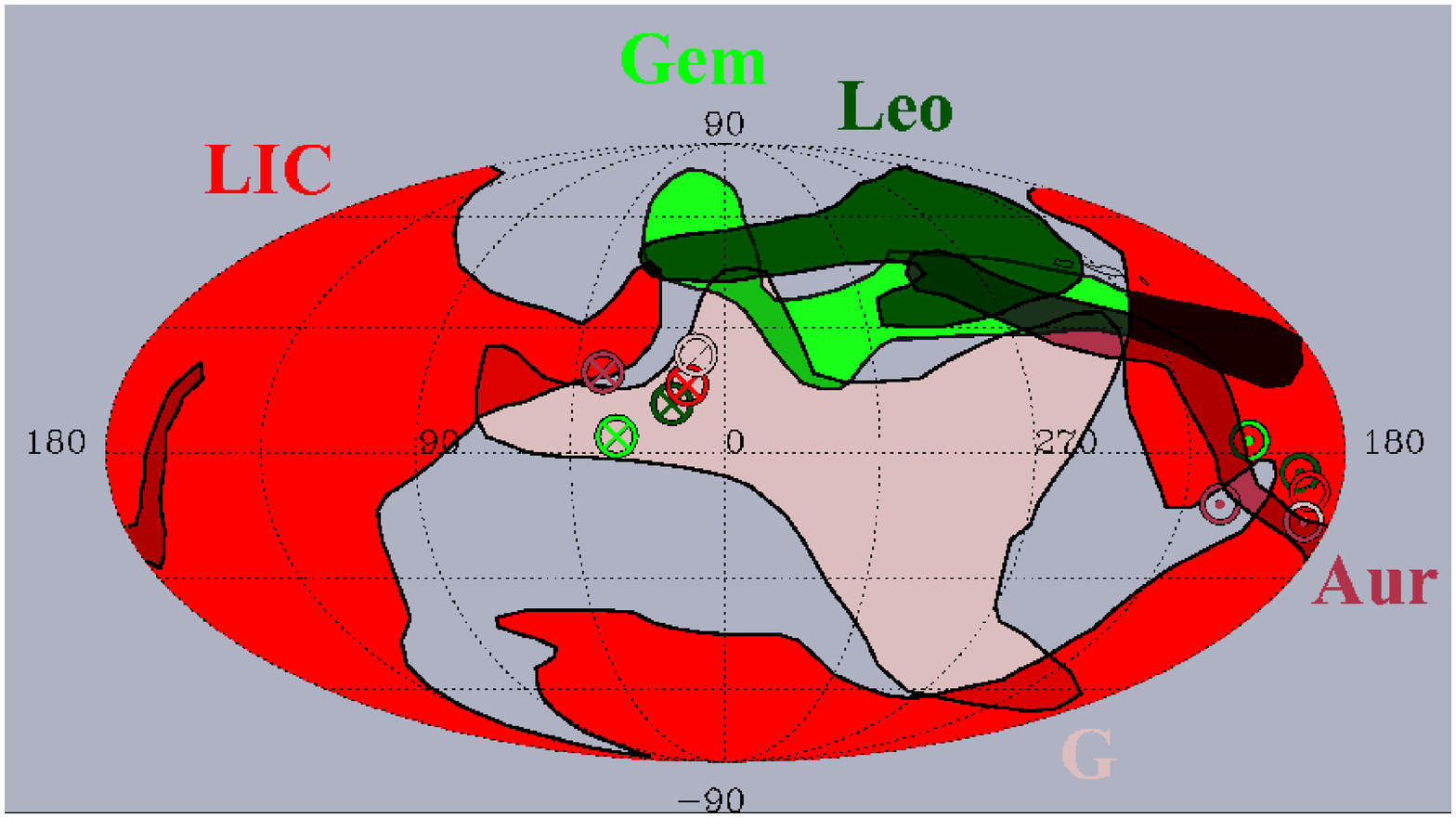}
\end{center}
\caption{Two subsets of dynamical cloud morphologies are overlaid and
colorcoded as in Figure~\ref{fig:vecs}.  The upwind heliocentric
direction of the velocity vectors are indicated by the $\otimes$
symbols, while the downwind heliocentric direction is indicated by the
$\odot$ symbols.  The {\it top} plot shows the projected morphological
similarities shared by the LIC, G, and Mic clouds, indicating that
clouds like the Mic could result from collisions of other clouds, in
this case the LIC and G clouds.  The {\it bottom} plot shows the
clouds in close angular proximity to the cold cloud (shown here are
\ion{H}{1} contours from \citet{heiles03b}, which are the series of
three small clumps centered at $l = 222^{\circ}$ and $b =
44^{\circ}$), which was identified to be within the Local Bubble by
\citet{meyer06}.  Note the alignment of the cold cloud matches well
with the alignment of the high-velocity Gem Cloud in the same
location.  The compressional macroscopic motions between the
surrounding warm dynamical clouds (e.g., the Gem Cloud with the slower
moving Leo, Aur, and LIC clouds), shown here may be the origin
mechanism of the observed cold material.
\label{fig:part}}
\end{figure}

One example of this interaction may be the Mic Cloud whose morphology
appears to mirror the adjacent sections of the LIC and G clouds, as
shown in Figure~\ref{fig:part}.  At positive Galactic latitudes, where
the projected morphologies of the LIC, G, and Mic clouds are
coincident, the median predicted radial velocity difference between
the LIC and G clouds is $\sim$5.5 km~s$^{-1}$.  The Mic Cloud may have
been created by the faster G Cloud colliding with the LIC, which is
moving $\sim$5.5 km~s$^{-1}$ slower in the radial direction.

\subsection{\it Cloud Interactions, Turbulence, and Shocks}

Except for the LIC, many clouds have only 3--5 sight lines with
measurements of physical properties, and six of the clouds have one or
no sight lines with measured physical properties. As a result, it is
difficult to explore the homogeneity or variation of properties across
an individual cloud.  Small-scale variations are not observed in the
warm LISM clouds, based on identical absorption properties of nearby
binary stars (e.g., $\alpha$ Cen A and B,
\citealt{linsky96,lallement95}; and $\alpha$ CMa A and B
\citealt{lallement94,hebrard99}), and by the lack of significant
variation among a sample of 18 closely spaced Hyades stars, down to
scales between 0.1--1 pc \citep{redfield01}.  Therefore, we may expect
that the physical properties of LISM clouds are relatively
homogeneous, or at least slowly varying within a cloud.

Table~\ref{tab:sumprop} lists the weighted mean values of physical
properties for all cloud members, as well as the weighted average
standard deviation, which gives an indication of how tightly the
values are scattered about the weighted mean.  For example, $D($Fe$)$ in
the LIC, even though there are 12 measurements, much more than any
other cloud, it has the lowest weighted average standard deviation,
indicating that little variation is detected across the LIC.  In
contrast, due to the a wide range of values in the Hyades and Mic
clouds, there is a large weighted average variance (although these
clouds have only a few measurements).  In particular, note the
anomalous depletion measurement of G191-B2B associated with the Hyades
Cloud, see Table~\ref{lism4_hyadestable}.  The $\sim$8.6~km~s$^{-1}$
component, consistent dynamically and spatially with the Hyades Cloud,
is detected in both low-ionization ions (e.g., \ion{D}{1}, \ion{N}{1},
\ion{O}{1}, \ion{Mg}{2}, etc, \citealt{lemoine96,redfield04sw}) and
high-ionization ions (e.g., \ion{C}{4}, \citealt{vennes01}).  The
nature of the absorbing material along this line of sight is not
well known, and the high-ionization material may be associated with
nebular circumstellar material surrounding G191-B2B
\citep{bannister03}.  Although such contamination may be present along
some sight lines in our sample, the need in this analysis to bring
together a large number of independent LISM measurements aids in
reducing and identifying anomalous data points.  Indeed, the high
weighted average standard deviation of depletion in the Hyades Cloud,
clearly identifies the G191-B2B sight line as anomalous.  Henceforth,
we assume all measurements sample the physical properties of the LISM,
although highly deviant data points may indicate interesting sight
lines that require additional observations and further attention.

If we assume we can estimate an individual cloud's mean properties by
assuming that clouds are approximately homogeneous, then we can use
the few available physical measurements to make a reasonable estimate
of the cloud properties, and search for possible correlations.  For
example, the Mic Cloud, already identified by its filamentary
morphology and location at the boundary of the LIC and G clouds, has
the highest temperature ($\langle T \rangle = 9900$ K) and one of the
highest turbulent velocities ($\langle \xi \rangle = 3.1$ \kms). These
properties support the argument that the Mic Cloud is the result of
the collision of the LIC and G Clouds.

\begin{figure}
\plottwo{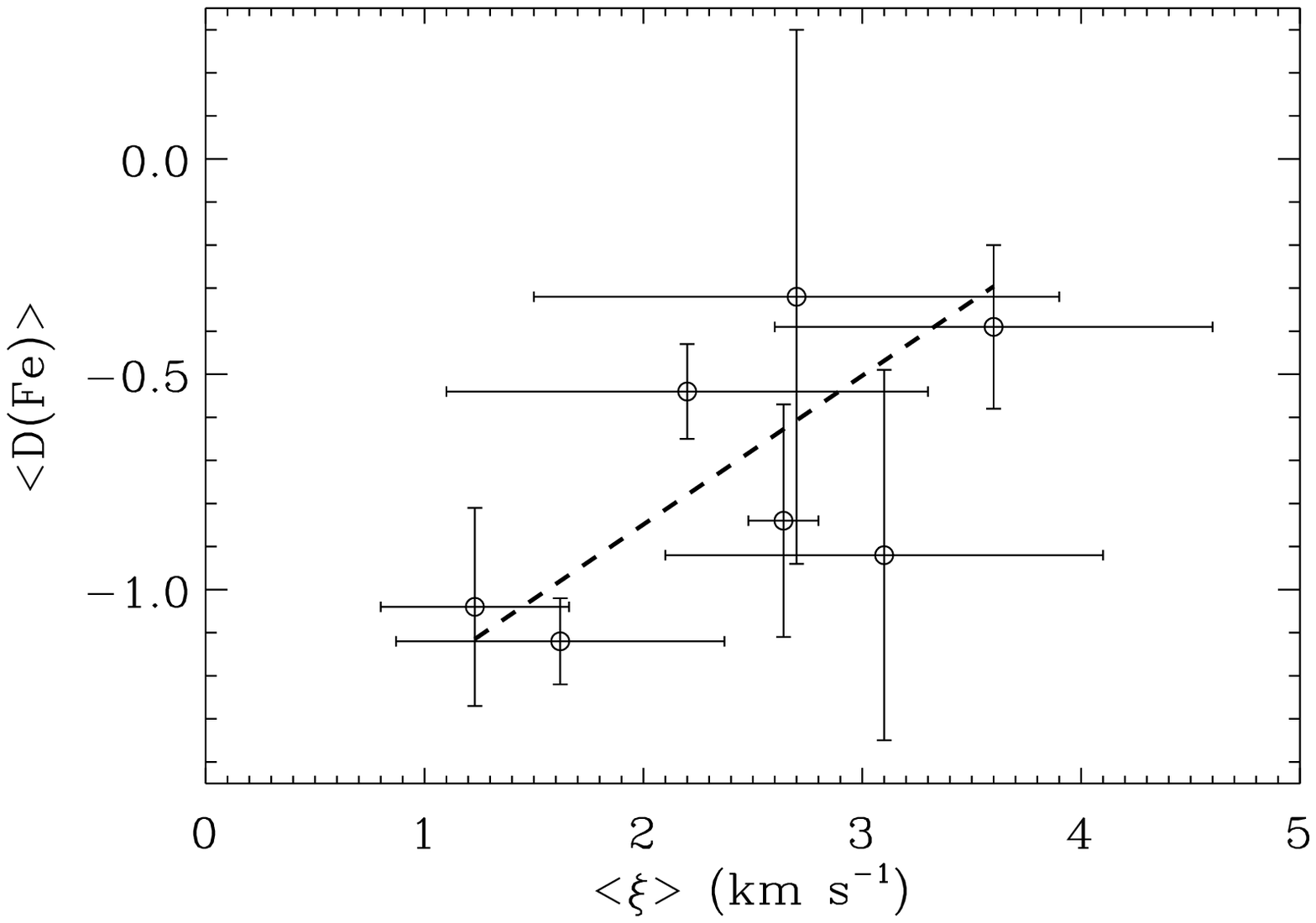}{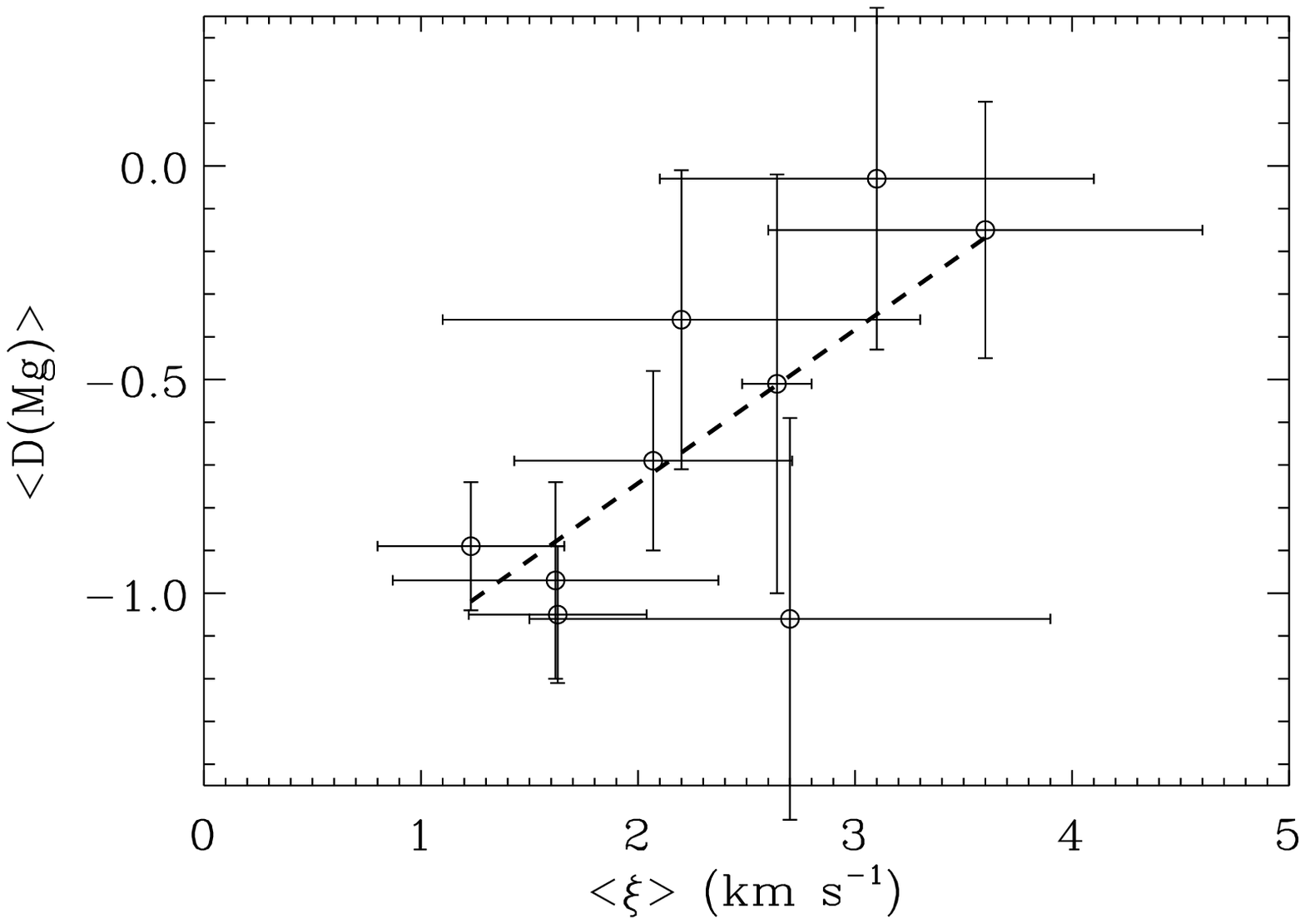}
\caption{Distribution of the weighted mean values of turbulent
velocity ($\xi$) and depletion of iron ($D($Fe$)$ and magnesium
($D($Mg$)$) for all clouds with more than one sight line having a
turbulent velocity measurement.  The errors are the dispersion about
the weighted mean.  The dashed line is a weighted minimum $\chi^2$
linear fit to the data.  A clear correlation exists between cloud
turbulent velocity and depletion for both elements, with a linear
correlation coefficient $r = 0.69$ and a probability that the
distribution could be drawn from an uncorrelated parent population of
only $P_c = 1.7\%$ for iron and $r = 0.73$ and $P_c = 1.2\%$ for
magnesium.  It is likely that regions of high turbulence result from
the dynamical interactions of clouds, which in turn produce shocks and
heat any dust, returning metal ions from the dust to the gas phase.
\label{fig:xidep}}
\end{figure}

The weighted mean depletions of iron [$D({\rm Fe})$] and magnesium
[$D({\rm Mg})$] are plotted versus the weighted mean turbulent
velocity ($\xi$) for the nine clouds in Figure~\ref{fig:xidep}.  A
clear correlation is evident.  For iron, the linear correlation
coefficient, $r = 0.69$, and the probability ($P_c$) that this
distribution could be drawn from an uncorrelated parent population is
1.7\%, while for magnesium, $r = 0.73$ and $P_c = 1.2\%$.

The correlation of small depletions with high turbulence suggests that
the destruction of dust grains has returned these ions to the gas
phase.  A possible alternative explanation is that statistically,
clouds with higher turbulence have higher percentage ionization of
hydrogen, since the depletions were computed assuming that hydrogen is
neutral.  However, using the ionization model of the LIC produced by
\citet{slavin02}, taking the ionization of hydrogen, magnesium, and
iron into account only produces a 0.05 to 0.10 decrease in the
measured depletion.  This adjustment is significantly less than the
typical 1$\sigma$ error for the weighted mean depletion for individual
clouds, and much less than the $\sim$1 dex variation seen over all LISM
clouds.  Therefore, since we currently have no evidence for a
correlation between turbulence and ionization structure and the
depletion adjustment using LIC ionization models is small, we consider
here possible dust destruction explanations for the observed
correlation.  Dust destruction in the warm partially ionized ISM is
often discussed in the context of shocks produced by supernovae
\citep{savage96,jones94}.  Shocks may also be produced in the LISM
from either turbulent motions possibly driven by shear flow
interactions between clouds, or from direct macroscopic collisions of
clouds.  The thermal sound speed $c_s = \sqrt{nkT/\rho}$, using LISM
densities from \citet{redfield06} and the mean LISM temperature from
\citet{redfield04tt}, is $\sim$8 \kms.  For the clouds with the
highest turbulence (e.g., Mic and Eri), the sight line-averaged
turbulent velocity is $\sim$3.5 \kms, which results in a turbulent
Mach number ($M_{\xi}$) of $\sim$0.4.  Although there may be regions
of enhanced turbulent motions, perhaps at the interaction boundaries
of clouds, average turbulent velocities are not high enough to produce
shocks.

\begin{figure}
\epsscale{1.1} 
\plotone{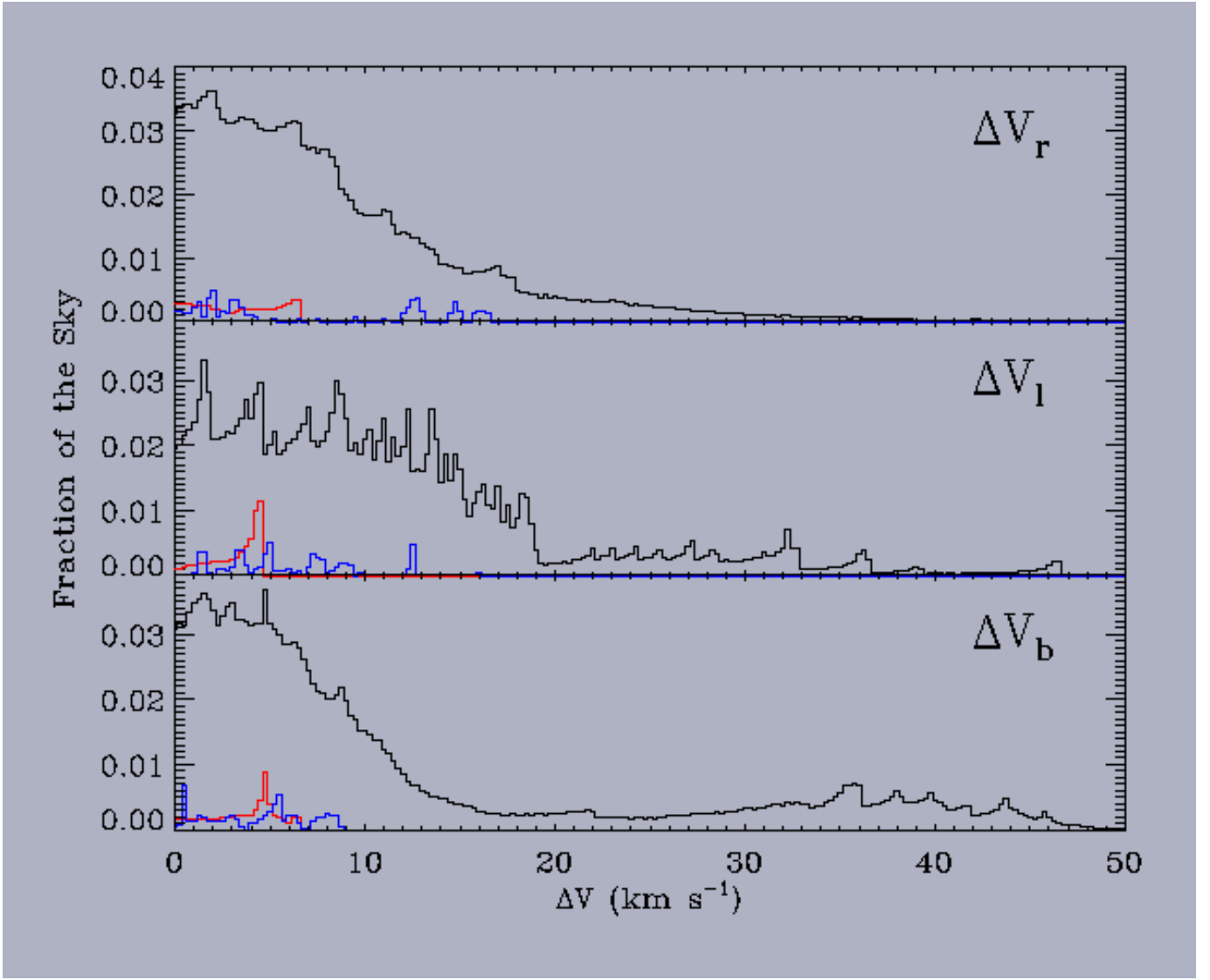}
\caption{ The distribution of cloud velocity differences between all
15 LISM clouds are shown here for all sight lines that are predicted
to traverse multiple clouds based on the spatial distribution of
clouds shown in Figure~\ref{fig:all}.  The predicted velocity
components of all 15 clouds were calculated for a uniform sample of
hypothetical lines of sight over the full sky.  Velocity differences
between the LIC and G clouds in directions in which both are observed
are indicated by the red histograms.  The mean thermal sound speed is
$\sim$8 km~s$^{-1}$.  A significant fraction of possible velocity
differences between LISM clouds include velocities greater than the
thermal sound speed.  Distributions in the radial ({\it top}), and
transverse ($l$, {\it middle}; $b$, {\it bottom}) Galactic directions
are shown.  The blue histogram (scaled by a factor of 50) represents
the macroscopic velocity differences for LISM material near the Leo
cold cloud.  Significant compressional velocities of the warm LISM
material in the radial direction may be a mechanism for the origin of
the cold dense material observed by \citet{meyer06}.
\label{fig:dvs}} 
\end{figure}

Macroscopic velocity differences between the 15 LISM clouds can be
significantly greater than the sound speed if they are interacting.
Figure~\ref{fig:dvs} shows the distribution of predicted velocity
differences ($\Delta$$V$) between the 15 LISM clouds when multiple
clouds are predicted to lie along a line of sight.  For a uniform
sample of hypothetical individual sight lines across the entire sky,
we predict the number of the 15 LISM clouds are predicted to lie along
the line of sight based on their boundaries shown in
Figures~\ref{fig:lic}--\ref{fig:cet}.  If multiple clouds are
predicted to lie along the line of sight, we calculate all possible
cloud velocity differences, which are shown in Figure~\ref{fig:dvs}.
For example, a hypothetical sight line that traverses two clouds
(e.g., at $l=270^{\circ}$ and $b=0^{\circ}$ where the G and Cet clouds
overlap) will provide one velocity difference measurement in
Figure~\ref{fig:dvs}, whereas, if three clouds are predicted along the
line of sight (e.g., at $l=165^{\circ}$ and $b=0^{\circ}$ where the
LIC, Aur and Hyades clouds overlap), three possible velocity
difference measurements are shown.  Calculating the radial and
transverse components of a velocity vector of LISM material along an
arbitrary line of sight requires projecting the velocity vector
($V_0$, $l_0$, $b_0$) along the radial and transverse unit vectors in
the arbitrary direction ($l$, $b$).  The magnitude of the radial and
transverse velocities are calculated from
\begin{eqnarray}
V_r = V_0 (\cos b \cos b_0 \cos (l_0-l) + \sin b_0 \sin b),\\
V_l = V_0 \cos b_0 \sin (l_0 - l),\phantom{(\cos b + \sin b_0 \sin b)}\\
V_b = V_0 (\sin b_0 \cos b - \cos b_0 \sin b \cos (l_0-l)).
\end{eqnarray}
Since we do not have a fully three-dimensional model of these LISM
clouds (i.e., we do not know which clouds are in fact adjacent and
actively interacting), we do not know whether all these
velocity differences are actually realized.  However, a large
percentage of the possible velocity differences are supersonic and
thus could cause shocks where adjacent clouds meet.

The question of grain destruction/erosion involves many factors
including grain size, composition, porosity, the relative speed of
collisions with other grains or particles, compression ratio in
shocks, grain charge, magnetic fields, and turbulence. This topic has
been addressed by a number of authors \citep*[e.g.,][]{jones96}.
\citet{frisch99} and others showed that the observed Fe and Mg ions in
warm gas like the LIC comes from grain destruction whether by shocks,
grain-grain collisions, or other physical processes. Since most of the
Fe and Mg in the LIC is locked up in grains, small differences in
grain destruction between or within clouds can produce large
differences in the gas phase abundances of these elements. High-speed
supernova-generated shocks (50--200 km~s$^{-1}$) are often cited as
the main grain destruction method, but grain-grain collisions with
relative velocities exceeding only 2.7 km~s$^{-1}$ for silicate grains
or 1.2 km~s$^{-1}$ for carbonaceous grains can lead to grain
shattering \citep{jones96}. Velocities exceeding these values are
typically found between two clouds along the same line of sight
(Figure~\ref{fig:dvs}) and are similar to the measured cloud turbulent
velocities. Given that interstellar dust grains are typically charged
and the ISM is magnetized and turbulent, \citet*{yan04} showed that
MHD turbulence can accelerate the grains through gyroresonance
interactions leading to supersonic grain speeds, grain-grain
collisions and shattering. This process could be the physical basis
for the observed correlation of low metal depletions with high
turbulent velocities.

\subsection{\it Connection with Radio Scintillation Screens}

For many years, radio observers have called attention to a rapid
variability of certain quasars and pulsars on hourly-to-yearly
timescales that has been attributed to interstellar scintillation.
The scattering screens responsible for the scintillation are generally
assumed to be turbulent regions of enhanced electron density.
Extensive monitoring of a source and measurement of time delays as
seen by widely separated radio telescopes provide critical data for
estimating the distance to the scattering screen, as well as its size,
transverse velocity, and shape.  In their VLA survey of northern sky
AGN for rapid intraday variability, \citet{lovell07} found that
56\% of the sources are variable on timescales of hours to several
days, but rapid variability indicative of nearby scattering screens is
rare indicating that nearby scattering screens cover only a small
fraction of the sky.

Studies of intraday variability of three quasars (J1819+3845, PKS
1257-326, and PKS 0405-385) and two pulsars (PSR J0437-4715 and PSR
B1133+16) find that some scattering screens lie within the LISM at
short distances from the Sun, although the distances have
significant uncertainty and are model dependent.  For example,
\citet{dennettthorpe03} and \citet{macquart06} estimate that the
scattering screen toward J1819+3845 is only 1--12 pc from the Sun.
\citet{bignall06} found that the scattering screen toward PKS
1257-3826 lies at a distance somewhat closer than 10 pc.
\citet*{rickett02} place the anisotropic scattering screen toward PKS
0405-385 at between 2 and 30 pc from the Sun with a preferred distance
of 25 pc, though analysis of more recent data by
\citet{kedziorachudczer06} suggests a distance of about 10 pc.
\citet*{smirnova06} show that the scattering screen toward PSR
J0437-4715 also lies at about 10 pc from the Sun and is likely the
same screen that causes the scintillation of PKS 0405-385, which is
only 10 degrees away.  If so, this scattering screen is extended
rather than very restricted in size.  \citet{putney06} present six
pulsars that show evidence of multiple scintillation screens along
their line of sight, the vast majority of which are located well
beyond the Local Bubble.  One of their nearest pulsars, PSR B1133+16, shows
evidence for a nearby scintillation screen only 21.6 pc from the Sun.
Several other scintillating quasars, or intraday variables, show
annual cycles that may provide constraints on the distance to the
scintillation screen, such as B0917+624 \citep{rickett01,jauncey01}
and PKS 1519-273 \citep{jauncey03}.  

We find that the five nearby scattering screens all lie close to the
edges of several of our dynamical clouds, as indicated in
Figure~\ref{fig:all}, where the direction of the radio scintillation
sources are indicated by star symbols.  In particular, three of the
five lie near the interface of the LIC and G clouds.  The radial
velocity differences between the LIC and G clouds in these directions
are generally quite small (i.e., $\sim$1 km s$^{-1}$).  However, the
transverse motion differences between the LIC and G clouds in these
directions can be quite substantial, reaching 6--7 km s$^{-1}$.  These
regions of significant transverse velocity differences could induce
shear flows and generate turbulence.  The annual variation of the
scintillation timescale of intraday variables, is a function of the
diffraction pattern of the screen and its transverse velocity.  With
our rigid velocity vectors of LISM clouds, we are able to calculate
the transverse motions of local clouds.  Even with just a handful of
sight lines, we can investigate the relationship between the LISM and
the scintillation screens, but many more radio scintillation and high
resolution LISM absorption line observations are needed together with
a fully three dimensional morphological and kinematic model of the
warm LISM, in order to fully explore the physical connection between
warm clouds and scintillation screens.  \citet*{linsky07} more fully
explore the relationship between scattering screens and LISM clouds,
including a direct comparison of the transverse velocities of the
screens and the clouds.

\subsection{\it Connection with Cold Dense Structures in the LISM}

In their 21-cm absorption line study of the warm and cold neutral
interstellar gas, \citet{heiles03b} mapped a region of cold gas
centered at $l = 225^{\circ}$, $b = 44^{\circ}$ that extends over
30$^{\circ}$ in Galactic longitude.  They found that the gas
temperature is about 25 K.  \citet{meyer06} observed narrow
\ion{Na}{1} absorption due to this cold cloud toward a series of
nearby stars and confirm a cloud temperature $\sim$20 K and turbulent
velocity of $\sim$0.4 km~s$^{-1}$.  Based on the well-known distances
of the observed stars, they were able to show that the distance to
this cold gas must be less than 41 pc, with a corresponding aspect
ratio (length perpendicular to the line of sight versus length along
the line of sight) of 70:1.  However, the cloud could be as close as 2
pc.  Thus the cold gas structure is located inside the LISM.  

The Galactic coordinates of the cold gas correspond to a region that
is not clearly a part of any individual warm cloud (see
Figure~\ref{fig:all}), but near the boundaries of several clouds,
including the LIC, G, Aur, Gem, and Leo clouds.  In particular, the
Gem Cloud has a high radial velocity in the direction of the cold
cloud, $\sim$24 km~s$^{-1}$, whereas the other clouds have modest
radial velocities, ranging from 8--12 km~s$^{-1}$.  The resulting high
radial velocity differences are indicated in Figure~\ref{fig:dvs}.
Along this line of sight, the high radial velocity ($\sim$24
km~s$^{-1}$) Gem cloud may be compressing material as it collides with
slow moving Leo and Aur ($\sim$12 km~s$^{-1}$), and ultimately the LIC
($\sim$10 km~s$^{-1}$).  The heliocentric velocity ($\sim$11.5
km~s$^{-1}$) of the cold material observed by \citet{meyer06} matches
well with the velocity of the slow moving Leo and Aur clouds, as
expected if it was formed by the compression of the Gem cloud against
the Leo and Aur clouds, and the cold material may actually be
physically associated with the warm material observed in the Leo and
Aur clouds.

The collision of warm gas clouds to produce small sheetlike cold
neutral clouds has been explored through detailed simulations of a
turbulent interstellar medium
\citep[e.g.,][]{vazquezsemadeni06,audit05}.  Also \citet{mckee77}
predicted that cold neutral clouds must be surrounded by warm clouds
(in pressure equilibrium in their model) to shield the cold gas from
UV and X-ray heating and ionization.  Our rigid velocity vector
solutions certainly indicate that in the direction of the cold cloud,
relatively large kinematic differences exist between clouds (e.g, the
Gem cloud is moving $\sim$12 km~s$^{-1}$ faster than the Leo and Aur
clouds, and $\sim$14 km~s$^{-1}$ faster than the LIC and G clouds).
Given that the farthest distance limit for the five clouds in the
proximity of the cold cloud line of sight is 11.1 pc (the Leo Cloud),
and that the limit on the Gem Cloud is 6.7 pc, these five clouds are
in very close proximity in distance as well, and collisions between
these warm clouds are likely.  In particular, it is critical to have a
collision of material along the radial direction to maximize the
chances of detection.  Because of the extreme aspect ratio of this
cloud, if it were oriented along the line of sight, the projection on
the sky would be extremely small.  Further work on the distances of
both the warm LISM clouds and the cold cloud is needed to determine
whether there is a spatial and dynamical connection between these
interstellar structures.

\subsection{\it The Transition between the LIC and G Clouds\label{sec:licg}}

Until now, we have assumed that individual clouds are rigid
structures, each with a simple velocity vector characterizing all of
the included sight lines within the radial velocity measurement
errors. This simple approximation may not be valid, as shown by two
sets of data. One, noted in Section~\ref{sec:coll}, is that the
velocity and, to a lesser degree of significance, the temperature of
interstellar gas flowing through the heliosphere are intermediate in
value between the corresponding quantities in the LIC and G clouds
(see Table~\ref{tab:vecs}), implying that the heliosphere lies in a
transition zone between the two clouds where there is a gradient in
properties. The other evidence is that \citet{redfield01} noted that
the components assigned to the LIC in the direction of the Hyades have
radial velocities 2.9 km~s$^{-1}$ smaller than predicted by the
\citet{lallement92} LIC vector.  These smaller radial velocities
suggest a deceleration of the LIC flow in the forward direction, where
it may be interacting with the Hyades Cloud. We now find that these
absorption components have radial velocities $\sim$1.0 km~s$^{-1}$
smaller than predicted by the new LIC vector. As a test we removed
these 16 components from the LIC vector calculation and found that the
vector velocity amplitude increased by only 0.39 km~s$^{-1}$. This
does not change our conclusion that the heliosphere is in a transition
zone between the LIC and G clouds, or that the LIC is decelerated at
its forward edge.

\begin{figure}
\includegraphics[angle=90,width=6.3in]{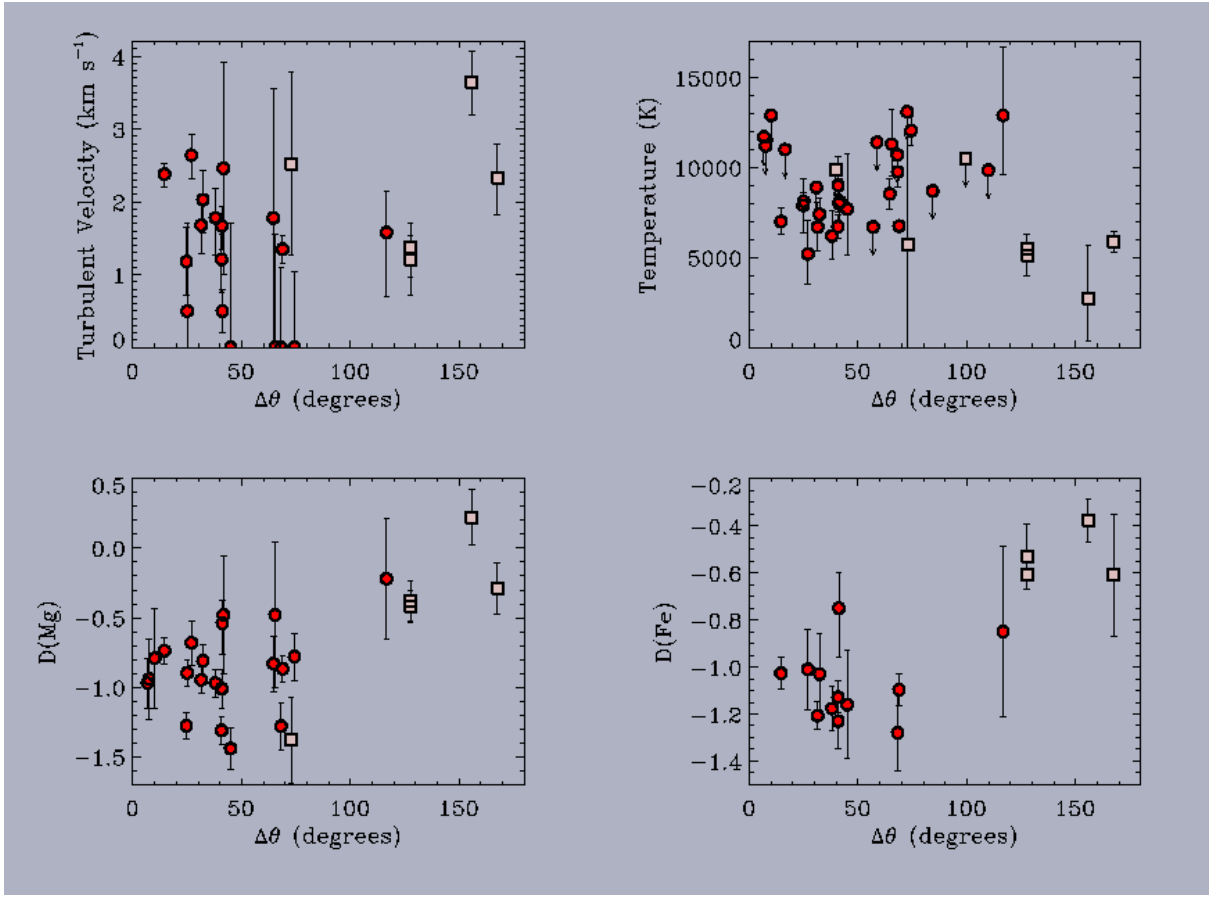}
\caption{Comparison of LIC and G sight line physical properties as a
function of angular distance from the LIC downwind direction.  Sight
lines through each cloud are distinguished by color (red: LIC; pale
pink: G) and symbol (circle: LIC; square: G).  Since the G Cloud is in
the upstream direction, all G Cloud sight lines are at high
$\Delta\theta$.  Except for one LIC value, all temperatures through
the LIC are higher than the G Cloud temperatures.  A correlation with
angle is evident in the depletion of magnesium ($r = 0.63$, $P_c =
0.012$\%) and iron ($r = 0.80$, $P_c = 0.0041$\%).
\label{fig:grad1}}
\end{figure}

\begin{figure}
\includegraphics[angle=90,width=6.3in]{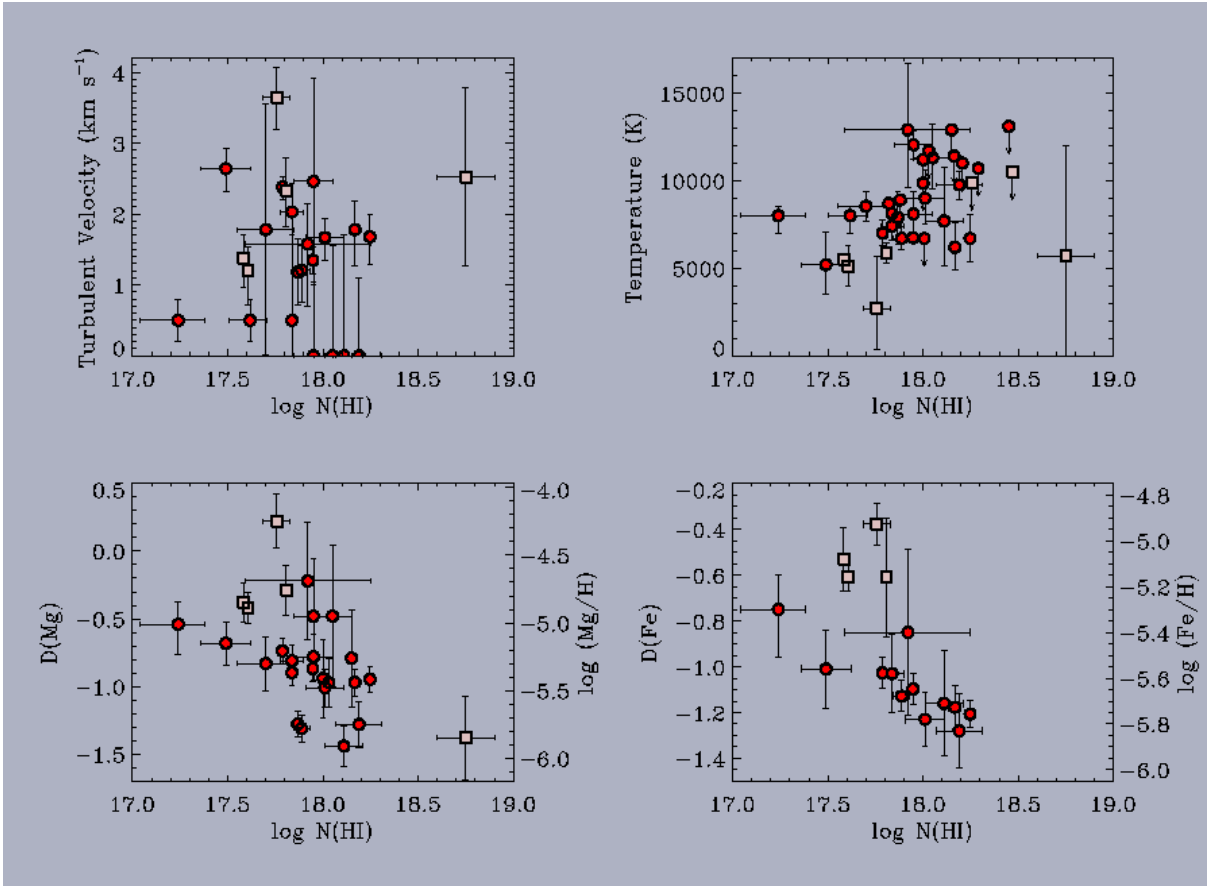}
\caption{Similar to Figure~\ref{fig:grad1}, but the comparison of LIC
and G sight line physical properties is shown as a function of
hydrogen column density.  Correlations are seen between hydrogen
column density and depletion of magnesium ($r = -0.53$, $P_c =
0.12$\%) and iron ($r = -0.65$, $P_c = 0.15$\%).
\label{fig:grad2}}
\end{figure}

We also considered whether the LIC and G Clouds are really one cloud
with a gradient of physical properties across their combined length.
We tested this hypothesis by plotting the physical parameters for the
LIC and G cloud sight lines with respect to angle relative to the
downwind direction (Figure~\ref{fig:grad1}) and with respect to the
hydrogen column density (Figure~\ref{fig:grad2}).  No correlation
exists between angle and hydrogen column density.  Since the LIC Cloud
is mostly in the downwind direction and the G Cloud mostly in the
upwind direction, the angle from the downwind direction is a
discriminant between the two clouds. With only one exception, the gas
temperatures for LIC sight lines are all larger than for the G Cloud
sight lines, implying that the two cloud approximation is valid. We
note, however, a trend of higher LIC Cloud temperatures with
increasing $N($\ion{H}{1}$)$ and toward the cross-wind direction where
the two Clouds meet.  The turbulent velocities do not show a definite
trend with angle or $N($\ion{H}{1}$)$.  If there were a velocity
gradient through these clouds, one would expect larger line broadening
(and thus higher turbulent velocity) in the downwind and upwind
directions, which is not seen. Finally, the metal depletions,
[$D($Mg$)$ and $D($Fe$)$], show clear trends of decreasing with larger
$N($\ion{H}{1}$)$ and increasing with angle, with both the LIC and G
Cloud sight lines fitting these trends.  On average, the LIC is
significantly more depleted than the G cloud.  The trend of decreasing
gas-phase abundances as a function of increasing $N($\ion{H}{1}$)$ is
well documented along distant sight lines (e.g., \citealt{wakker00};
\citealt*{jenkins86}).  This implies that the processes that remove
and replace ions to the gas phase occur on scales smaller than the LIC
and G clouds, and that the volume density of the LIC and G clouds may
not be constant.  Alternatively, decreasing $D($Mg$)$ and $D($Fe$)$
with increasing $N($\ion{H}{1}$)$ may be explained if \ion{H}{1} is
photoionized at the edges of clouds and is increasingly neutral with
increasing \ion{H}{1} through the centers of clouds due to shielding
of UV radiation from the cloud itself.  Thus, while the fraction of
magnesium and iron in the gas phase relative to the total amount of
hydrogen remains the same throughout, $D($Mg$)$ and $D($Fe$)$ will
decrease as hydrogen becomes predominately neutral.  On the basis of
the temperature and turbulent velocity data, we conclude that the
evidence supports the idea that the LIC and G Clouds are separate
entities with their own distinct properties, but there is likely a
narrow transition zone between the two clouds where the heliosphere is
now located.

\section{CONCLUSIONS}

We have created a database consisting of interstellar radial
velocities and gas physical properties for 157 sight lines toward
stars within 100~pc of the Sun. The data were extracted from
high-resolution UV spectra obtained with the GHRS and STIS instruments
on the {\em HST} and ground-based \ion{Ca}{2} spectra. This database
has allowed us to create a dynamical model of the local interstellar
medium\footnote{Projected and transverse velocities, along with
probable cloud membership based on the projected cloud morphologies,
can be calculated for any sight line at
http://cobalt.as.utexas.edu/$\sim$sredfield/LISMdynamics.html.}
including 15 warm gas clouds, which we define as contiguous parcels of
interstellar gas with consistent kinematical properties.  Although
measurements of physical properties are sparse, for the LIC, which has
the most such measurements, the properties seem to be homogeneous.
Using this database, we find that:

\begin{enumerate}

\item The flow velocity vectors for these 15 clouds fit 81.2\% of the
velocity components in the database to within the radial velocity
measurement errors.  These clouds all lie within 15~pc of the Sun. The
remaining velocity components may be produced in more distant clouds
that subtend smaller angles with less than the four lines of sight
needed to compute a useful velocity vector.

\item The directions of most of these velocity vectors are roughly
parallel with their flow from the Scorpio-Centaurus Association.  The
velocity amplitudes have a considerable range, leading us to compute
relative velocities between adjacent clouds that are often supersonic
and therefore capable of producing shocks.

\item About one-third of the clouds appear to have filamentary
structures.

\item All of the clouds for which we have physical properties along
three or more sight lines are warm with mean temperatures in the range
of 5300--9900~K, although the uncertainties in these measurements are
often large.  We estimate that between 5.5\% and 19\% of the LISM
within 15~pc of the Sun is filled with warm gas clouds.

\item We find a strong correlation of low metal depletion with large
turbulent velocity. Since high turbulence suggests the presence or
recent existence of shocks, this correlation could be explained by
shock dissipation of dust grains that returns the metals to the gas
phase.

\item Contrary to previous work, the heliosphere appears to be located
in a transition zone between the LIC and G Clouds. The evidence for
this is that the temperature and velocity of the interstellar gas
flowing through the heliosphere are both intermediate between these
quantities measured in the LIC and G Clouds.  The deviation in
velocity ranges from 2.4--2.9$\sigma$ and in temperature the deviation
is less significant ranging from 0.9--1.4$\sigma$.  Additional
observations of multiple ions are required to increase the number of
temperature measurements of the LIC and G clouds in order to increase
the significance of any possible deviation between these clouds and
{\it in situ} measurements.  Previous work based on much smaller
velocity data sets placed the heliosphere inside but near the edge of
the LIC.

\item The G Cloud is surrounded by and likely interacting with a
number of other clouds. We refer to this active boundary as the ``Ring
of Fire''.  The filamentary-shaped Mic Cloud has the same shape as the
boundary of the G and LIC clouds and may be indicative of an
interaction between these two clouds.  The high temperature and
turbulence of the Mic Cloud support this conclusion.

\item The nearby scintillation screens toward three quasars and two
pulsars are located near cloud boundaries, and three of the five are
in directions where the LIC and G Clouds may be interacting.  The
large transverse relative velocities between these two clouds could
produce the turbulence that is the cause of the scintillation.

\item The nearby cold cloud recently observed by \citet{heiles03b} and
\citet{meyer06} is in a direction where it could be surrounded by
several warm clouds.  We find evidence for significant compression
based on large macroscopic velocity differences between warm clouds in
the direction of the cold cloud.  The alignment of the cold cloud
matches well with the alignment of the high-velocity Gem Cloud, which
may be colliding with the slower moving Leo, Aur, and LIC clouds.
Compression of warm material may be the origin mechanism for such an
isolated cold cloud in the Local Bubble.  

\end{enumerate}

\acknowledgements 

We thank the referee for the careful and thoughtful comments, they
contributed much to the quality of the paper.   We would also like
to thank Dr.\ Brian Wood for his useful comments.  S.R. would like to
acknowledge support for this work provided by NASA through Hubble
Fellowship grant HST-HF-01190.01 awarded by the Space Telescope
Science Institute, which is operated by the Association of
Universities for Research in Astronomy, Inc., for NASA, under contract
NAS 5-26555.  Support for {\it HST} observing programs \#9525 and
\#10236 was provided by NASA through a grant from the Space Telescope
Science Institute.  This research has made use of NASA's Astrophysics
Data System Bibliographic Services and the SIMBAD database, operated
at CDS, Strasbourg, France.  Some of the data presented in this paper
were obtained from the Multimission Archive at the Space Telescope
Science Institute (MAST).

{\it Facilities:} \facility{HST (GHRS, STIS)}



\clearpage

\clearpage

\clearpage

\clearpage

\clearpage

\clearpage

\clearpage

\clearpage

\clearpage

\clearpage

\clearpage

\clearpage

\clearpage

\clearpage

\clearpage

\clearpage

\clearpage

\clearpage

\clearpage

\clearpage

\clearpage

\clearpage

\clearpage

\clearpage

\begin{deluxetable}{lcccccccccl}
\tablewidth{0pt}
\tabletypesize{\tiny}
\tablecaption{LIC Cloud Sight Line Properties \label{tab:licmem}}
\tablehead{Star & HD & $d$ & $v$ & $\sigma$\tablenotemark{a} & $\log N({\rm H~I})$ & $T$ & $\xi$ & $D($Fe$)$ & $D($Mg$)$ & Other \\ 
Name & \# & (pc) & (\kms) & & (cm$^{-2}$) & (K) & (\kms) & & & Clouds\tablenotemark{b} \\}
\startdata
$\alpha$ CMa A & 48915 & 2.6 & $ 18.55$ & 0.12 & 17.2 &
 8000$^{+500}_{-1000}$ & 0.5 $\pm$ 0.3 & $-0.75^{+0.15}_{-0.21}$ & $-0.54^{+0.17}_{-0.22}$ & \\
$\alpha$ CMa B & 48915B &  2.6 & $ 17.60$ & 0.51 & 17.6 &
 8000$^{+500}_{-1000}$ & 0.5 $\pm$ 0.3 &         &      & \\
$\epsilon$ Eri & 22049 &  3.2 & $ 18.73$ & 0.70 & 17.8 &
 7410$^{+860}_{-830}$ & 2.03$^{+0.41}_{-0.45}$ & $-1.03 \pm 0.17$ & $-0.81 \pm 0.12$ & \\
 $\alpha$ CMi  & 61421 &  3.5 & $ 19.76$ & 0.78 & 17.9 &
 6710$^{+660}_{-630}$ & 1.21$^{+0.35}_{-0.45}$ & $-1.13 \pm 0.07$ & $-1.31 \pm 0.10$ & Aur\\
$\epsilon$ Ind & 209100 &  3.6 & $ -9.20$ & 0.02 & 18.0 &
$<$9850 & &         &        & Vel\\
   $\tau$ Cet  & 10700 &  3.6 & $ 12.34$ & 0.20 & 18.0 &
$<$6700 & &         &        & G, Blue\\
     40 Eri A  & 26965 &  5.0 & $ 21.73$ & 0.58 & 17.8 &
$8120 \pm 450$ & 0.5$^{+1.2}_{-0.5}$ &         & $-0.90 \pm 0.10$ & \\
 $\eta$ Cas A  & 4614 &  6.0 & $ 11.18$ & 0.67 &        &
      & &         &        &  \\
  $\alpha$ PsA & 216956 &  7.7 & $ -5.87$ & 0.42 &        &
      & &         &        & Mic\\
  $\alpha$ Lyr & 172167 &  7.8 & $-12.90$ & 0.30 &        &
      & &         &        & G\\
 $\chi^1$ Ori  & 39587 &  8.7 & $ 23.08$ & 0.38 & 17.8 &
 7000$^{+730}_{-680}$ & 2.38$^{+0.15}_{-0.17}$ & $-1.03 \pm 0.07$ & $-0.74 \pm 0.09$ & \\
 $\delta$ Eri  & 23249 &  9.0 & $ 19.60$ & 0.07 & 17.9 &
$<$8900 & &         &        & Blue, Hyades\\
$\kappa^1$ Cet & 20630 &  9.2 & $ 20.84$ & 0.31 & 17.5 &
 5200$^{+1900}_{-1700}$ & 2.64$^{+0.28}_{-0.32}$ & $-1.01 \pm 0.17$ & $-0.68 \pm 0.16$ & \\
       CF UMa  & 103095 &  9.2 & $  2.05$ & 1.19 &        &
      & &         &        & \\
  $\beta$ Gem  & 62509 & 10.3 & $ 19.65$ & 0.60 & 18.0 &
 9000$^{+1600}_{-1500}$ & 1.67$^{+0.27}_{-0.32}$ & $-1.23 \pm 0.12$ & $-1.01 \pm 0.14$ & \\
       EP Eri  & 17925 & 10.4 & $ 19.50$ & 0.88 & 18.0 &
$8100 \pm 1300$ & 2.46 $\pm$ 1.45 &         & $-0.48 \pm 0.42$ & G, Blue\\
 $\gamma$ Ser  & 142806 & 11.1 & $-18.19$ & 0.93 &        &
      & &         &        & Mic, Leo\\
       13 Per  & 16895 & 11.2 & $ 16.45$ & 0.46 &        &
      & &         &        & \\
       HR 1925 & 37394 & 12.2 & $ 17.50$ & 1.25 & 18.3 &  
$<$9900 & &       &       & Aur\\
 $\alpha$ Aur  & 34029 & 12.9 & $ 21.48$ & 0.82 & 18.2 &
 6700$^{+1400}_{-1300}$ & 1.68$^{+0.32}_{-0.39}$ & $-1.21 \pm 0.06$ & $-0.95 \pm 0.10$ & \\
         HR 8  & 166 & 13.7 & $  6.50$ & 0.44 & 18.3 &
$<$10700 & &         &        & Eri, Hyades\\
  $\alpha$ Oph & 159561 & 14.3 & $-22.57$ & 1.68 &        &
      & &         &        & \\
       72 Her  & 157214 & 14.4 & $-15.51$ & 0.33 &        &
      & &         &        & \\
  $\alpha$ Cep & 203280 & 15.0 & $  0.20$ & 0.54 &        &
      & &         &        & \\
 $\sigma$ Boo  & 128167 & 15.5 & $ -2.58$ & 0.63 &        &
      & &         &        & \\
       99 Her  & 165908 & 15.7 & $-17.43$ & 1.54 &        &
      & &         &        & G\\
  $\beta$ Cas  & 432 & 16.7 & $  9.15$ & 0.69 & 18.2 &
 9760$^{+800}_{-880}$ & 0.0$^{+1.1}_{-0.0}$ & $-1.28 \pm 0.16$ & $-1.28 \pm 0.17$ & \\
       DX Leo  & 82443 & 17.7 & $ 11.00$ & 0.22 & 17.7 &
$8540 \pm 850$ & 1.78 $\pm$ 1.77 &         & $-0.83 \pm 0.20$ & \\
 $\tau^6$ Eri  & 23754 & 17.9 & $ 16.99$ & 0.04 &        &
      & &         &        & \\
     V368 Cep  & 220140 & 19.7 & $  6.00$ & 0.19 & 18.0 &
12050$^{+820}_{-790}$ & 0.0$^{+1.0}_{-0.0}$ &         & $-0.78 \pm 0.17$ & \\
  $\alpha$ Tri & 11443 & 19.7 & $ 17.89$ & 1.84 & 18.1 &
 7700$^{+3100}_{-2600}$ & 0.0$^{+1.7}_{-0.0}$ & $-1.16 \pm 0.23$ & $-1.44 \pm 0.15$ & \\
      HR 4345  & 97334 & 21.7 & $  4.30$ & 0.14 & 17.8 &
$<$8700 & &         &        & NGP\\
       PW And  & 1405 & 21.9 & $  8.50$ & 0.13 & 18.1 &
11300$^{+1900}_{-1800}$ & 0.0$^{+1.6}_{-0.0}$ &         & $-0.48 \pm 0.52$ & Eri, Hyades\\
   SAO 136111  & 73350 & 23.6 & $ 12.00$ & 0.50 & 18.2 &
$<$11400 & &         &        & G, Aur\\
 $\delta$ UMa  & 106591 & 25.0 & $  3.80$ & 1.65 &        &
      & &         &        & \\
  $\beta$ Aur  & 40183 & 25.2 & $ 22.30$ & 1.74 &        &
      & &         &        & \\
 $\gamma$ UMa  & 103287 & 25.6 & $  4.40$ & 1.54 &        &
      & &         &        & \\
 $\lambda$ And & 222107 & 25.8 & $  6.50$ & 0.03 & 18.5 &
$<$13100 & &         &       & Hyades \\
 $\sigma$ Cet  & 15798 & 25.8 & $ 15.99$ & 0.81 &        &
      & &         &        & \\
 $\tau^3$ Eri  & 18978 & 26.4 & $ 15.90$ & 0.82 &        &
      & &         &        & \\
       HR 860  & 17948 & 26.5 & $ 15.10$ & 0.31 &        &
      & &         &        & \\
    SAO 32862  & 198084 & 27.1 & $ -2.60$ & 0.71 &        &
      & &         &        & \\
      HR 1099  & 22468 & 29.0 & $ 21.90$ & 0.88 & 17.9 &
$7900 \pm 1500$ & 1.18 $\pm$ 0.47 & $<-1.12$ & $-1.28 \pm 0.10$ & \\
 $\theta$ Peg  & 210418 & 29.6 & $ -4.20$ & 1.30 &        &
      & &         &        & Eri\\
    SAO 85045  & 157466 & 29.8 & $-19.02$ & 0.96 &        &
      & &         &        & \\
   $\eta$ Ari  & 13555 & 30.1 & $ 16.99$ & 0.50 &        &
      & &         &        & Hyades\\
 $\delta$ Cas  & 8538 & 30.5 & $ 13.05$ & 1.06 &        &
      & &         &        & \\
  $\alpha$ Gru & 209952 & 31.1 & $-10.93$ & 0.58 &        &
      & &         &        & Vel\\
  $\alpha$ Lac & 213558 & 31.4 & $  3.50$ & 1.11 &        &
      & &         &        & Hyades\\
       DK UMa  & 82210 & 32.4 & $  9.41$ & 0.06 & 17.9 &
 6750$ \pm 240$ & 1.35$^{+0.18}_{-0.20}$ & $-1.10 \pm 0.07$ & $-0.87 \pm 0.09$ & \\
$\epsilon$ Gru & 215789 & 39.7 & $ -7.30$ & 0.43 &        &
      & &         &        & Vel\\
    SAO 76593  & 27808 & 40.9 & $ 23.10$ & 0.17 &        &
      & &         &        & Aur\\
    SAO 93981  & 28568 & 41.2 & $ 23.90$ & 0.39 & 18.0 &
$<$11700 & &         & $-0.97 \pm 0.18$ & Aur\\
   SAO 111879  & 28736 & 43.2 & $ 21.60$ & 1.51 &        &
      & &         &        & Aur\\
      101 Tau  & 31845 & 43.3 & $ 22.40$ & 1.44 &        &
      & &         &        & Aur \\
    SAO 94033  & 29225 & 43.5 & $ 22.50$ & 1.10 &        &
      & &         &        & Aur \\
    SAO 93982  & 28608 & 43.6 & $ 23.20$ & 0.20 &        &
      & &         &        & Aur\\
    SAO 76683  & 29419 & 44.2 & $ 23.20$ & 0.25 &        &
      & &         &        & Aur\\
     V993 Tau  & 28205 & 45.8 & $ 23.30$ & 0.14 & 18.0 &
$<$11200 & &         & $-0.94 \pm 0.29$ & Aur\\
    SAO 93963  & 28406 & 46.3 & $ 22.10$ & 1.37 &        &
      & &         &        & Aur\\
    SAO 76609  & 28033 & 46.4 & $ 23.60$ & 0.28 & 18.2 &
$<$12900 & &         & $-0.79 \pm 0.36$ & Aur\\
     V471 Tau  & \nodata & 46.8 & $ 20.90$ & 0.52 & 18.2 &
$<$11000 & &         &        & Aur, Hyades\\
    SAO 93945  & 28237 & 47.2 & $ 22.40$ & 0.98 &        &
      & &         &        & Aur\\
    SAO 93831  & 26784 & 47.4 & $ 23.00$ & 0.05 &        &
      & &         &        & Aur\\
 $\gamma$ Aqr  & 212061 & 48.4 & $ -4.52$ & 0.76 &        &
      & &         &        & \\
    SAO 56530  & 21847 & 48.9 & $ 21.10$ & 0.39 &        &
      & &         &        & \\
    SAO 93885  & 27561 & 51.4 & $ 22.20$ & 1.11 &        &
      & &         &        & Aur\\
 $\kappa$ And  & 222439 & 52.0 & $  7.60$ & 1.06 &        &
      & &         &        & Hyades\\
 $\delta$ Cyg  & 186882 & 52.4 & $ -9.60$ & 1.47 &        &
      & &         &        & G\\
    SAO 93913  & 27848 & 53.4 & $ 22.40$ & 0.98 &        &
      & &         &        & Aur\\
      HR 1608  & 32008 & 54.7 & $ 21.60$ & 0.07 & 17.8 &
      & &         &        & Blue\\
       45 Aur  & 43905 & 57.0 & $ 18.47$ & 0.13 &        &
      & &         &        & \\
  $\zeta$ Peg  & 214923 & 64.1 & $ -2.20$ & 0.34 &        &
      & &         &        & Eri\\
  $\iota$ Cap  & 203387 & 66.1 & $-12.06$ & 0.06 & 17.9 &
12900$^{+3800}_{-3300}$ & 1.58$^{+0.56}_{-0.89}$ & $-0.85 \pm 0.36$ & $-0.22 \pm 0.43$ & \\
   $\eta$ Aur  & 32630 & 67.2 & $ 23.00$ & 0.94 &        &
      & &         &        & \\
     G191-B2B  & \nodata & 68.8 & $ 19.19$ & 0.09 & 18.2 &
 6200$^{+1400}_{-1300}$ & 1.78$^{+0.40}_{-0.51}$ & $-1.18 \pm 0.09$ & $-0.97 \pm 0.10$ & \\
  $\iota$ Oph  & 152614 & 71.7 & $-21.50$ & 0.13 &        &
      & &         &        & \\
     Feige 24  & \nodata & 74.4 & $ 17.60$ & 0.17 & 18.1 &
      & &         &        & Hyades\\
 $\gamma$ Ori  & 35468 & 74.5 & $ 25.40$ & 1.88 &        &
      & &         &        & Aur\\
\enddata
\tablenotetext{a}{$\sigma = (|v_0 - v_{\star}|) / \sigma_v$, where we have imposed a minimum $\sigma_v$ of 1 \kms for all high resolution data, and a minimum $\sigma_v$ of 3 \kms for all medium resolution data.}
\tablenotetext{b}{Other dynamical clouds for which the observed velocity is within 3$\sigma$ of the predicted value and the sight line is closer than 10$^{\circ}$ to the cloud.  These are less likely, but possible dynamical assignments for these sight lines.}
\end{deluxetable}  

\begin{deluxetable}{lcccccccccl}
\tablewidth{0pt}
\tabletypesize{\tiny}
\tablecaption{G Cloud Sight Line Properties\label{lism4_gtable}}
\tablehead{Star & HD & $d$ & $v$ & $\sigma$\tablenotemark{a} & $\log N({\rm H~I})$ & $T$ & $\xi$ & $D($Fe$)$ & $D($Mg$)$ & Other\\ 
Name & \# & (pc) & (\kms) & & (cm$^{-2}$) & (K) & (\kms) & & & Clouds\tablenotemark{b}\\}
\startdata
$\alpha$ Cen B & 128621 &  1.3 & $-18.14$ & 0.02 & 17.6 &
 5500$^{+330}_{-320}$ & 1.37$^{+0.34}_{-0.41}$ & $-0.53 \pm 0.14$ & $-0.38 \pm 0.14$ & \\
$\alpha$ Cen A & 128620 &  1.3 & $-18.45$ & 0.33 & 17.6 &
 5100$^{+1200}_{-1100}$ & 1.21$^{+0.33}_{-0.49}$ & $-0.61 \pm 0.06$ & $-0.42 \pm 0.11$ & \\
       70 Oph  & 165341 &  5.1 & $-26.50$ & 0.07 & 17.8 &
 2700$^{+3000}_{-2300}$ & 3.64$^{+0.42}_{-0.44}$ & $-0.38 \pm 0.09$ & $ 0.22 \pm 0.20$ & Mic\\
     36 Oph A  & 155886 &  5.5 & $-28.40$ & 0.17 & 17.8 &
$5870 \pm 560$ & 2.33$^{+0.46}_{-0.51}$ & $-0.61 \pm 0.26$ & $-0.29 \pm 0.18$ & \\
 $\alpha$ Lyr  & 172167 &  7.8 & $-16.86$ & 1.33 &        &
      &  &         &         & Mic\\
       AB Dor  & 36705 & 14.9 & $  5.19$ & 0.15 &        &
      &  &         &         & Blue, Dor\\
       LQ Hya  & 82558 & 18.3 & $  6.50$ & 0.23 & 18.8 &
 5700$^{+6300}_{-5700}$ & 2.52 $\pm$ 1.26 &         & $-1.38 \pm 0.31$ & LIC, Aur\\
 $\alpha$ Hyi  & 12311 & 21.9 & $  4.90$ & 0.76 &        &
      &  &         &         & Cet\\
 $\delta$ Vel  & 74956 & 24.4 & $  1.30$ & 1.18 &        &
      &  &         &         & \\
  $\zeta$ Aql  & 177724 & 25.5 & $-21.30$ & 0.30 &        &
      &  &         &         & Eri\\
 $\sigma$ Cet  & 15798 & 25.8 & $ 21.85$ & 0.39 &        &
      &  &         &         & \\
 $\tau^3$ Eri  & 18978 & 26.4 & $ 20.90$ & 0.12 &        &
      &  &         &         & \\
      HR 4023  & 88955 & 31.5 & $ -1.70$ & 0.36 &        &
      &  &         &         & \\
    SAO 68491  & 184499 & 32.0 & $-14.28$ & 0.41 &        &
      &  &         &         & LIC\\
  $\beta$ Car  & 80007 & 34.1 & $ -4.32$ & 1.55 &        &
      &  &         &         & \\
    $\mu$ Vel  & 93497 & 35.3 & $ -4.38$ & 0.41 & 18.5 &
$<$10500 &  &         &         & Cet\\
  $\beta$ Lib  & 135742 & 49.1 & $-26.90$ & 0.61 &        &
      &  &         &         & \\
      $o$ Ser  & 160613 & 51.5 & $-29.00$ & 0.33 &        &
      &  &         &         & Oph\\
   SAO 159459  & 140283 & 57.3 & $-28.55$ & 0.02 &        &
      &  &         &         & Gem\\
    $\nu$ Ser  & 156928 & 59.3 & $-27.70$ & 1.49 &        &
      &  &         &         & Mic\\
    HD 141569  & 141569 & 99.0 & $-28.70$ & 0.22 &        &
      &  &         &         & Gem\\
\enddata
\tablenotetext{a}{$\sigma = (|v_0 - v_{\star}|) / \sigma_v$, where we have imposed a minimum $\sigma_v$ of 1 \kms for all high resolution data, and a minimum $\sigma_v$ of 3 \kms for all medium resolution data.}
\tablenotetext{b}{Other dynamical clouds for which the observed velocity is within 3$\sigma$ of the predicted value and the sight line is closer than 10$^{\circ}$ to the cloud.  These are less likely, but possible dynamical assignments for these sight lines.}
\end{deluxetable}

\begin{deluxetable}{lcccccccccl}
\tablewidth{0pt}
\tabletypesize{\tiny}
\tablecaption{Blue Cloud Sight Line Properties\label{lism4_bluetable}}
\tablehead{Star & HD & $d$ & $v$ & $\sigma$\tablenotemark{a} & $\log N({\rm H~I})$ & $T$ & $\xi$ & $D($Fe$)$ & $D($Mg$)$ & Other\\ 
Name & \# & (pc) & (\kms) & & (cm$^{-2}$) & (K) & (\kms) & & & Clouds\tablenotemark{b}\\}
\startdata
$\alpha$ CMa A & 48915 &  2.6 & $ 12.70$ & 0.04 & 17.2 &
 3000$^{+2000}_{-1000}$ & 2.7 $\pm$ 0.3 & $-0.95^{+0.15}_{-0.21}$ & $-0.77^{+0.17}_{-0.22}$ & \\
$\alpha$ CMa B & 48915B &  2.6 & $ 11.70$ & 0.63 &        &
 3000$^{+2000}_{-1000}$ & 2.7 $\pm$ 0.3 &         &         & \\
       EP Eri  & 17925 & 10.4 & $  9.00$ & 0.65 &        &
      &  &         &         & LIC\\
  $\zeta$ Dor  & 33262 & 11.7 & $  8.41$ & 0.25 & 17.8 &
 7700$^{+2300}_{-2100}$ & 2.34$^{+0.38}_{-0.48}$ & $-0.52 \pm 0.30$ & $-0.05 \pm 0.27$ & G \\
      HR 2225  & 43162 & 16.7 & $ 14.00$ & 0.44 & 17.9 &
$<$10400 &  &         &         & LIC\\
  $\beta$ Pic  & 39060 & 19.3 & $ 10.12$ & 0.81 &        &
      &  &         &         & \\
      HR 2882  & 59967 & 21.8 & $  8.40$ & 0.47 & 18.5 &
$<$15100 &  &         &         & \\
 $\sigma$ Cet  & 15798 & 25.8 & $  9.74$ & 0.38 &        &
      &  &         &         & \\
  $\beta$ Car  & 80007 & 34.1 & $  3.57$ & 0.24 &        &
      &  &         &         & Vel, Cet\\
  $\alpha$ Car & 45348 & 95.9 & $  8.19$ & 0.58 &        &
      &  &         &         & Dor\\
\enddata
\tablenotetext{a}{$\sigma = (|v_0 - v_{\star}|) / \sigma_v$, where we have imposed a minimum $\sigma_v$ of 1 \kms for all high resolution data, and a minimum $\sigma_v$ of 3 \kms for all medium resolution data.}
\tablenotetext{b}{Other dynamical clouds for which the observed velocity is within 3$\sigma$ of the predicted value and the sight line is closer than 10$^{\circ}$ to the cloud.  These are less likely, but possible dynamical assignments for these sight lines.}
\end{deluxetable}

\begin{deluxetable}{lcccccccccl}
\tablewidth{0pt}
\tabletypesize{\tiny}
\tablecaption{Aql Cloud Sight Line Properties
\label{lism4_aqltable}}
\tablehead{Star & HD & $d$ & $v$ & $\sigma$\tablenotemark{a} & $\log N({\rm H~I})$ & $T$ & $\xi$ & $D($Fe$)$ & $D($Mg$)$ & Other\\ 
Name & \# & (pc) & (\kms) & & (cm$^{-2}$) & (K) & (\kms) & & & Clouds\tablenotemark{b}\\}
\startdata
     61 Cyg A  & 201091 &  3.5 & $ -3.00$ & 1.28 & 17.8 &
$6850 \pm 880$ & 2.08 $\pm$ 0.64 &         & $-0.95 \pm 0.22$ & LIC\\
 $\alpha$ Aql  & 187642 &  5.1 & $-20.90$ & 0.14 & 17.9 &
$12600 \pm 2400$ & 0.63$^{+0.90}_{-0.63}$ & $-0.96 \pm 0.54$ & $-0.69 \pm 0.55$ & G, Eri\\
       70 Oph  & 165341 &  5.1 & $-43.34$ & 0.92 & 17.1 &
$3300 \pm 2100$ & 2.31 $\pm$ 0.37 &         & $-0.57 \pm 0.15$ & \\
      HR 6748  & 165185 & 17.4 & $-29.20$ & 0.36 & 18.1 &
$<$15700 &  &         &         & G \\
  $\zeta$ Aql  & 177724 & 25.5 & $-30.20$ & 0.79 &        &
      &  &         &         & Oph\\
 $\lambda$ Aql & 177756 & 38.4 & $-30.70$ & 1.19 &        &
      &  &         &         & Mic\\
       BO Mic  & 197890 & 44.4 & $  0.00$ & 0.31 &        &
      &  &         &         & \\
 $\delta$ Cyg  & 186882 & 52.4 & $-18.80$ & 0.71 &        &
      &  &         &         & Mic\\
  $\alpha$ Del & 196867 & 73.8 & $-10.40$ & 1.01 &        &
      &  &         &         & \\
\enddata
\tablenotetext{a}{$\sigma = (|v_0 - v_{\star}|) / \sigma_v$, where we have imposed a minimum $\sigma_v$ of 1 \kms for all high resolution data, and a minimum $\sigma_v$ of 3 \kms for all medium resolution data.}
\tablenotetext{b}{Other dynamical clouds for which the observed velocity is within 3$\sigma$ of the predicted value and the sight line is closer than 10$^{\circ}$ to the cloud.  These are less likely, but possible dynamical assignments for these sight lines.}
\end{deluxetable}

\begin{deluxetable}{lcccccccccl}
\tablewidth{0pt}
\tabletypesize{\tiny}
\tablecaption{Eri Cloud Sight Line Properties 
\label{lism4_eritable}}
\tablehead{Star & HD & $d$ & $v$ & $\sigma$\tablenotemark{a} & $\log N({\rm H~I})$ & $T$ & $\xi$ & $D($Fe$)$ & $D($Mg$)$ & Other\\ 
Name & \# & (pc) & (\kms) & & (cm$^{-2}$) & (K) & (\kms) & & & Clouds\tablenotemark{b}\\}
\startdata
     61 Cyg A  & 201091 &  3.5 & $ -9.00$ & 0.27 & 17.8 &
$6850 \pm 880$ & 2.08 $\pm$ 0.64 &         & $-0.35 \pm 0.22$ & \\
 $\alpha$ Aql  & 187642 &  5.1 & $-17.10$ & 1.04 & 17.9 &
12300$^{+2000}_{-2200}$ & 0.0$^{+1.2}_{-0.0}$ & $-0.86 \pm 0.27$ & $-0.67 \pm 0.28$ & G \\
 $\lambda$ Aql & 177756 & 38.4 & $-21.90$ & 0.35 &        &
      &  &         &         & G \\
  $\alpha$ Peg & 218045 & 42.8 & $ -0.90$ & 0.12 &        &
      &  &         &         & LIC\\
     V376 Peg  & 209458 & 47.1 & $ -6.60$ & 0.00 & 18.4 &
$<$15900 &  &         &         & \\
$\upsilon$ Peg & 220657 & 53.1 & $  1.73$ & 0.13 & 17.9 &
 1700$^{+1100}_{-900}$ & 3.93 $\pm$ 0.22 & $-0.35 \pm 0.08$ & $ 0.01 \pm 0.12$ & LIC\\
  $\alpha$ Del & 196867 & 73.8 & $-14.00$ & 0.14 &        &
      &  &         &         & \\
 $\theta$ Aql  & 191692 & 88.0 & $-17.70$ & 0.15 &        &
      &  &         &         & Aql\\
\enddata
\tablenotetext{a}{$\sigma = (|v_0 - v_{\star}|) / \sigma_v$, where we have imposed a minimum $\sigma_v$ of 1 \kms for all high resolution data, and a minimum $\sigma_v$ of 3 \kms for all medium resolution data.}
\tablenotetext{b}{Other dynamical clouds for which the observed velocity is within 3$\sigma$ of the predicted value and the sight line is closer than 10$^{\circ}$ to the cloud.  These are less likely, but possible dynamical assignments for these sight lines.}
\end{deluxetable}

\begin{deluxetable}{lcccccccccl}
\tablewidth{0pt}
\tabletypesize{\tiny}
\tablecaption{Aur Cloud Sight Line Properties
\label{tab:aurmem}}
\tablehead{Star & HD & $d$ & $v$ & $\sigma$\tablenotemark{a} & $\log N({\rm H~I})$ & $T$ & $\xi$ & $D($Fe$)$ & $D($Mg$)$ & Other\\ 
Name & \# & (pc) & (\kms) & & (cm$^{-2}$) & (K) & (\kms) & & & Clouds\tablenotemark{b}\\}
\startdata
 $\alpha$ CMi  & 61421 &  3.5 & $ 23.00$ & 1.03 & 17.6 &
 6710$^{+660}_{-630}$ & 1.21$^{+0.35}_{-0.45}$ & $-1.13 \pm 0.07$ & $-0.79 \pm 0.10$ & \\
       LQ Hya  & 82558 & 18.3 & $ 14.00$ & 0.19 &        &
      &  &         &         & LIC, G\\
  $\alpha$ Tau & 29139 & 20.0 & $ 20.62$ & 1.29 & 17.8 &
      &  &         &         & LIC\\
    SAO 93801  & 26345 & 43.1 & $ 21.10$ & 0.51 &        &
      &  &         &         & LIC\\
    SAO 93973  & 28483 & 50.2 & $ 20.80$ & 0.31 &        &
      &  &         &         & LIC\\
 $\gamma$ Crv  & 106625 & 50.6 & $ -2.00$ & 0.32 &        &
      &  &         &         & \\
       45 Aur  & 43905 & 57.0 & $ 11.87$ & 0.60 &        &
      &  &         &         & \\
   $\eta$ Aur  & 32630 & 67.2 & $ 17.44$ & 0.87 &        &
      &  &         &         & \\
      HR 2324  & 45320 & 70.1 & $ 26.00$ & 1.18 &        &
      &  &         &         & \\
\enddata
\tablenotetext{a}{$\sigma = (|v_0 - v_{\star}|) / \sigma_v$, where we have imposed a minimum $\sigma_v$ of 1 \kms for all high resolution data, and a minimum $\sigma_v$ of 3 \kms for all medium resolution data.}
\tablenotetext{b}{Other dynamical clouds for which the observed velocity is within 3$\sigma$ of the predicted value and the sight line is closer than 10$^{\circ}$ to the cloud.  These are less likely, but possible dynamical assignments for these sight lines.}
\end{deluxetable}

\begin{deluxetable}{lccccccccccl}
\tablewidth{0pt}
\tabletypesize{\tiny}
\tablecaption{Hyades Cloud Sight Line Properties
\label{lism4_hyadestable}}
\tablehead{Star & HD & $d$ & $v$ & $\sigma$\tablenotemark{a} & $\log N({\rm H~I})$ & $T$ & $\xi$ & $D($Fe$)$ & $D($Mg$)$ & Other\\ 
Name & \# & (pc) & (\kms) & & (cm$^{-2}$) & (K) & (\kms) & & & Clouds\tablenotemark{b}\\}
\startdata
       EV Lac  & \nodata &  5.0 & $  7.30$ & 0.12 & 18.0 &
$<$8300 &  &         &         & LIC, Eri\\
$\kappa^1$ Cet & 20630 &  9.2 & $ 13.35$ & 1.11 & 17.5 &
 3600$^{+2900}_{-2200}$ & 2.17$^{+0.34}_{-0.52}$ & $-1.29 \pm 0.25$ & $-0.79 \pm 0.19$ & \\
  $\alpha$ Tri & 11443 & 19.7 & $ 13.65$ & 0.16 & 17.8 &
 8900$^{+3900}_{-3400}$ & 1.3$^{+1.7}_{-1.3}$ & $-0.97 \pm 0.23$ & $-0.58 \pm 0.15$ & LIC \\
      HR 1099  & 22468 & 29.0 & $ 14.80$ & 0.63 & 17.6 &
 8800$^{+800}_{-1100}$ & 0.0$^{+0.9}_{-0.0}$ & $<-0.82$ & $-1.36 \pm 0.0.09$ & \\
    SAO 93801  & 26345 & 43.1 & $ 13.60$ & 0.08 &        &
      &  &         &         & \\
   SAO 111879  & 28736 & 43.2 & $ 13.40$ & 0.19 &        &
      &  &         &         & \\
    SAO 76683  & 29419 & 44.2 & $ 12.30$ & 0.54 &        &
      &  &         &         & \\
     V993 Tau  & 28205 & 45.8 & $ 14.80$ & 0.84 &        &
      &  &         &         & \\
    SAO 93831  & 26784 & 47.4 & $ 15.50$ & 1.62 &        &
      &  &         &         & \\
    SAO 93885  & 27561 & 51.4 & $ 14.40$ & 0.86 &        &
      &  &         &         & \\
$\upsilon$ Peg & 220657 & 53.1 & $  8.80$ & 0.51 & 17.4 &
 1000$^{+1900}_{-1000}$ & 3.4$^{+0.61}_{-0.63}$6 & $-0.87 \pm 0.45$ & $-1.07 \pm 0.30$ & \\
       45 Aur  & 43905 & 57.0 & $  8.14$ & 0.84 &        &
      &  &         &         & \\
   $\eta$ Aur  & 32630 & 67.2 & $ 10.70$ & 0.02 &        &
      &  &         &         & \\
     G191-B2B  & \nodata & 68.8 & $  8.61$ & 0.74 & 17.4 &
 4400$^{+2800}_{-2400}$ & 3.27$^{+0.37}_{-0.39}$ & $ 0.05\pm 0.11$ & $ 0.87 \pm 0.54$ & \\
\enddata
\tablenotetext{a}{$\sigma = (|v_0 - v_{\star}|) / \sigma_v$, where we have imposed a minimum $\sigma_v$ of 1 \kms for all high resolution data, and a minimum $\sigma_v$ of 3 \kms for all medium resolution data.}
\tablenotetext{b}{Other dynamical clouds for which the observed velocity is within 3$\sigma$ of the predicted value and the sight line is closer than 10$^{\circ}$ to the cloud.  These are less likely, but possible dynamical assignments for these sight lines.}
\end{deluxetable}

\begin{deluxetable}{lcccccccccl}
\tablewidth{0pt}
\tabletypesize{\tiny}
\tablecaption{Mic Cloud Sight Line Properties\label{lism4_mictable}}
\tablehead{Star & HD & $d$ & $v$ & $\sigma$\tablenotemark{a} & $\log N({\rm H~I})$ & $T$ & $\xi$ & $D($Fe$)$ & $D($Mg$)$ & Other\\ 
Name & \# & (pc) & (\kms) & & (cm$^{-2}$) & (K) & (\kms) & & & Clouds\tablenotemark{b}\\}
\startdata
 $\alpha$ Aql  & 187642 &  5.1 & $-25.02$ & 0.22 & 17.5 &
12500$^{+2700}_{-2400}$ & 1.4$^{+0.7}_{-1.4}$ & $-0.61 \pm 0.23$ & $-0.38 \pm 0.24$ & \\
  $\alpha$ PsA & 216956 &  7.7 & $-10.64$ & 0.07 &        &
      &  &         &         & \\
  $\alpha$ Lyr & 172167 &  7.8 & $-19.40$ & 0.19 &        &
      &  &         &         & \\
       AU Mic  & 197481 &  9.9 & $-21.45$ & 0.05 & 18.2 &
$8700 \pm 1200$ & 4.30 $\pm$ 0.93 &         & $-0.55 \pm 0.19$ & \\
  $\alpha$ Oph & 159561 & 14.3 & $-26.23$ & 0.05 &        &
      &  &         &         & G \\
       99 Her  & 165908 & 15.7 & $-22.90$ & 0.50 &        &
      &  &         &         & \\
  $\alpha$ CrB & 139006 & 22.9 & $-17.40$ & 0.21 &        &
      &  &         &         & Oph, Leo\\
  $\beta$ Cet  & 4128 & 29.4 & $  1.63$ & 0.15 & 16.9 &
$12400 \pm 2800$ & 2.29 $\pm$ 0.44 &         & $ 0.23 \pm 0.09$ & \\
    SAO 68491  & 184499 & 32.0 & $-19.73$ & 0.74 &        &
      &  &         &         & \\
 $\lambda$ Aql & 177756 & 38.4 & $-26.50$ & 1.38 &        &
      &  &         &         & G \\
  $\beta$ Ser  & 141003 & 46.9 & $-20.70$ & 0.47 &        &
      &  &         &         & LIC, Oph, Leo\\
 $\lambda$ Oph & 148857 & 50.9 & $-24.80$ & 0.07 &        &
      &  &         &         & LIC, Oph\\
 $\delta$ Cyg  & 186882 & 52.4 & $-16.30$ & 0.27 &        &
      &  &         &         & \\
  $\iota$ Cap  & 203387 & 66.1 & $-20.48$ & 0.41 & 18.1 &
11700$^{+4100}_{-3600}$ & 3.82$^{+0.37}_{-0.44}$ & $-1.22 \pm 0.23$ & $-0.47 \pm 0.21$ & \\
 $\theta$ CrB  & 138749 & 95.3 & $-15.70$ & 0.29 &        &
      &  &         &         & Oph\\
\enddata
\tablenotetext{a}{$\sigma = (|v_0 - v_{\star}|) / \sigma_v$, where we have imposed a minimum $\sigma_v$ of 1 \kms for all high resolution data, and a minimum $\sigma_v$ of 3 \kms for all medium resolution data.}
\tablenotetext{b}{Other dynamical clouds for which the observed velocity is within 3$\sigma$ of the predicted value and the sight line is closer than 10$^{\circ}$ to the cloud.  These are less likely, but possible dynamical assignments for these sight lines.}
\end{deluxetable}

\begin{deluxetable}{lcccccccccl}
\tablewidth{0pt}
\tabletypesize{\tiny}
\tablecaption{Oph Cloud Sight Line Properties
\label{lism4_ophtable}}
\tablehead{Star & HD & $d$ & $v$ & $\sigma$\tablenotemark{a} & $\log N({\rm H~I})$ & $T$ & $\xi$ & $D($Fe$)$ & $D($Mg$)$ & Other\\ 
Name & \# & (pc) & (\kms) & & (cm$^{-2}$) & (K) & (\kms) & & & Clouds\tablenotemark{b}\\}
\startdata
       70 Oph  & 165341 &  5.1 & $-32.53$ & 1.30 & 17.5 &
 1700$^{+2100}_{-1700}$ & 3.3 $\pm$ 1.1 &         & $-0.84 \pm 0.34$ & \\
 $\gamma$ Ser  & 142806 & 11.1 & $-22.12$ & 0.28 &        &
      &  &         &         & Mic\\
  $\alpha$ Oph & 159561 & 14.3 & $-28.40$ & 1.17 &        &
      &  &         &         & G, Mic\\
       72 Her  & 157214 & 14.4 & $-25.95$ & 0.32 &        &
      &  &         &         & NGP\\
 $\gamma$ Oph  & 161868 & 29.1 & $-29.90$ & 0.62 &        &
      &  &         &         & G, Mic\\
    SAO 68491  & 184499 & 32.0 & $-27.85$ & 0.02 &        &
      &  &         &         & \\
\enddata
\tablenotetext{a}{$\sigma = (|v_0 - v_{\star}|) / \sigma_v$, where we have imposed a minimum $\sigma_v$ of 1 \kms for all high resolution data, and a minimum $\sigma_v$ of 3 \kms for all medium resolution data.}
\tablenotetext{b}{Other dynamical clouds for which the observed velocity is within 3$\sigma$ of the predicted value and the sight line is closer than 10$^{\circ}$ to the cloud.  These are less likely, but possible dynamical assignments for these sight lines.}
\end{deluxetable}

\begin{deluxetable}{lcccccccccl}
\tablewidth{0pt}
\tabletypesize{\tiny}
\tablecaption{Gem Cloud Sight Line Properties
\label{lism4_gemtable}}
\tablehead{Star & HD & $d$ & $v$ & $\sigma$\tablenotemark{a} & $\log N({\rm H~I})$ & $T$ & $\xi$ & $D($Fe$)$ & $D($Mg$)$ & Other\\ 
Name & \# & (pc) & (\kms) & & (cm$^{-2}$) & (K) & (\kms) & & & Clouds\tablenotemark{b}\\}
\startdata
  $\xi$ Boo A  & 131156 &  6.7 & $-17.69$ & 0.33 & 17.9 &
$5310 \pm 830$ & 1.68 $\pm$ 0.23 &         & $-0.92 \pm 0.10$ & Mic, NGP\\
  $\beta$ Gem  & 62509 & 10.3 & $ 31.84$ & 0.00 & 17.8 &
 6100$^{+3100}_{-2600}$ & 1.93$^{+0.79}_{-0.59}$ & $-1.29 \pm 0.12$ & $-1.19 \pm 0.14$ & \\
  $\alpha$ Boo & 124897 & 11.3 & $-13.89$ & 0.50 &        &
      &  &         &         & NGP\\
 $\delta$ Crv  & 108767 & 26.9 & $ -0.50$ & 0.76 &        &
      &  &         &         & Aur\\
      HR 4803  & 109799 & 34.6 & $ -0.37$ & 1.02 &        &
      &  &         &         & Aur\\
 $\sigma$ Gem  & 62044 & 37.5 & $ 32.26$ & 0.55 & 17.7 &
 7200$^{+1000}_{-1200}$ & 0.0$^{+1.1}_{-0.0}$ &         & $-1.18 \pm 0.14$ & \\
  $\beta$ Ser  & 141003 & 46.9 & $-23.30$ & 1.61 &        &
      &  &         &         & Oph, NGP\\
  $\beta$ Lib  & 135742 & 49.1 & $-23.60$ & 0.17 &        &
      &  &         &         & \\
 $\gamma$ Crv  & 106625 & 50.6 & $  1.60$ & 0.73 &        &
      &  &         &         & \\
    c$^2$ Cen  & 129685 & 63.5 & $-16.90$ & 0.36 &        &
      &  &         &         & \\
\enddata
\tablenotetext{a}{$\sigma = (|v_0 - v_{\star}|) / \sigma_v$, where we have imposed a minimum $\sigma_v$ of 1 \kms for all high resolution data, and a minimum $\sigma_v$ of 3 \kms for all medium resolution data.}
\tablenotetext{b}{Other dynamical clouds for which the observed velocity is within 3$\sigma$ of the predicted value and the sight line is closer than 10$^{\circ}$ to the cloud.  These are less likely, but possible dynamical assignments for these sight lines.}
\end{deluxetable}

\begin{deluxetable}{lcccccccccl}
\tablewidth{0pt}
\tabletypesize{\tiny}
\tablecaption{NGP Cloud Sight Line Properties
\label{lism4_ngptable}}
\tablehead{Star & HD & $d$ & $v$ & $\sigma$\tablenotemark{a} & $\log N({\rm H~I})$ & $T$ & $\xi$ & $D($Fe$)$ & $D($Mg$)$ & Other\\ 
Name & \# & (pc) & (\kms) & & (cm$^{-2}$) & (K) & (\kms) & & & Clouds\tablenotemark{b}\\}
\startdata
       61 Vir  & 115617 &  8.5 & $-16.50$ & 0.01 & 17.9 &
$<$8500 &  &         &         & Leo\\
  $\alpha$ Oph & 159561 & 14.3 & $-32.74$ & 0.84 &        &
      &  &         &         & \\
   $\chi$ Her  & 142373 & 15.9 & $-12.90$ & 0.76 & 18.2 &
$<$10200 &  &         &         & Mic\\
  $\iota$ Cen  & 115892 & 18.0 & $-18.20$ & 0.31 &        &
      &  &         &         & G\\
    SAO 28753  & 116956 & 21.9 & $  3.20$ & 0.57 & 18.2 &
$<$12600 &  &         &         & LIC\\
  $\iota$ Leo  & 99028 & 24.2 & $  4.99$ & 0.65 &        &
      &  &         &         & Leo\\
   $\eta$ UMa  & 120315 & 30.9 & $ -3.02$ & 0.11 & 17.9 &
 8900$^{+2500}_{-2300}$ & 1.34$^{+0.24}_{-0.31}$ & $-0.93 \pm 0.06$ & $-0.78 \pm 0.10$ & \\
  $\iota$ Dra  & 137759 & 31.3 & $ -6.81$ & 1.74 &        &
      &  &         &         & LIC\\
        HZ 43  & \nodata & 32.0 & $ -6.52$ & 0.40 & 18.0 &
 7500$^{+2100}_{-2000}$ & 1.7$^{+0.8}_{-1.7}$ & $-1.42 \pm 0.11$ & $-1.08 \pm 0.17$ & \\
$\alpha^2$ CVn & 112413 & 33.8 & $ -1.90$ & 0.77 &        &
      &  &         &         & LIC\\
      HR 4803  & 109799 & 34.6 & $-11.41$ & 0.55 &        &
      &  &         &         & \\
    c$^2$ Cen  & 129685 & 63.5 & $-26.30$ & 0.33 &        &
      &  &         &         & G\\
       GD 153  & \nodata & 70.5 & $ -5.04$ & 0.44 & 17.9 &
 7000$^{+2900}_{-2800}$ & 1.2$^{+1.1}_{-1.2}$ & $-1.18 \pm 0.13$ &         & \\
       31 Com  & 111812 & 94.2 & $ -3.37$ & 0.03 & 18.0 &
 8200$^{+1000}_{-1400}$ & 0.0$^{+1.0}_{-0.0}$ & $-0.85 \pm 0.15$ & $-0.98 \pm 0.14$ & \\
 $\theta$ CrB  & 138749 & 95.3 & $-19.90$ & 0.72 &        &
      &  &         &         & Oph, Gem\\
\enddata
\tablenotetext{a}{$\sigma = (|v_0 - v_{\star}|) / \sigma_v$, where we have imposed a minimum $\sigma_v$ of 1 \kms for all high resolution data, and a minimum $\sigma_v$ of 3 \kms for all medium resolution data.}
\tablenotetext{b}{Other dynamical clouds for which the observed velocity is within 3$\sigma$ of the predicted value and the sight line is closer than 10$^{\circ}$ to the cloud.  These are less likely, but possible dynamical assignments for these sight lines.}
\end{deluxetable}

\begin{deluxetable}{lcccccccccl}
\tablewidth{0pt}
\tabletypesize{\tiny}
\tablecaption{Leo Cloud Sight Line Properties
\label{lism4_leotable}}
\tablehead{Star & HD & $d$ & $v$ & $\sigma$\tablenotemark{a} & $\log N({\rm H~I})$ & $T$ & $\xi$ & $D($Fe$)$ & $D($Mg$)$ & Other\\ 
Name & \# & (pc) & (\kms) & & (cm$^{-2}$) & (K) & (\kms) & & & Clouds\tablenotemark{b}\\}
\startdata
  $\beta$ Leo  & 102647 & 11.1 & $  0.51$ & 0.05 &        &
      &  &         &         & NGP\\
      HR 4657  & 106516 & 22.6 & $ -3.40$ & 0.42 & 18.6 &
$<$27500 &  &         &         & Aur, Gem, NGP\\
  $\alpha$ Leo & 87901 & 23.8 & $ 10.50$ & 0.76 &        &
      &  &         &         & LIC\\
  $\iota$ Leo  & 99028 & 24.2 & $  1.97$ & 0.50 &        &
      &  &         &         & \\
      HR 4803  & 109799 & 34.6 & $ -7.52$ & 0.40 &        &
      &  &         &         & \\
  $\beta$ Ser  & 141003 & 46.9 & $-16.70$ & 1.05 &        &
      &  &         &         & LIC\\
  $\alpha$ Vir & 116658 & 80.4 & $-11.99$ & 0.80 &        &
      &  &         &         & \\
\enddata
\tablenotetext{a}{$\sigma = (|v_0 - v_{\star}|) / \sigma_v$, where we have imposed a minimum $\sigma_v$ of 1 \kms for all high resolution data, and a minimum $\sigma_v$ of 3 \kms for all medium resolution data.}
\tablenotetext{b}{Other dynamical clouds for which the observed velocity is within 3$\sigma$ of the predicted value and the sight line is closer than 10$^{\circ}$ to the cloud.  These are less likely, but possible dynamical assignments for these sight lines.}
\end{deluxetable}

\begin{deluxetable}{lcccccccccl}
\tablewidth{0pt}
\tabletypesize{\tiny}
\tablecaption{Dor Cloud Sight Line Properties
\label{lism4_dortable}}
\tablehead{Star & HD & $d$ & $v$ & $\sigma$\tablenotemark{a} & $\log N({\rm H~I})$ & $T$ & $\xi$ & $D($Fe$)$ & $D($Mg$)$ & Other\\ 
Name & \# & (pc) & (\kms) & & (cm$^{-2}$) & (K) & (\kms) & & & Clouds\tablenotemark{b}\\}
\startdata
  $\zeta$ Dor  & 33262 & 11.7 & $ 13.90$ & 0.35 & 18.1 &
 7000$^{+3500}_{-3000}$ & 5.47$^{+0.39}_{-0.41}$ & $-0.80 \pm 0.30$ & $-0.65 \pm 0.31$ & \\
 $\tau^6$ Eri  & 23754 & 17.9 & $ 41.40$ & 0.12 &        &
      &  &         &         & \\
$\epsilon$ Gru & 215789 & 39.7 & $ 12.20$ & 0.66 &        &
      &  &         &         & \\
  $\alpha$ Eri & 10144 & 44.1 & $ 21.20$ & 0.24 &        &
      &  &         &         & \\
\enddata
\tablenotetext{a}{$\sigma = (|v_0 - v_{\star}|) / \sigma_v$, where we have imposed a minimum $\sigma_v$ of 1 \kms for all high resolution data, and a minimum $\sigma_v$ of 3 \kms for all medium resolution data.}
\tablenotetext{b}{Other dynamical clouds for which the observed velocity is within 3$\sigma$ of the predicted value and the sight line is closer than 10$^{\circ}$ to the cloud.  These are less likely, but possible dynamical assignments for these sight lines.}
\end{deluxetable}

\begin{deluxetable}{lcccccccccl}
\tablewidth{0pt}
\tabletypesize{\tiny}
\tablecaption{Vel Cloud Sight Line Properties
\label{lism4_veltable}}
\tablehead{Star & HD & $d$ & $v$ & $\sigma$\tablenotemark{a} & $\log N({\rm H~I})$ & $T$ & $\xi$ & $D($Fe$)$ & $D($Mg$)$ & Other\\ 
Name & \# & (pc) & (\kms) & & (cm$^{-2}$) & (K) & (\kms) & & & Clouds\tablenotemark{b}\\}
\startdata
       AB Dor  & 36705 & 14.9 & $ 14.22$ & 0.11 &        &
      &  &         &         & Cet\\
   SAO 254993  & 254993 & 20.5 & $-13.10$ & 1.24 & 18.8 &
$<$23000 &  &         &         & LIC, G\\
  $\alpha$ Hyi & 12311 & 21.9 & $  9.80$ & 0.31 &        &
      &  &         &         & \\
 $\delta$ Vel  & 74956 & 24.4 & $ 11.80$ & 0.18 &        &
      &  &         &         & \\
  $\alpha$ Eri & 10144 & 44.1 & $ 11.00$ & 0.45 &        &
      &  &         &         & \\
       BO Mic  & 197890 & 44.4 & $-24.20$ & 0.28 & 18.3 &
$10600 \pm 2700$ & 3.48 $\pm$ 1.87 &         & $-0.03 \pm 0.72$ & Mic\\
  $\alpha$ Pav & 193924 & 56.2 & $-19.60$ & 0.63 &       &
      &  &         &         & \\
\enddata
\tablenotetext{a}{$\sigma = (|v_0 - v_{\star}|) / \sigma_v$, where we have imposed a minimum $\sigma_v$ of 1 \kms for all high resolution data, and a minimum $\sigma_v$ of 3 \kms for all medium resolution data.}
\tablenotetext{b}{Other dynamical clouds for which the observed velocity is within 3$\sigma$ of the predicted value and the sight line is closer than 10$^{\circ}$ to the cloud.  These are less likely, but possible dynamical assignments for these sight lines.}
\end{deluxetable}

\begin{deluxetable}{lcccccccccl}
\tablewidth{0pt}
\tabletypesize{\tiny}
\tablecaption{Cet Cloud Sight Line Properties
\label{tab:cetmem}}
\tablehead{Star & HD & $d$ & $v$ & $\sigma$\tablenotemark{a} & $\log N({\rm H~I})$ & $T$ & $\xi$ & $D($Fe$)$ & $D($Mg$)$ & Other\\ 
Name & \# & (pc) & (\kms) & & (cm$^{-2}$) & (K) & (\kms) & & & Clouds\tablenotemark{b}\\}
\startdata
 $\sigma$ Boo  & 128167 & 15.5 & $-12.90$ & 1.10 &        &
      &  &         &         & \\
 $\delta$ Vel  & 74956 & 24.4 & $ 15.60$ & 0.79 &        &
      &  &         &         & \\
  $\beta$ Cet  & 4128 & 29.4 & $  9.14$ & 0.53 & 18.5 &
$6300 \pm 2900$ & 1.31 $\pm$ 0.76 &         & $ 0.21 \pm 0.33$ & LIC\\
      HR 4023  & 88955 & 31.5 & $ 10.50$ & 0.48 &        &
      &  &         &         & \\
  $\alpha$ Eri & 10144 & 44.1 & $  7.60$ & 1.80 &        &
      &  &         &         & G, Vel\\
\enddata
\tablenotetext{a}{$\sigma = (|v_0 - v_{\star}|) / \sigma_v$, where we have imposed a minimum $\sigma_v$ of 1 \kms for all high resolution data, and a minimum $\sigma_v$ of 3 \kms for all medium resolution data.}
\tablenotetext{b}{Other dynamical clouds for which the observed velocity is within 3$\sigma$ of the predicted value and the sight line is closer than 10$^{\circ}$ to the cloud.  These are less likely, but possible dynamical assignments for these sight lines.}
\end{deluxetable}  

\begin{deluxetable}{lccccc}
\tablewidth{0pt}
\tabletypesize{\footnotesize}
\tablecaption{LISM Cloud Heliocentric Velocity Vectors\label{tab:vecs}}
\tablehead{Cloud & Number of & $V_0$ & $l_0$ & $b_0$ & $\chi_{\nu}^2$\\ 
Name & Sight Lines & (\kms) & ($^{\circ}$) & ($^{\circ}$) & }
\startdata
LIC    & 79 & 23.84 $\pm$ 0.90 & 187.0 $\pm$ 3.4 & --13.5 $\pm$ 3.3 & 2.2\\
G      & 21 & 29.6 $\pm$ 1.1 & 184.5 $\pm$ 1.9 & --20.6 $\pm$ 3.6 & 1.3\\
Blue   & 10 & 13.89 $\pm$ 0.89 & 205.5 $\pm$ 4.3 & --21.7 $\pm$ 8.3 & 2.4\\
Aql    & ~9 & 58.6 $\pm$ 1.3 & 187.0 $\pm$ 1.5 & --50.8 $\pm$ 1.0 & 2.6\\
Eri    & ~8 & 24.1 $\pm$ 1.2 & 196.7 $\pm$ 2.1 & --17.7 $\pm$ 2.6 & 0.3\\
Aur    & ~9 & 25.22 $\pm$ 0.81 & 212.0 $\pm$ 2.4 & --16.4 $\pm$ 3.6 & 2.1\\
Hyades & 14 & 14.69 $\pm$ 0.81 & 164.2 $\pm$ 9.4 & --42.8 $\pm$ 6.1 & 1.3\\
Mic    & 15 & 28.45 $\pm$ 0.95 & 203.0 $\pm$ 3.4 & --03.3 $\pm$ 2.3 & 0.5\\
Oph    & ~6 & 32.25 $\pm$ 0.49 & 217.7 $\pm$ 3.1 &  +00.8 $\pm$ 1.8 & 3.9\\
Gem    & 10 & 36.3 $\pm$ 1.1 & 207.2 $\pm$ 1.6 & --01.2 $\pm$ 1.3 & 1.7\\
NGP    & 15 & 37.0 $\pm$ 1.4 & 189.8 $\pm$ 1.7 & --05.4 $\pm$ 1.1 & 3.8\\
Leo    & ~7 & 23.5 $\pm$ 1.6 & 191.3 $\pm$ 2.8 & --08.9 $\pm$ 1.8 & 1.5\\
Dor    & ~4 & 52.94 $\pm$ 0.88 & 157.3 $\pm$ 1.5 & --47.93 $\pm$ 0.63 & 0.8\\
Vel    & ~7 & 45.2 $\pm$ 1.8 & 195.4 $\pm$ 1.1 & --19.1 $\pm$ 1.0 & 0.8\\
Cet    & ~5 & 60.0 $\pm$ 2.0 & 197.11 $\pm$ 0.56 & --08.72 $\pm$ 0.50 & 8.9\\
\tableline
LIC\tablenotemark{a} &
         ~9 & 25.7 $\pm$ 0.5 & 186.1             & --16.4          & \nodata\\
LIC\tablenotemark{b} &
         16 & 26 $\pm$ 1     & 186 $\pm$ 3       & --16 $\pm$ 3    & \nodata\\
LIC\tablenotemark{c} &
         63 & 24.20 $\pm$ 1.05 & 187.0 $\pm$ 3.1 & --13.5 $\pm$ 3.0 & 2.1\\
G\tablenotemark{a}   &
   \nodata  & 29.4           & 185.5             & --20.5          & \nodata\\
Helio\tablenotemark{d} &
   \nodata  & 26.24 $\pm$ 0.45 & 183.4 $\pm$ 0.4 & --15.9 $\pm$ 0.4 &\nodata\\
(LIC+G)/2\tablenotemark{e} &
   \nodata  & 26.74 $\pm$ 0.71 & 185.7 $\pm$ 3.4   & --16.95 $\pm$ 3.6 &\nodata\\
\enddata
\tablenotetext{a}{\citet{lallement92}}
\tablenotetext{b}{\citet{lallement95}}
\tablenotetext{c}{{LIC} flow vector deleting the 16 lines of sight near the
decelerated leading edge of the LIC in the direction of the Hyades Cloud.}
\tablenotetext{d}{Flow vector for interstellar helium gas in the heliosphere.
Temperature is 6303 $\pm$ 390~K: \citet{mobius04}.  See temperatures for individual dynamical clouds in Table~\ref{tab:sumprop}.}
\tablenotetext{e}{Average of the LIC and G vectors. Average temperature of 
the LIC and G Clouds is 6500 $\pm$ 680~K.  The {\it in situ} ``Helio'' measurement is closer to the average LIC and G temperature, than either cloud individually, see Table~\ref{tab:sumprop}.}
\end{deluxetable}  

\begin{deluxetable}{lccccccccl}
\tablewidth{0pt}
\tabletypesize{\tiny}
\tablecaption{Unassigned Sight Line Properties
\label{tab:unass}}
\tablehead{Star & HD & $d$ & $v$ & $\log N({\rm H~I})$ & $T$ & $\xi$ & $D($Fe$)$ & $D($Mg$)$ & Possible\\ 
Name & \# & (pc) & (\kms) & (cm$^{-2}$) & (K) & (\kms) & & & Clouds\tablenotemark{a}\\}
\startdata
             AD Leo &           &    4.7 &  13.13 & 18.5 & $<$12300 &      &       &        & LIC, Leo\\
       $\eta$ Cas A &       4614&    6.0 &   5.80 &       &         &      &       &        & \\
     $\kappa^1$ Cet &      20630&    9.2 &   7.36 & 17.4 &  5800 $\pm$ 2700 & 1.48 $\pm$ 0.92 &       & $-1.51 \pm 0.23$  & \\
        $\beta$ Leo &     102647&   11.1 &  11.80 &       &         &      &       &        & \\
             AB Dor &      36705&   14.9 &  19.38 &       &         &      &       &        & \\
       $\tau^6$ Eri &      23754&   17.9 &  27.93 &       &         &      &       &        & \\
       $\tau^6$ Eri &      23754&   17.9 &   8.82 &       &         &      &       &        & \\
             PW And &       1405&   21.9 &   2.00 &       &         &      &       &        & LIC, Eri\\
         SAO 158720 &     128987&   23.6 & --22.00 & 18.1 & $<$12400 &      &       &        & G, Gem, NGP, Leo\\
       $\delta$ Her &     156164&   24.1 & --19.50 &       &         &      &       &        & LIC\\
          SAO 32862 &     198084&   27.1 & --12.88 &       &         &      &       &        & \\
            HR 1099 &      22468&   29.0 &   8.20 & 17.2 &  7100 $\pm$ 1400 & 2.30 $\pm$ 0.25 &       & $-1.21 \pm 0.12$  & \\
       $\gamma$ Oph &     161868&   29.1 & --33.00 &       &         &      &       &        & Oph, NGP\\
       $\alpha$ And &        358&   29.8 &  13.00 &       &         &      &       &        & Hyades\\
         $\eta$ UMa &     120315&   30.9 &   2.60 & 16.6 &     0$^{+4400}_{-0}$ & 5.6$^{+0.9}_{-1.1}$ &       & $-0.78 \pm 0.15$  & NGP\\
       $\alpha$ Gru &     209952&   31.1 &  --6.40 &       &         &      &       &        & LIC, Vel \\
       $\alpha$ Gru &     209952&   31.1 & --21.40 &       &         &      &       &        & LIC, Vel\\
       $\sigma$ Gem &      62044&   37.5 &  21.77 & 17.9 &  8600 $\pm$ 1600 & 2.46 $\pm$ 0.45 &       & $-0.92 \pm 0.14$  & LIC\\
            HR 2298 &      44769&   39.4 &   7.00 &       &         &      &       &        & \\
     $\epsilon$ Gru &     215789&   39.7 &   6.80 &       &         &      &       &        & \\
     $\epsilon$ Gru &     215789&   39.7 &  --1.70 &       &         &      &       &        & \\
     $\epsilon$ Gru &     215789&   39.7 & --12.30 &       &         &      &       &        & \\
     $\epsilon$ Gru &     215789&   39.7 & --22.40 &       &         &      &       &        & \\
          SAO 93981 &      28568&   41.2 &  16.50 &       &         &      &       &        & Hyades\\
         SAO 111879 &      28736&   43.2 &  --4.30 &       &         &      &       &        & \\
          SAO 93982 &      28608&   43.6 &  15.80 &       &         &      &       &        & Hyades\\
       $\alpha$ Eri &      10144&   44.1 &  18.90 &       &         &      &       &        & Dor\\
          SAO 93945 &      28237&   47.2 &  15.60 &       &         &      &       &        & Hyades\\
        $\beta$ Lib &     135742&   49.1 & --33.70 &       &         &      &       &        & \\
          SAO 94162 &      30738&   51.8 &  20.30 &       &         &      &       &        & Aur\\
       $\kappa$ And &     222439&   52.0 &   0.80 &       &         &      &       &        & \\
     $\upsilon$ Peg &     220657&   53.1 &  --7.48 & 17.4 &  3600$^{+4400}_{-3000}$ & 1.7$^{+0.7}_{-1.5}$ & $-0.83 \pm 0.2$ & $-0.54 \pm 0.21$  & \\
          SAO 93913 &      27848&   53.4 &  16.40 &       &         &      &       &        & Aur, Hyades\\
            HR 1608 &      32008&   54.7 &   6.10 & 17.7 & $<$11500 &      &       &        & Blue, Hyades\\
       $\alpha$ Pav &     193924&   56.2 & --18.60 &       &         &      &       &        & Vel\\
         $\eta$ Aqr &     213998&   56.3 &  --2.10 &       &         &      &       &        & LIC\\
         SAO 159459 &     140283&   57.3 & --22.48 &       &         &      &       &        & Gem, Leo\\
        $\iota$ Cap &     203387&   66.1 &  --2.22 & 18.4 &  5500$^{+10600}_{-5500}$ & 3.7$^{+0.7}_{-1.1}$ & $-0.63 \pm 0.37$ & $-0.28 \pm 0.38$  & \\
         $\eta$ Aur &      32630&   67.2 &   6.80 &       &         &      &       &        & \\
        $\iota$ Oph &     152614&   71.7 & --11.00 &       &         &      &       &        & \\
        $\iota$ Oph &     152614&   71.7 & --30.80 &       &         &      &       &        & NGP\\
           Feige 24 &           &   74.4 &   3.10 & 18.2 &         &      &       &        & \\
       $\gamma$ Ori &      35468&   74.5 &  16.00 &       &         &      &       &        & \\
       $\gamma$ Ori &      35468&   74.5 &  20.20 &       &         &      &       &        & \\
       $\gamma$ Ori &      35468&   74.5 &  27.80 &       &         &      &       &        & \\
       $\alpha$ Vir &     116658&   80.4 &  --5.40 &       &         &      &       &        & Gem\\
       $\alpha$ Vir &     116658&   80.4 &  --3.68 &       &         &      &       &        & \\
       $\theta$ Aql &     191692&   88.0 & --22.60 &       &         &      &       &        & Mic\\
         $\tau$ Her &     147394&   96.4 & --33.50 &       &         &      &       &        & \\
         $\tau$ Her &     147394&   96.4 & --38.00 &       &         &      &       &        & \\
          HD 141569 &     141569&   99.0 & --13.70 &       &         &      &       &        & \\
\enddata
\tablenotetext{a}{Although they do not fit the criteria for membership in any of the dynamical clouds, we list here possible membership where the observed velocity is within 3$\sigma$ of a predicted cloud value and the sight line is closer than 10$^{\circ}$ to the cloud. }
\end{deluxetable}  

\clearpage
\begin{deluxetable}{lcccccccccccccc}
\rotate
\tablewidth{0pt}
\tabletypesize{\tiny}
\tablecaption{Summary of Cloud Properties\label{tab:sumprop}}
\tablehead{Cloud & \# of & \multicolumn{2}{c}{Central Coord.} &
Closest & Surface Area & $\langle T \rangle$ & \# of & $\langle \xi \rangle$ & \# of & $\langle D({\rm Fe}) \rangle$ & \# of & $\langle D({\rm Mg}) \rangle$ & \# of & Morphology \\
Name & Sight Lines & $l(^{\circ})$ & $b(^{\circ})$ & Star (pc) & (sq. deg) & (K) & Stars &  (km s$^{-1}$) & Sight Lines & & Sight Lines & & Sight Lines & \\}
\startdata
LIC    & 79 & 170 & $-10$ & ~2.6 & 18270 & ~7500 $\pm$ 1300 & 19 & 1.62 $\pm$ 0.75 & 19 & --1.12 $\pm$ 0.10 & 12 & --0.97 $\pm$ 0.23 & 21 & compact \\
G      & 21 & 315 & $+00$ & ~1.3 & 8230 & ~5500 $\pm$ 400 & ~5 & 2.2 $\pm$ 1.1 & ~5 & --0.54 $\pm$ 0.11 & 4 & --0.36 $\pm$ 0.35 & ~5 & compact \\
Blue   & 10 & 250 & $-30$ & ~2.6 & 2310 & ~3900 $\pm$ 2300 & ~3 & 2.64 $\pm$ 0.16 & ~3 & --0.84 $\pm$ 0.27 & 2 & --0.51 $\pm$ 0.49 & ~2  & compact \\
Aql    & ~9 & ~40 & $-05$ & ~3.5 & 2960 & ~7000 $\pm$ 2800 & ~3 & 2.07 $\pm$ 0.64 & ~3 & (--0.96)\tablenotemark{a} & 1 & --0.69 $\pm$ 0.21 & ~3 & compact \\
Eri    & ~8 & ~70 & $-20$ & ~3.5 & 1970 & ~5300 $\pm$ 4000 & ~3 & 3.6 $\pm$ 1.0 & ~3 & --0.39 $\pm$ 0.19 & 2 & --0.15 $\pm$ 0.30 & ~3 & compact \\
Aur    & ~9 & 210 & $+10$ & ~3.5 & 1640 & ~(6710)\tablenotemark{a} & ~1 & (1.2)\tablenotemark{a} & ~1 &
 (--1.13)\tablenotemark{a} & 1 & (--0.79)\tablenotemark{a} & ~1 & filamentary \\
Hyades & 14 & 180 & $-20$ & ~5.0 & 1810 & ~6200 $\pm$ 3800 & ~5 & 2.7 $\pm$ 1.2 & ~5 & --0.32 $\pm$ 0.62 & 4 & --1.06 $\pm$ 0.47 & ~5 & filamentary \\
Mic    & 15 & ~40 & $+15$ & ~5.1 & 3550 & ~9900 $\pm$ 2000 & ~4 & 3.1 $\pm$ 1.0 & ~4 & --0.92 $\pm$ 0.43 & 2 & --0.03 $\pm$ 0.40 & ~4 & filamentary \\
Oph    & ~6 & ~45 & $+25$ & ~5.1 & 1360 & ~(1700)\tablenotemark{a} & ~1 & (3.3)\tablenotemark{a} & ~1 &
 \nodata & 0 & (--0.84)\tablenotemark{a} & ~1  & compact \\
Gem    & 10 & 300 & $+40$ & ~6.7 & 3300 & ~6000 $\pm$ 1100 & ~3 & 1.63 $\pm$ 0.41 & ~3 & (--1.29)\tablenotemark{a} & 1 & --1.05 $\pm$ 0.16 & ~3 & filamentary \\
NGP    & 15 & ~~5 & $+75$ & ~8.5 & 4020 & ~8000 $\pm$ 600 & ~4 & 1.23 $\pm$ 0.43 & ~4 & --1.04 $\pm$ 0.23 & 4 & --0.89 $\pm$ 0.15 & ~3 & compact \\
Leo    & ~7 & 270 & $+55$ & 11.1 & 2400 & \nodata & ~0 & \nodata & ~0 & \nodata  & ~0 & \nodata & ~0  & compact \\
Dor    & ~4 & 270 & $-50$ & 11.7 & 1550 & ~(7000)\tablenotemark{a} & ~1 & (5.5)\tablenotemark{a} & ~1 & 
(--0.80)\tablenotemark{a} & 1 & (--0.65)\tablenotemark{a} & ~1 & compact \\
Vel    & ~7 & 300 & $-45$ & 14.9 & 2190 & (10600)\tablenotemark{a} & ~1 & (3.5)\tablenotemark{a} & ~1 & 
\nodata & 0 & (--0.03)\tablenotemark{a} & ~1 & compact\\
Cet    & ~5 & 290 & $-40$ & 15.5 & 2270 & ~(6300)\tablenotemark{a} & ~1 & (1.3)\tablenotemark{a} & ~1 & 
\nodata & 0 & (0.21)\tablenotemark{a} & ~1 & filamentary \\

\enddata
\tablenotetext{a}{Those clouds with only one sight line with a physical measurement are indicated in parentheses.  Since a weighted average is not possible, the listed properties should be considered uncertain.}

\end{deluxetable}

\end{document}